\documentclass[desactivate]{aa}  

\usepackage{graphicx}
\usepackage{float}
\usepackage{natbib}
\bibpunct{(}{)}{;}{a}{}{,}
\usepackage{txfonts}

\usepackage{color}
\usepackage{multirow}
\usepackage{caption} 
\usepackage{lscape}
\usepackage{placeins}
\usepackage{gensymb}
\usepackage{enumitem}
\defcitealias{alvarez25}{A25}

\usepackage[colorlinks=true, linkcolor=blue, citecolor=blue, urlcolor=blue]{hyperref}

\begin{document} 

\titlerunning{Cosmography with DESI-DR1 Cosmic Chronometers}
\authorrunning{C. A. Álvarez et al.}

   \title{Cosmography with DESI-DR1 Cosmic Chronometers:\\
   Direct $H(z)$ measurements from Luminous Red Galaxy ages}

   \author{Carlos A. Álvarez \inst{1}
          \and
          Marcos M. Cueli \inst{2}
          \and
          Balakrishna S. Haridasu \inst{1, 3}
          \and
          Michele Moresco \inst{4, 5}
          \and
          Martina Torsello \inst{1, 6}
          \and
          Alessandro Bressan \inst{1, 7}
          \and
          Lumen Boco \inst{8}
          \and
          Luigi Danese \inst{1}
          \and
          Andrea Lapi \inst{1,3,6,9}
          }

   \institute{Scuola Internazionale Superiore di Studi Avanzati (SISSA), Via Bonomea 265, 34136 Trieste (TS), Italy\\
              \email{calonsoa@sissa.it}
         \and
    Facultad de Ciencias, Universidad de Oviedo, Leopoldo Calvo Sotelo, 18, 33007 Oviedo, Asturias, Spain
         \and
    Institute for Fundamental Physics of the Universe (IFPU), Via Beirut 2, 34014 Trieste, Italy
        \and
    Dipartimento di Fisica e Astronomia “Augusto Righi”–Università di Bologna, via Piero Gobetti 93/2, I-40129 Bologna, Italy
        \and
    INAF - Osservatorio di Astrofisica e Scienza dello Spazio di Bologna, via Piero Gobetti 93/3, I-40129 Bologna, Italy
        \and
    IRA-INAF, Via Gobetti 101, 40129 Bologna, Italy
        \and
    Purple Mountain Observatory, Chinese Academy of Sciences, Nanjing 210023, People’s Republic of China
        \and
    Universitat Heidelberg, Zentrum fur Astronomie, Institut fur theoretische Astrophysik, Albert-Ueberle-Str. 3, 69120 Heidelberg, Germany
        \and
    INFN-Sezione di Trieste, via Valerio 2, 34127 Trieste, Italy
        }

\abstract
   {With robust redshifts for almost 3 million luminous red galaxies (LRGs), the Dark Energy Spectroscopic Instrument (DESI) offers a unique opportunity to test the expansion rate of the Universe with independent approaches.}
   {We apply the cosmic chronometer (CC) method to the DESI LRGs to derive new, cosmology-independent constraints on the Hubble parameter at intermediate redshifts from the differential age evolution of passively evolving galaxies.}
   {We selected the spectra through catalogue-level and spectroscopic cuts that ensure the purity of the sample, and stacked them to obtain stable, high signal-to-noise (S/N) spectra, a procedure that also provides a democratic binning of the $t-z$ plane. We measured Lick indices on the stacks and fitted them with a theoretical stellar population synthesis (SPS) model to estimate ages. From the resulting $t-z$ relations, we derived $H(z)$ with two independent approaches: a pivotal-redshift cosmography and the original CC finite-difference approach.}
   {The cosmographic fit yields kinematic parameters, $\{H_{z_0}, q_{z_0}, j_{z_0}\}$, compatible with currently favoured cosmologies and a competitive constraint on $H(z)$ around the pivotal redshift. We provide the maximum-a-posteriori (MAP) $H(z)$ reconstruction, the median of the credible region sampled over $0.36\leq z\leq 0.80$, and its covariance matrix. Exploiting the different redshift distributions of the velocity dispersion groups, we also obtain two independent finite-difference measurements, $H(z) \approx - \Delta z/[(1+z)\Delta t]$. The one from the \textit{reddest} envelope of CCs reads $H(z \approx 0.64) = 87.5 ^{+11.4}_{-7.2} \text{ (stat.) } \pm 14.3 \text{ (syst.) km s}^{-1} \text{ Mpc}^{-1}$. The systematic uncertainties are estimated directly from all the methodological choices of the data treatment, to which we add an estimated budget for those that could not be directly assessed.}
   {Lick-index ages of stacked DESI LRG spectra provide competitive, cosmology-independent constraints on $H(z)$. Applying this approach at intermediate redshifts requires accounting for several limitations of current Lick-index SPS models. We address them through redshift- and quality-based selections, whose impact is included in our systematic budget, and further developments of these models will help to fully exploit forthcoming CC datasets.}

   \keywords{Cosmography --
                Cosmology with Early-Type Galaxies --
                Stellar Ages -- Expansion Rate of the Universe
               }

   \maketitle

\section{Introduction}

In the quest for independent measurements of the Hubble parameter, cosmic chronometers (CC) have emerged as a promising way to probe the expansion history of the Universe without assuming a specific cosmological model. By dating well-defined groups of passively evolving galaxies at different redshifts \citep{jimenez02}, CCs trace the time evolution of the Universe. These measurements have been widely used to test alternative cosmologies \citep{zheng16, verde17, leaf17, sultana22, lozanotorres23, gonzalez23, gomezvalent24}, to refine the distance ladder calibration \citep{favale23, favale24, favale26} and to explore new explanations for the Hubble tension \citep{ruchika26, peronaci26}.

The approach relies on the accurate identification of massive elliptical galaxies \citep{shimasaku01, strateva01, baldry04, bell04, sanchezblazquez09, ilbert13} that formed in short star-formation bursts at high redshift and have evolved passively ever since, the so-called downsizing scenario \citep{COWIE96, KAUFFMANN03, GALLAZZI05, THOMAS05, thomas10, johansson12, CONROY14}. If the redshift interval between the groups is small enough, the finite-difference approximation, $H(z) \approx -\Delta z /\Delta t / (1+z)$, provides a reliable independent measurement. This has been the framework of most studies to date, which populate the Hubble diagram up to $z\approx 2$. The most recent constraints have come from full spectral fitting (FSF) \citep{simon05, stern10, zhang14, ratsimbazafy17, jiao22, tomasetti23, tomasetti26}, which fits the whole spectrum with synthetic stellar populations, and from the D$4000$/D$_\text{n}4000$ feature \citep{moresco12, moresco15, moresco16a, loubser25, loubser25b, lisiecki26}, which does not provide absolute ages but directly constrains $dt/dz$ through its differential behaviour.

Lick indices, a well-calibrated set of absorption features sensitive to age, metallicity and chemical abundances \citep{worthey92, worthey94a, worthey97}, offer an alternative that focuses on the features characteristic of old, passively evolving populations and isolates spurious features in other parts of the spectrum. They were previously used to support the downsizing scenario \citep{clemens06, clemens09, carson10}, and \cite{borghi22a} and \cite{alvarez25} (hereafter \citetalias{alvarez25}) were the first to use them to derive CC ages. Methods combining photometry with the well-understood spectra of a training set through machine learning \citep{jimenez23} have also been proposed for cases where good spectra are not available.

Fitting either full spectra or a set of indices requires a stellar population synthesis (SPS) model. Several SPS models are widely used \citep{BRUZUAL03, thomas03, conroy10, maraston11, vazdekis15}, which differ in their ability to constrain age, metallicity, chemical abundances, dust content or star formation history. For Lick index studies, the \citet{thomas11} model (hereafter TMJ) is regarded as one of the most complete, although recent models, such as $\alpha$-MC \citep{park25} or that of \cite{knowles23}, could complement it.

Stacked spectra are commonly used to enhance the signal-to-noise ratio (S/N) and the information extracted from large datasets such as SDSS or DESI \citep{labarbera13, fitzpatrick15, eftekhari19}. \citetalias{alvarez25} showed an additional virtue of stacking in the CC context: it provides a democratic redshift binning. This choice arises when the data span a large redshift range rather than the ideal \textit{two ensembles of passively evolving galaxies at somewhat different redshifts} of \cite{jimenez02}, and it is a source of systematic uncertainty in \cite{borghi22b} and \cite{jiao22}.

In this work, we refine the stacking methodology of \citetalias{alvarez25} and apply it to the Dark Energy Spectroscopic Instrument Data Release 1 \citep[][DESI-DR1]{ADAME25}. Its Luminous Red Galaxy (LRG) sample \citep{zhou23}, with nearly 3 million objects with robust redshifts and wide spectral coverage, is an ideal CC target, expected to yield the most stringent CC results at $0.5<z<1$ \citep{moresco22}, where statistical uncertainties become subdominant with respect to systematic ones. DESI-DR1 was first exploited for CCs by \cite{loubser25, loubser25b} with the $\text{D}4000$ feature: from $\sim 360\,000$ ETGs, they obtained three $H(z)$ measurements at $0.3<z<1.0$, which highlights the large data cost of these studies. We follow an independent approach, estimating ages from Lick indices and performing a direct cosmographic analysis of the $t-z$ relations, which provides a set of $H(z)$ measurements with their full covariance, evenly populating $0.36<z<0.80$, independently of cosmological assumptions. Whereas in \citetalias{alvarez25} the cosmographic expansion \citep{capozziello12} was anchored at $z=0$, which was reasonable for SDSS, here we adopt the pivotal-redshift approach of \cite{fazzari25}, suitable for the DESI LRG redshift window and for future $z$-extended CC datasets. This work thus combines the DESI redshift coverage, Lick-index ages and a pivotal-redshift cosmography, together with the release of the corresponding covariance. Aware of the interest of the community in statistically independent $H(z)$ estimates, we also provide two local, uncorrelated measurements from the slope of the $t-z$ relations of the intermediate ($200 < \sigma \ [\text{km s}^{-1}] < 280$) and high ($280 < \sigma \ [\text{km s}^{-1}] < 355$) velocity dispersion groups. Since they come from physically different objects, they should not be regarded as a redshift evolution of $H$.

The paper is structured as follows. Section \ref{sec:CC} introduces the cosmographic framework. Section \ref{sec:Sample} details the sample selection, including the redshift cut applied to each velocity dispersion group to account for the response of the TMJ model at low ages, which we call the archaeological coherence cut. Section \ref{sec:DATATREATMENT} describes the measurement of spectral features for the almost $2$ million LRGs that pass our catalogue-level selection, the stacking, the SPS fit, the quality flags, the cosmographic fit, and the assessment of systematic uncertainties. Section \ref{sec:RESULTS} presents the main results: the cosmographic reconstruction of $H(z)$, sampled from $z = 0.36$ to $0.80$ with its covariance, and the two direct, independent $H(z)$ measurements. Section \ref{sec:conclusions} discusses the limitations of the methodology employed and summarises our conclusions.

\section{How to measure $H$: pivotal cosmography\label{sec:CC}}

The most common way to analyse CC data is the finite-difference approximation of the Hubble parameter,
\begin{equation}
    H\equiv \frac{1}{a}\frac{\text{d}a}{\text{d}t_{\text{\scriptsize{U}}}}=-\frac{1}{1+z}\frac{\text{d}z}{\text{d}t_{\text{\scriptsize{U}}}} \approx -\frac{1}{1+z}\frac{\Delta z}{\Delta t},
    \label{eq:Hubble}
\end{equation}
which holds for short redshift intervals. Here, $t_U$ denotes the cosmic time, whereas $t$ denotes stellar ages ($t_v$ for each velocity dispersion group later in the text). Beyond this approximation, a cosmographic expansion of the Hubble parameter \citep{cattoen07, capozziello12, gruber14, dunsby16} provides a continuous reconstruction of $H(z)$ without assuming a cosmological model, a midpoint between the finite-difference approximation and parametric cosmological fits, which \citetalias{alvarez25} applied to SDSS CCs.

In this work, we expand $H(z)$ around a pivotal redshift, $z_0$, instead of $z=0$, trading a direct measurement of $H_0$ for precision on $H(z)$ around $z_0$; this approximation reliably reproduces $H(z)$ in the vicinity of any pivotal redshift \citep{bargiacchi21, fazzari25}. We start from the age difference, $\Delta t$ \citep{jimenez02}, between groups of passively evolving stellar populations at different redshifts, obtained by inverting and integrating eq. (\ref{eq:Hubble}),
\begin{equation}
    \Delta t = t_{z_1}-t_{z_2} = \int_{z_1} ^{z_2} \frac{1}{1+z}\frac{\text{d}z}{H(z)} \equiv \int_{z_1}^{z_2} f(z)\,\text{d}z, \qquad z_2 > z_1,
    \label{eq:DtdezdeHz}
\end{equation}
where $f(z) \equiv [(1+z)H(z)]^{-1}$. We Taylor-expand $H(z)$ and $f(z)$ around $z_0$,
\begin{align}
    H(z) &= \sum_{n=0}^{\infty} a_n \, (z-z_0)^n, \qquad a_n \equiv \frac{H^{(n)}(z_0)}{n!}, \label{eq:Han}\\
    f(z) &= \sum_{m=0}^{\infty} b_m \, (z-z_0)^m , \qquad b_m \equiv \frac{f^{(m)}(z_0)}{m!}, \label{eq:fbm}
\end{align}
where $^{(n)}$ denotes the $n$-th derivative with respect to $z$, so that eq.~(\ref{eq:DtdezdeHz}) integrates to
\begin{equation}
    \Delta t = \sum_{m=0}^{\infty} \frac{b_m}{m+1} \, (z-z_0)^{m+1} .
    \label{eq:Dtbm}
\end{equation}
The coefficients $a_n$ are related to the deceleration and jerk parameters \citep{visser04, fazzari25},
\begin{equation}
    q(z) = -1+(1+z)\frac{H'(z)}{H(z)}, \qquad j(z) = q^2(z) + (1+z)^2\frac{H''(z)}{H(z)},
    \label{eq:qyj}
\end{equation}
where primes denote derivatives with respect to $z$. The full derivation is given in Appendix \ref{app:theoretical_cosmography}, although it can also be read from \cite{fazzari25}. That work addresses other observables (in particular the luminosity distance, $D_L$), so we present our own for the specific case of $t(z)$. Introducing
\begin{equation}
    x = (z-z_0)/(1+z_0),
    \label{eq:ztox}
\end{equation}
$H$ takes the convenient form
\begin{equation}
    H(x) = H_{z_0}\left(1 + (1+q_{z_0})x + \frac{1}{2}(j_{z_0}-q^2_{z_0})x^2+\mathcal{O}(x^3)\right),
    \label{eq:Hexplicit}
\end{equation}
and, specialising eq. (\ref{eq:DtdezdeHz}) to $z_1=z_0$ and $z_2=z$, $\Delta t(z,z_0) \equiv t(z_0) - t(z)$ is given through $G(x;q_{z_0},j_{z_0}) \equiv H_{z_0}\,\Delta t(z,z_0)$, which isolates the cosmological evolution of differential ages:
\begin{equation}
    \begin{aligned}
    G(x; q_{z_0}, j_{z_0}) = \ & x -(1+q_{z_0}/2) x^2 \\
     \ &+\left(1 + q_{z_0} + \frac{1}{2}q^2_{z_0} - \frac{1}{6}j_{z_0}\right)x^3 + \mathcal{O}\left(x^4\right).
    \end{aligned}
    \label{eq:Dtzz0}
\end{equation}

The notation differs from that of \citetalias{alvarez25} only because the expansion point moves from $z=0$ to $z_0$: for $z_0=0$, $x=z$, $\{H_{z_0}, q_{z_0}, j_{z_0}\}\to\{H_0, q_0, j_0\}$, and eq.~(\ref{eq:Dtzz0}) reduces to the standard expansion of the lookback time used in that work. We tested eq. (\ref{eq:Hexplicit}) against fiducial cosmologies, finding residuals below $1\%$ over $0.4<z<0.8$ for a pivotal redshift within that range. This agrees with \cite{fazzari25}, the only available assessment of this approximation to our knowledge, who found, for flat $\Lambda$CDM, truncation errors over $0.3 \lesssim z \lesssim 1.0$ typically at the $\sim 0.1\%$ level, and below $1\%$ for any pivot within the interval, with the best estimates of $H(z)$ for pivots at the centre of the data. We therefore adopt as $z_0$ the median redshift of the final dataset (shown in Fig. \ref{fig:tzmain}).

\section{Sample selection \label{sec:Sample}}

A rigorous selection of massive, passively evolving galaxies is essential to minimise contamination by star-forming outliers, which can bias the stellar ages and the derived cosmological parameters \citep{moresco18}. Combinations of morphological \citep{shimasaku01, clemens06}, spectroscopic \citep{dressler83, dressler92, moresco11, borghi22a, tomasetti23}, and photometric \citep{strateva01, masters11, white11, ilbert13, maraston13, mclure18} cuts have been proposed to maximise the purity of such samples. The DESI LRG selection already meets many morphological and photometric conditions \citep{zhou23}, which we reinforce with a catalogue-level selection (Sec. \ref{sec:selection}) and a spectroscopic selection (Sec. \ref{sec:specsel}) in the fashion of \citetalias{alvarez25} and other recent works \citep{moresco11, moresco12, borghi22a, tomasetti23}.

\subsection{Catalogue selection\label{sec:selection}}

We use the public FastSpecFit value-added catalogue (VAC) of the iron spectroscopic production\footnote{Publicly available to be downloaded from \url{https://data.desi.lbl.gov/public/dr1/vac/dr1/fastspecfit/iron/}.}, which provides uniformly reprocessed spectra and physical parameters for the first 13 months of the DESI main survey. The DESI LRG target selection \citep{zhou23, abdulkarim26} includes a strong D$4000$ break, high mass-to-light ratios and sliding colour-magnitude cuts ($\text{r}-\text{W}_1$ vs. $\text{W}_1$), optimised to select the most massive systems, dominated by old stars. We use the VAC flags to apply the following cuts:
\begin{enumerate}[label=\roman*)]
    \item {\fontfamily{qcr}\selectfont METADATA/SPECTYPE} = {\fontfamily{qcr}\selectfont GALAXY}. $8\,490\,171$ galaxy targets, hereafter the \textit{parent} sample, as in \citetalias{alvarez25}.
    \item {\fontfamily{qcr}\selectfont METADATA/DESI\_TARGET} flagged as LRG but not as Emission Line Galaxy (ELG) or Quasar (QSO). These targets already passed the morphological and photometric conditions of \cite{zhou23}; $2\,735\,943$ parent galaxies meet this criterion.
    \item Reliable redshift: {\fontfamily{qcr}\selectfont METADATA/ZWARN} $= 0$ and agreement within $1\%$ between the {\fontfamily{qcr}\selectfont redrock} ({\fontfamily{qcr}\selectfont METADATA/Z\_RR}) and the adopted ({\fontfamily{qcr}\selectfont METADATA/Z}) redshifts. $8\,669\,569$ objects pass this cut, $2\,680\,813$ of which meet the previous ones.
    \item Velocity dispersion ({\fontfamily{qcr}\selectfont FASTSPEC/VDISP}) $200 < \sigma \ [\text{km s}^{-1}] < 400$, as commonly adopted in CC works \citep[][\citetalias{alvarez25}]{moresco12, ratsimbazafy17}. The lower limit selects the most massive \citep{faber76} and oldest passive galaxies \citep{moresco24, moresco26} and prevents contamination from rotation-dominated objects \citep{diteodoro16, veale17}; galaxies with $160<\sigma \ [\text{km s}^{-1}] <200$ are also passive, but formed too recently \citep{renzini06, thomas10} for the mean redshift of our sample ($\sim 0.5-0.6$). The upper limit is an observational physical limit for elliptical galaxies \citep{bernardi03, shen03, sheth03, taylor10, bezanson11, trujillo11, bolton12, thomas13}. $2\,467\,194$ objects have $\sigma$ in this range, $1\,855\,845$ of which pass all the previous cuts.
\end{enumerate}

We downloaded the spectra of these almost $2$ million LRGs from the {\fontfamily{qcr}\selectfont SPARCL} database \citep{juneau25} of the NOIRLab Astro Data Lab, using the {\fontfamily{qcr}\selectfont METADATA/TARGETID} column. After shifting them to the rest frame and degrading them to the MILES resolution ($\sigma_\text{MILES} = 1.06$ \AA, slightly coarser than that of DESI; see appendix \ref{sec:app_VD}), we measured $33$ spectral features with {\fontfamily{qcr}\selectfont pyLick} \citep{borghi22a}: the $25$ Lick indices \citep{worthey94b, worthey97, trager98}, the $[\text{OII}]\lambda 3727$, $[\text{OIII}]\lambda 5007$ and $\text{H}\alpha$ lines, the $\text{H}\beta_0$ definition of the Balmer-$\beta$ line, the CaII K and CaII H lines and the two definitions of the $4000$ \AA \ break, $\text{D}4000$ and $\text{D}4000_\text{n}$. After rejecting a small fraction of suboptimal spectra, $1\,728\,366$ sources remain, which we call the \textit{good LRGs} (or simply LRGs) sample, since the {\fontfamily{qcr}\selectfont DESI\_TARGET} cut is the most restrictive one. Its purity is similar to that of the \textit{query} sample of \citetalias{alvarez25}, prior to the final spectroscopic selection.

\begin{figure*}
    \centering
    \includegraphics[width=\textwidth]{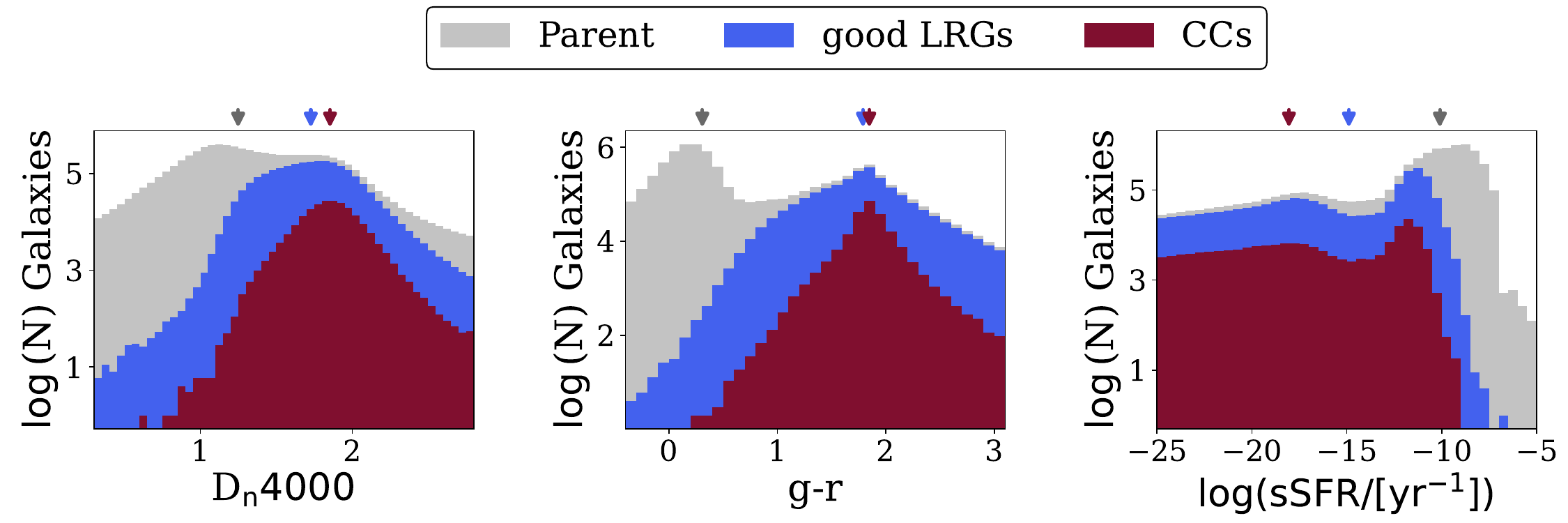}
    \caption{Distribution of the $\text{D}_\text{n}4000$ index, observed $\text{g}-\text{r}$ colour and sSFR estimated with {\fontfamily{qcr}\selectfont FastSpecFit} \protect\citep{moustakas23}, with the colour code of a similar figure in \protect\citetalias{alvarez25}: DESI galaxies (\textit{parent sample}) in grey, \textit{good LRGs} in blue, and spectroscopically selected CCs in maroon.}
    \label{fig:distribslines}
\end{figure*}

\subsection{Spectroscopic selection\label{sec:specsel}}

The spectroscopic selection is applied to the indices corrected for velocity dispersion (appendix \ref{sec:app_VD}):
\begin{enumerate}[label=\roman*)]
    \item We reject sources with significant emission in $[\text{OII}]\lambda3727$, $\text{H}\beta$, $[\text{OIII}]\lambda5007$ or $\text{H}\alpha$ (the latter only at $z \lesssim 0.49$, since the DESI spectra end at $\sim 9800$ \AA). That is, with equivalent widths (EWs; negative for emission) below $-5$ \AA, or between $-5$ and $-2$ \AA \ with $\text{EW}/\sigma_\text{EW}>2$. These lines are measured on the individual spectra with our generalisation of {\fontfamily{qcr}\selectfont pyLick} (\citetalias{alvarez25}). We found that $861\,513$ sources pass the [OII] cut, $1\,552\,560$ the H$\beta$ cut, $1\,373\,084$ the [OIII] cut and $155\,121$ (out of the $226\,997$ with measurable H$\alpha$) the H$\alpha$ cut. The first three cuts together leave $707\,114$ sources, showing that the most stringent one ([OII]) must be complemented with the others, and $699\,372$ remain after the H$\alpha$ cut when applicable. Indirectly, this cut sets a maximum redshift at $z \approx 0.9$, since the requirement to fully observe the [OIII] line removes the sources at higher redshifts, where the line lies at wavelengths redder than $9800$ \AA.
    \item We require $\text{H}/\text{K}<1.2$. The H$\epsilon$ line overlaps with CaII H ($\lambda3969$) and deepens in the presence of B- and A-type stars ($\lesssim 1$ Gyr), a signature of recent star formation such as that of post-starburst galaxies, while CaII K ($\lambda3934$) is unaffected \citep{rose85, moresco18}. In purely passive populations, the K line is deeper than the H line, and the inversion ($\text{H}/\text{K}|_{F_\text{min}} > 1$) occurs for EW ratios, H/K, above $1.2-1.5$ \citep{borghi22a}. $1\,220\,942$ sources pass this cut alone; combined with the emission-line cuts, it yields the final spectroscopic sample of $527\,287$ sources, slightly less than a third of the \textit{good LRGs}.
\end{enumerate}

\subsection{Archaeological coherence cut\label{sec:archaeo}}

The strict spectroscopic selection, analogous to those in the literature, ends here. In addition, we apply a redshift cut to address the poor match between the observed indices and those predicted by the TMJ model at low ages (appendix \ref{sec:app_TMJ}), which mostly affects the lower velocity dispersion groups, whose ages would fall below $\sim2.5$ Gyr at $z\gtrsim 0.5-0.6$. This cut combines two prescriptions: first, a scaling relation, $\sigma \to t$, that gives the maximum age expected for each velocity dispersion group; then, a cosmological prescription, $t \to z$, that translates this maximum age into a maximum redshift. In between, a margin of $\Delta t = 2.5$ Gyr is included ($t_\text{max} = t(\sigma)-2.5$ Gyr) to account for the poor index response of the TMJ model at low ages (appendix \ref{sec:app_TMJ}). We refer to this process as the \textit{archaeological coherence cut}. It ensures that the model never yields ages exceeding those predicted by the age scaling relations of similar works (\citealt{thomas10}; \citealt{johansson12}; \citetalias{alvarez25}), while accounting for its poor performance at (expected) low ages.

\begin{table}[H]
\centering
\renewcommand{\arraystretch}{1.15}
\begin{tabular}{|c|c|c|c|}
\hline
$\sigma$ group [km s$^{-1}$] & $z_\text{max}$ & N (CCs) & N (arch.) \\ \hline
$200-225$ & $0.4823$ & $86414$ & $10521$ \\
$225-250$ & $0.6166$ & $109071$ & $40584$ \\
$250-280$ & $0.7995$ & $131524$ & $101108$ \\
$280-320$ & $1.0648$ & $120290$ & $120290$ \\
$320-355$ & $1.4911$ & $51860$ & $51860$ \\
$355-400$ & $2.3206$ & $28128$ & $28128$ \\ \hline
All groups & -- & $527\, 287$ & $352\,491$\\
\hline
\end{tabular}
\caption{Archaeological coherence cut for each velocity dispersion group: maximum redshift and number of CCs before (third column) and after (last column) the cut. Groups with $\sigma>280 \text{ km s}^{-1}$ are not affected owing to the redshift coverage.}
\label{tab:archaeologycuttabla}
\end{table}

Table \ref{tab:archaeologycuttabla} gives, for each group, the resulting maximum redshift, $z_\text{max}$, and the number of CCs before and after the cut, for the baseline prescriptions (the \citetalias{alvarez25} $t(\sigma)$ relation and the \cite{planck18} $z(t)$ rule). The cut is conservative, since $z_\text{max}$ is computed from the lower $\sigma$ bound of each group; a finer stratification would discard fewer galaxies, but would lengthen the redshift window covered by each stack. Groups with $\sigma>280 \text{ km s}^{-1}$ are not affected, since the [OIII] cut already imposes a stricter redshift limit. The $t \to z$ prescription only limits the observed redshift window and provides no prior on the $t-z$ relations, so it does not bias the cosmographic results. The small systematic effect of $z_\text{max}$ is assessed with different prescriptions \citep{planck18, RIESS22} and included in the systematic budget (Sec. \ref{sec:systematics}). In the baseline, $352\,491$ sources (individual galaxies) contribute to the stacks.

Figure \ref{fig:distribslines} shows three empirical diagnostics of the purity of our sample, commonly shown in CC works (\citealt{borghi22a}, \citealt{jiao22}, \citealt{tomasetti23}, \citetalias{alvarez25}, \citealt{loubser25b}). All three distributions are bimodal, particularly $\text{g}-\text{r}$. Their medians (upper axes) shift steadily with purity (rightwards for D$_\text{n}4000$ and $\text{g}-\text{r}$, leftwards for $\log (\text{sSFR})$). The $\text{g}-\text{r}$ colour, which is already selected in the LRG sample \citep{zhou23}, does not change on average from the LRGs to the CCs, whereas the sSFR and D$_\text{n}4000$ still separate CCs from other LRGs. The $\text{D}_\text{n}4000$ distribution can also be compared with those of \cite[][\citetalias{alvarez25}]{borghi22a, jiao22}: its peak follows the expected inverse trend with redshift, since $\text{D}_\text{n}4000$ traces the differential age \citep{moresco12}. Indeed, $\langle z_\text{A25}\rangle \approx 0.15 < \langle z_\text{this work}\rangle \approx 0.61 < \langle z_\text{Borghi+22a}\rangle \approx 0.74$, while $\langle \text{D}_\text{A25}\rangle \approx 1.93 > \langle \text{D}_\text{this work} \rangle \approx 1.87 > \langle \text{D}_\text{Borghi+22a}\rangle \approx 1.69$. Other indicators of passive evolution available in the VAC support the same picture.

\section{Data treatment and methodology\label{sec:DATATREATMENT}}

This section describes the stacking procedure, an optimised version of that of \citetalias{alvarez25} that reaches a target spectral S/N to within $1\%$, and the SPS fit, which is unchanged except for new, tailor-made velocity dispersion corrections (Appendix \ref{sec:app_VD}). As learnt from SDSS, the strict selection of sources must be complemented with a limit on the redshift span of each stack, and the different performance of the SPS fit in different age regimes (Appendix \ref{sec:app_fulltz}) requires a further selection on the S/N of the ages. We then present the pivotal cosmographic fit and the systematic uncertainties introduced by the methodology; as in \citetalias{alvarez25}, those due to the SPS model are not assessed directly.

\subsection{Stacking spectra}

As shown in \citetalias{alvarez25}, the CC method performs better with homogeneous, high-S/N spectra, owing to the better response of the SPS models in the age determination. This is naturally achieved by stacking low-S/N individual spectra in groups of velocity dispersion and redshift, with a fixed stratification in velocity dispersion, which traces different formation epochs, and an adaptive grouping in redshift. We adopt the bin edges $\{200, 225, 250, 280, 320, 355, 400\} \ [\text{km s}^{-1}]$, roughly a logarithmic spacing of $0.05$, as in \cite{loubser25b} and twice as fine as in \citetalias{alvarez25}. The $355 < \sigma \ [\text{km s}^{-1}] < 400$ group is excluded from the cosmographic fit because of its small age offset with respect to the $320 < \sigma \ [\text{km s}^{-1}] < 355$ group and of other doubts on the reliability of these objects as CCs (Sec. \ref{sec:RESULTS}).

\begin{figure}[H]
    \centering
    \includegraphics[width=\linewidth]{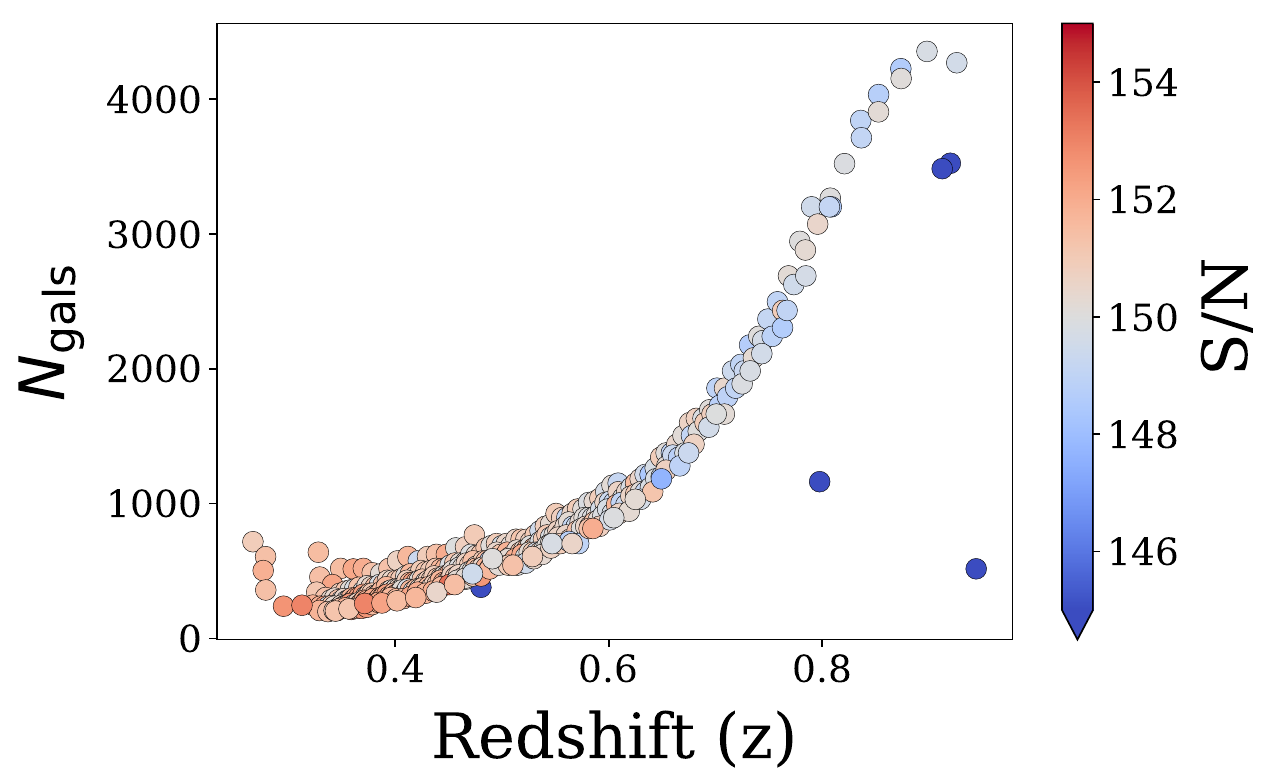}
    \caption{Number of galaxies against redshift for the stacks of the baseline. Blue data points represent the final stack in each velocity dispersion group, which have spectral S/N ratios well below the lower limit of the grading bar.}
    \label{fig:snnz}
\end{figure}

The numerical procedure to normalise and combine the spectra into a single stack, based on the median \citep{borghi22a}, is identical to that of \citetalias{alvarez25}, whereas the selection of the galaxies forming each stack uses a new, optimised algorithm based on the bisection method: within each velocity dispersion group, the galaxies ordered by redshift are fast-stacked by iteratively doubling and/or bisecting their number until the stack reaches the required S/N within $1\%$. The attainable S/N results from a trade-off between the redshift spacing of the stacks and the memory limit of the {\fontfamily{qcr}\selectfont NumPy.median} routine applied to an M$\times$N array, with M the length of the spectra (fixed to $31\,002$ pixels to preserve the spectral information) and N the number of galaxies, which caps the stacks at slightly more than $5\,000$ galaxies. Using the arithmetic mean instead, through successive weighted means, would remove this limit.

\cite{borghi22a} found that index S/N values $\text{S}/\text{N}(I_i)\gtrsim 25$ ensure reliable estimates of age, metallicity and $\alpha$-enhancement, and we find that a spectral S/N of $150$ yields median index S/N values above that threshold for all indices. As in \citetalias{alvarez25}, the spectral S/N level of the stacks is a possible source of systematic uncertainty, so we also built stacks at S/N levels from $50$ to $140$ in steps of $10$ and ran the full pipeline on each set (Sec. \ref{sec:systematics}).

Figure \ref{fig:snnz}, similar to the right panel of figure 3 in \citetalias{alvarez25}, shows the number of galaxies per stack as a function of redshift for the baseline S/N of $150$. All stacks reach a spectral S/N within $1\%$ of $150$, except for the last one of each velocity dispersion group (off-trend blue points), for which not enough sources remain to complete the algorithm.

\subsection{Lick indices, corrections and SPS fit\label{sec:Lick}}

We measure the spectral features of the rest-frame stacks with {\fontfamily{qcr}\selectfont pyLick}, as for single galaxies (Sec. \ref{sec:selection}); only absorption features are relevant here, although emission lines are also computed. To compare them with the TMJ predictions, the indices must be taken to the same instrumental resolution and to zero velocity dispersion (\citetalias{alvarez25}). Since DESI has a better resolution than the MILES library on which TMJ is built, the resolution match is implemented directly in the index measurement. For the velocity dispersion, we need correction functions, $C_I(\sigma)$, giving the ratio (difference for molecular indices) between an index and its value at a given velocity dispersion \citep{carson10}; SPS models that predict spectra \citep{vazdekis15, knowles23} can instead be broadened before measuring the indices. We computed these corrections with the MILES v9.1 spectra \citep{falconbarroso11}, also for the $8$ additional features used for selection or comparison purposes. They take the form of third-degree polynomials, $C_I(\sigma) = \Sigma^3 _{n = 0} a_n \sigma^n$ (appendix \ref{sec:app_VD}).

The corrected indices, $I_i$, with uncertainties $\sigma_{I_i}$, are compared with those predicted by the TMJ model to estimate the age ($t$), metallicity ($[Z/H]$) and abundance of $\alpha$-elements ($[\alpha / \text{Fe}]$). TMJ directly predicts Lick indices and is optimised to produce reliable and stable age trends \citep{thomas11, borghi22a}. It combines the MILES empirical stellar library \citep{sanchezblazquez06, falconbarroso11} with theoretical modifications for non-solar abundance ratios, and is based on the \cite{maraston05} evolutionary synthesis code, with the \cite{cassisi97} and, for high metallicities, the \cite{girardi00} isochrones. As an SSP model, it assumes no cosmology or galaxy formation history, so that the derived ages are purely stellar-evolution ages. We restrict the metallicity prior to $[Z/H]\leq 0.26$ ($\sim 1.8\,Z_\odot$). This limit is not a physical prior on the metallicity of LRGs: it was adopted in \citetalias{alvarez25} to compare TMJ with the \cite{knowles23} models, based on the extended sMILES (semi-empirical MILES) library and limited to that metallicity, and we keep it for consistency. It also limits the age--metallicity degeneracy, since allowing the full metallicity range produces, for part of the stacks, a high-metallicity, low-age solution with an irregular redshift evolution. As the limit lies above the average metallicities of comparable CC samples \citep{borghi22a, jiao22}, it mainly removes the high-metallicity branch of the degeneracy (appendix \ref{sec:app_TMJ}).

We estimate the stellar parameters by running an MCMC sampler, assuming that the stacked spectra are well approximated by simple stellar populations, with the \textit{emcee} \citep{FOREMAN13} implementation of the \cite{GOODMAN10} ensemble sampler and the log-likelihood
\begin{equation}
    \log{\mathcal{L}} \propto -\sum_i \left(\frac{I_i - \hat{I}_i(t, [Z/H], [\alpha/\text{Fe}])}{\sigma_{I_i}}\right)^2
    ,
    \label{eq:loglike_Lick}
\end{equation}
where $\hat{I}_i$ are the model indices and $i$ runs over the chosen Lick indices. We use the same indices as \citetalias{alvarez25}: the baseline set of \cite{johansson12} plus four indices that, according to \cite{thomas11}, reproduce well the stellar populations of Milky Way globular clusters. These are five Balmer indices (H$\beta$, H$\delta_{\text{A}/\text{F}}$ and H$\gamma_{\text{A}/\text{F}}$), five iron indices (Fe$4383$, Fe$4531$, Fe$5270$, Fe$5335$ and Fe$5406$), two magnesium indices (Mgb and Mg$_2$) and G$4300$, whose combination is essential to break the age-metallicity degeneracy \citep{worthey94a, trager00, terlevich02}. Balmer lines trace the effective temperature of the main-sequence turnoff, which makes them the primary age indicators \citep{worthey94a, beasley00}. Iron lines are sensitive to $[\text{Fe}/\text{H}]$, which traces the delayed chemical enrichment by Type Ia supernovae \citep{matteucci86, nomoto06}, whereas magnesium features trace the $\alpha$-elements produced by core-collapse supernovae of short-lived massive stars; together, the metallic lines determine $[\alpha/\text{Fe}]$ \citep{trager00, thomas03}, as seen in diagnostics such as the $\text{Mg}_\text{b} - \text{Fe}5270$ or $\text{Mg}_\text{b} - \text{Fe}5335$ planes \citep{vazdekis15, knowles23}. G$4300$ captures the CH molecular band of the cool stars of the red giant branch and of the main-sequence turnoff of old populations \citep{worthey94a}; in passive populations, it constrains $[\text{C}/\text{Fe}]$ and correlates tightly with age.

As in \citetalias{alvarez25}, we adopt uniform priors $t \in [0.1, 14.0]$ Gyr (in steps of $0.1$ Gyr), $[Z/H] \in [-0.35, 0.26]$ and $[\alpha / \text{Fe}] \in [0.0, 0.4]$ (in steps of $0.01$ dex), on a grid interpolated from the coarser native grid of the model; a functional interpolator proved computationally impractical.

\subsection{Quality flags in the $t-z$ plane\label{sec:QFs}}

In \citetalias{alvarez25}, a cut in $\sigma_z$ removed age outliers at high redshift, where the stacks averaged galaxies over wide redshift ranges and showed odd locations in the $t-z$ plane. Here, we upgrade it with a cut in $\text{S/N}(z) \equiv z/\sigma_z$, requiring $\text{S}/\text{N}(z) > 50.0$ and $\sigma_z < 0.01$. In addition, at the low-redshift end of every velocity dispersion group, the ages and their uncertainties are much larger than elsewhere, and $\text{S}/\text{N}(t) \equiv t/\sigma_t$ drops sharply. This is consistent with the flattening of the model index--age relations at older ages (Appendix~\ref{sec:app_TMJ}), which widens the age posteriors at fixed spectral S/N, and points to a systematic effect that would drive the cosmographic fit at low redshift regardless of the low $\text{S}/\text{N}(t)$. Our \textit{baseline} therefore also requires $\text{S}/\text{N}(t) > 5.0$ and $\sigma_t < 1.0$ Gyr. The full $t-z$ relations are shown in Appendix \ref{sec:app_fulltz}.

\subsection{Cosmographic fit\label{sec:COSMOmodel}}

For each velocity dispersion group, the age at a redshift $z$ in the vicinity of $z_0$ is modelled as the group age at the pivotal redshift minus the cosmographic term of eq. (\ref{eq:Dtzz0}),
\begin{equation}
    t_v(z) = t_{z_0, v} - \frac{1}{H_{z_0}} G(x; q_{z_0}, j_{z_0}),
    \label{eq:tvdez}
\end{equation}
with $x = (z-z_0)/(1+z_0)$ (eq. \ref{eq:ztox}). We fit $\{H_{z_0}, q_{z_0}, j_{z_0}\}$ and the pivotal ages with an MCMC sampler and the log-likelihood
\begin{equation}
    \log{\mathcal{L}} \propto - \sum_{i} \left(\frac{t_{i} - \hat{t}(z_{i}; z_0 ; t_{z_0, v}; H_{z_0}, q_{z_0}, j_{z_0})}{\sigma_{\text{t}_{i}}}\right)^2,
    \label{eq:loglike_cosmo}
\end{equation}
where $i$ runs over all the stacks and $t_{z_0, v}$ is the pivotal age of the group to which each stack belongs. We use uniform priors: $t_{z_0,v}\in[0, 20]$ Gyr (they rarely exceed $5$ Gyr at $z_0 \gtrsim 0.5$), $H_{z_0}\in[10, 160]$ km s$^{-1}$ Mpc$^{-1}$, and $q_{z_0}, j_{z_0}\in[-10, 10]$, which are loose compared with the variations allowed in beyond-$\Lambda$CDM cosmologies. The $q_{z_0}$ posterior closes within this prior, whereas that of $j_{z_0}$ only becomes approximately Gaussian for a prior of $\sim [-250, 250]$. Since $j_{z_0}$ values departing from unity by more than a few tenths would challenge current cosmological models, the constraints on the secondary parameters must be read as measures of the non-linearity of the $t-z$ trends. Unlike \citetalias{alvarez25}, we impose no prior on $\text{d}H/\text{d}z$ nor any relation between $q$ and $j$ (such as that following from $0<\Omega_M<1$), so our credible interval for $H(z)$ is conservative, as it includes $q$ and $j$ combinations beyond the currently studied cosmologies.

\subsection{Systematic uncertainties\label{sec:systematics}}

Systematic effects are the main limitation of astrophysical probes. The systematic budget that we compute directly is limited to the choices made in the data treatment, which we group into \textit{pre-SPS-fit} (archaeological coherence cut and spectral S/N level) and \textit{post-SPS-fit} (quality flags) systematics. Dedicated works assess the impact of the SPS model \citep{moresco20} and of flaws in the selection \citep{moresco18}. The star formation history (SFH) also contributes \citep{borghi22b}, owing to the degeneracy between the duration of the star formation episode and the ages, although our selection of very massive ETGs ($\sigma > 200$ km s$^{-1}$) reduces this effect. Following \cite{moresco20} and \cite{borghi22b}, we show in Sec. \ref{sec:RESULTS} how these systematics would enlarge our credible intervals. The effect of the metallicity prior, already restricted in \citetalias{alvarez25} to stabilise the $t-z$ relations, could be included in the systematic uncertainty due to the choice of SPS model; it is discussed in Appendix \ref{sec:app_TMJ}. \cite{borghi22b} also include the effect of the set of Lick indices and of the binning. We work with a fixed set of indices (\citetalias{alvarez25}), whose effect is included among the systematics estimated from the literature (Sec.~\ref{sec:jointcos}), while the binning is replaced in our case by the choice of the spectral S/N level, which results from a trade-off between the sampling in redshift, the number of galaxies per stack (reducing spurious biases), and the reliability of the posterior distribution of the SPS parameters.

\begin{figure*}
    \centering
    \includegraphics[width=\textwidth]{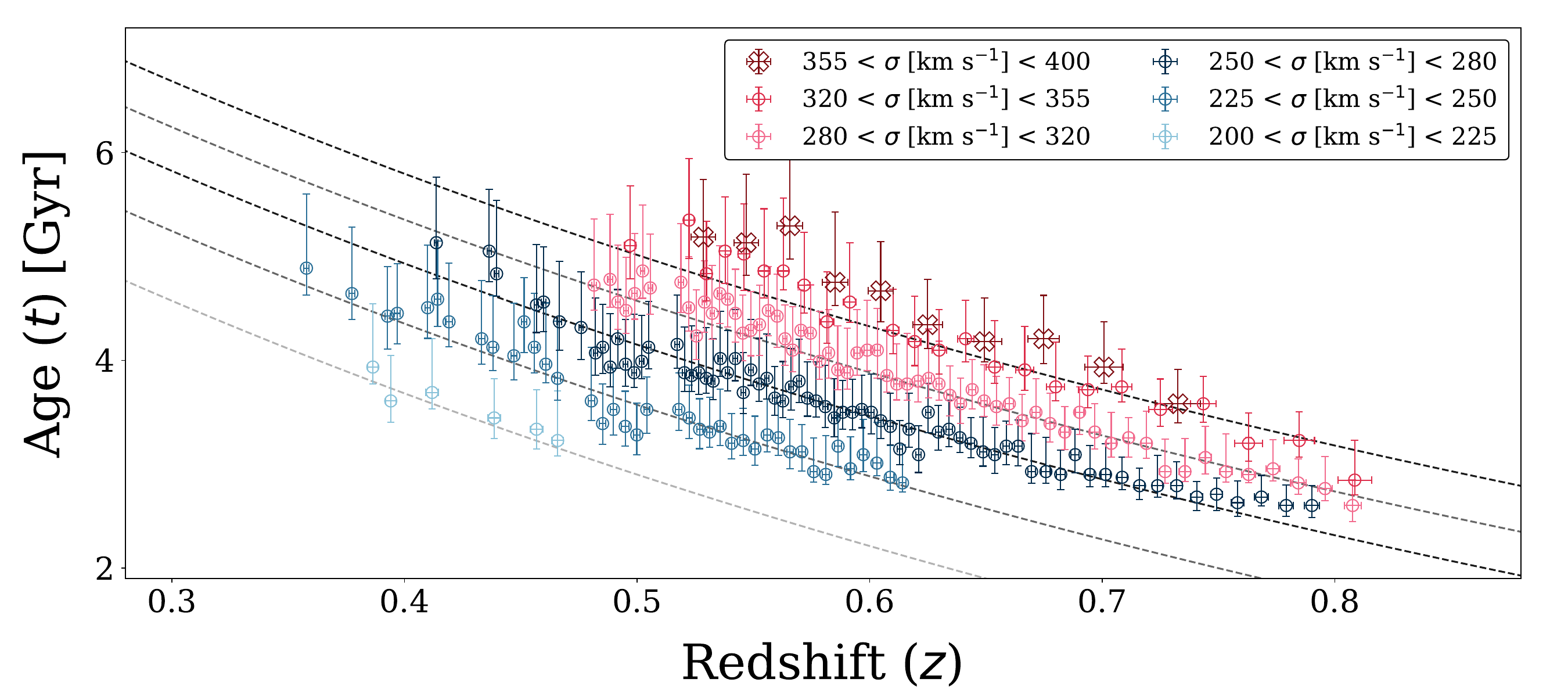}
    \caption{$t-z$ relations of the velocity dispersion groups for the stacks that pass the quality flags of Sec.~\ref{sec:QFs} ($\text{S}/\text{N}(z) > 50.0$, $\sigma_z < 0.01$, $\text{S}/\text{N} (t) > 5.0$ and $\sigma_t < 1.0$ Gyr). Blue and red tones indicate the super-groups used for the local $H(z)$ measurements (Sec. \ref{sec:localcos}). Dashed lines show, for illustration, the $\Lambda$CDM trends for the \protect\cite{planck18} parameters, offset to the modal pivotal age of each group, $t_{z_0}$, in the cosmographic fit (Sec. \ref{sec:jointcos}; appendix \ref{sec:app_scalrel}).}
    \label{fig:tzmain}
\end{figure*}

In practice, the directly assessed systematic uncertainty is computed as follows:\begin{enumerate}[label=\roman*)]
\item for the pre-SPS-fit choices, we build the stacks for all the spectral $\text{S}/\text{N}$ levels, from $50$ to $150$ in steps of $10$, and all the combinations of the prescriptions of the archaeological coherence cut (Sec. \ref{sec:archaeo}), namely the $t(\sigma)$ relations of \cite{thomas10}, \cite{johansson12} and \citetalias{alvarez25}, and the $z(t)$ rules of \cite{planck18} and \cite{RIESS22},
\item for the post-SPS-fit choices, we consider different combinations of the quality flags in the $t-z$ plane,
\item for each case, we run the full analysis (SPS fit, cosmographic and finite-difference fits) and keep the best-fit $H(z)$,
\item at each redshift, the systematic uncertainty is the standard deviation of the best-fit $H(z)$ among all cases, the pre- and post-SPS-fit contributions (Fig.~\ref{fig:statsyst}) being obtained by varying only the corresponding choices.
\end{enumerate}The baseline corresponds to a spectral $\text{S}/\text{N}$ of $150$, the archaeological coherence cut of Table \ref{tab:archaeologycuttabla} and all the quality flags.

For the two finite-difference measurements of Sec.~\ref{sec:localcos}, the redshift window must be large enough for the age evolution to exceed the statistical scatter, yet small enough to keep the approximation accurate and to limit evolutionary effects such as progenitor bias \citep{vandokkum00, moresco12, moresco16a, ratsimbazafy17, jiao22, moresco24}. The slope of the weighted linear fit (eq.~\ref{eq:newtmodel}, with a common slope $m$) is dominated by the stacks with the largest lever arm and smallest age uncertainties, so it does not trace $H$ at the median redshift. Expanding $t(z)$ to second order within the window, $\hat{m}=-\mathrm{d}t/\mathrm{d}z$ holds exactly at
\begin{equation}
z_\text{eff} = \frac{\sum_i w_i \left(z_i-\bar{z}\right)^2\left(z_i+\bar{z}\right)}{2\sum_i w_i\left(z_i-\bar{z}\right)^2},
\label{eq:zeff}
\end{equation}
where $w_i=1/\sigma_{t_i}^2$ and $\bar{z} = \sum_iw_iz_i/\sum_iw_i$, so that $H(z_\text{eff})=[(1+z_\text{eff})\,\hat{m}]^{-1}$, with no cosmological assumption. Fitting noiseless $t(z)$ relations at our stack redshifts for several cosmologies (flat $\Lambda$CDM with $\Omega_\text{m}=\{0.25,0.35\}$ and a $w_0w_a$CDM model), we find that assigning the measurements of the $200<\sigma\ [\text{km s}^{-1}]<280$ and $280<\sigma\ [\text{km s}^{-1}]<355$ super-groups to their median redshifts ($0.55$ and $0.61$) would bias $H$ high by $2$--$3\%$, almost independently of the cosmology ($<0.2\%$ variation), whereas at $z_\text{eff}=0.57$ and $0.64$ the residual bias is below $0.5\%$. We therefore quote both measurements at $z_\text{eff}$, with no additional systematic contribution for non-linearity. The standard deviations of the redshift distributions of the two super-groups, $\sigma_z = \{0.10, 0.09\}$, are similar to those of recent works using this approximation: $\Delta z \approx 0.15$ \citep{ratsimbazafy17}, $\sigma_z \approx 0.12$ \citep{borghi22b, jiao22} and $\sigma_z \approx \{0.07, 0.06, 0.07\}$ \citep{loubser25b}.

Progenitor bias is further limited by our restrictive spectroscopic and velocity dispersion selections (Sects. \ref{sec:specsel} and \ref{sec:selection}) and by the stratification in velocity dispersion. Moreover, for the most massive LRGs in BOSS and DESI at $0.15<z<0.8$, \citet{ditrani26} find no evolution in [Fe/H] and [$\alpha$/Fe], indicating that the systems added to the population below $z\sim0.8$ were already quenched and chemically mature, so that progenitor bias is negligible in this redshift range \citep[see also][]{trojeiro12}.

\section{Results\label{sec:RESULTS}}

In this section, we present the $t-z$ relations and the cosmographic fit performed on them. The derived ages, shown in Fig. \ref{fig:tzmain}, display a very parallel redshift evolution for the different velocity dispersion groups. The stacks of the $\sigma>355$ km s$^{-1}$ group are shown as empty crosses, since this group is excluded from the cosmographic fit. Such systems would be extremely massive galaxies, and the possible contamination by superpositions of objects with $\sigma>350$ km s$^{-1}$ has been discussed in the literature \citep{bernardi06}. In addition, we observe that this group overlaps in age with the group immediately below in velocity dispersion, which is at odds with the downsizing scenario for galaxy formation.

For illustrative purposes, Fig. \ref{fig:tzmain} also shows the theoretical $\Lambda$CDM relations for the cosmological parameters of \cite{planck18}, shifted to the pivotal-redshift age, $t_{z_0}$, of each velocity dispersion group, whose full posterior distributions are shown in appendix \ref{sec:app_scalrel}. These pivotal ages are estimated jointly (eq. \ref{eq:loglike_cosmo}) with the cosmographic parameters, which are shown in Fig. \ref{fig:posteriorsjoint}. The good visual agreement between the \cite{planck18} predictions and our observed $t-z$ relations already anticipates that our $H(z)$ estimate (Fig. \ref{fig:Hzjoint}) is compatible with $\Lambda$CDM, as discussed in Sec. \ref{sec:jointcos}.

The absence of stacks with ages above $\sim 6.5$ Gyr in Fig. \ref{fig:tzmain} is due to the quality flags introduced in Sec. \ref{sec:QFs} (the combination of $\text{S}/\text{N}(t) \equiv t/\sigma_t > 5.0$ and $\sigma_t < 1.0$ Gyr), and it produces an unexpected increase of $\langle z\rangle$ with $\sigma$, where $\langle z \rangle$ is the average redshift of each velocity dispersion group in Fig. \ref{fig:tzmain}. This allows us to use the velocity dispersion groups as redshift tracers, at the expense of some constraining power on the cosmographic parameters, and to apply the classic finite-difference approximation, $H\approx -\Delta z/[(1+z)\Delta t]$. This could be done for each velocity dispersion group separately but, to gain statistical power, we group together the three $200< \sigma \ [\text{km s}^{-1}] <280$ and the two $280<  \sigma \ [\text{km s}^{-1}] <355$ groups, which provide two independent, direct measurements of $H$ at the effective redshifts $z_\text{eff} = \{0.57, 0.64\}$ (see Sec. \ref{sec:systematics} for the definition of these redshifts).

Another relevant consequence of the cleaning in the $t-z$ plane is the absence of stacks in the $0.49\lesssim z\lesssim0.53$ window. As shown in appendix \ref{sec:app_fulltz}, the $t-z$ relations display a strong oscillation there, similar to those observed in \citetalias{alvarez25} and \cite{moresco16a}, which is already visible in the observed indices of the stacks, most clearly in Fe$4383$ (Fig.~\ref{fig:indicesSDSSDESI}). The oscillation increases the ages (Fig. \ref{fig:fulltz}), and their uncertainties increase even more, so that $\text{S}/\text{N}(t)$ falls below the threshold of $5$ and these stacks are excluded. More generally, we stress that the evolution of the age uncertainties with redshift (or, equivalently, with age) reflects the response of the TMJ model in different age regimes. We also note that the empty regions at $z \lesssim 0.47$ do not correspond to large redshift averages, but to stacks excluded because their quality parameters cross the acceptance thresholds. The effect of the quality flags, including the removal of the stacks around the $z\approx 0.5$ oscillation, is accounted for in the systematic budget and is subdominant with respect to other sources of uncertainty. Further independent tests show that including the stacks involved in the oscillation changes the $H(z)$ estimate only within the statistical noise, so the cosmographic estimate is robust against this oscillation, as also found in \citetalias{alvarez25} (see also appendix \ref{sec:app_fulltz}).

\subsection{Cosmography \label{sec:jointcos}}

The central result of this work is a joint cosmographic fit to all the stacks shown in Fig. \ref{fig:tzmain}, which directly translates our observable, the $t-z$ relations, into a redshift-continuous credible interval for the Hubble parameter, in the fashion of \citetalias{alvarez25}. The expansion presented in Sec. \ref{sec:CC} provides an accurate model-independent reconstruction of $H(z)$ when the pivotal redshift is chosen within the range sampled by the data \citep{fazzari25}. Therefore, we choose the median redshift of the stacks as our pivot, $z_0 \approx 0.57$. The cosmographic parameters are common to all the velocity dispersion groups, while each group has its own pivotal-redshift age, $t_{z_0,v}$ (eq.~\ref{eq:tvdez}), so that velocity dispersion only enters the fit through these nuisance parameters.

\begin{figure}
	\includegraphics[width=\columnwidth]{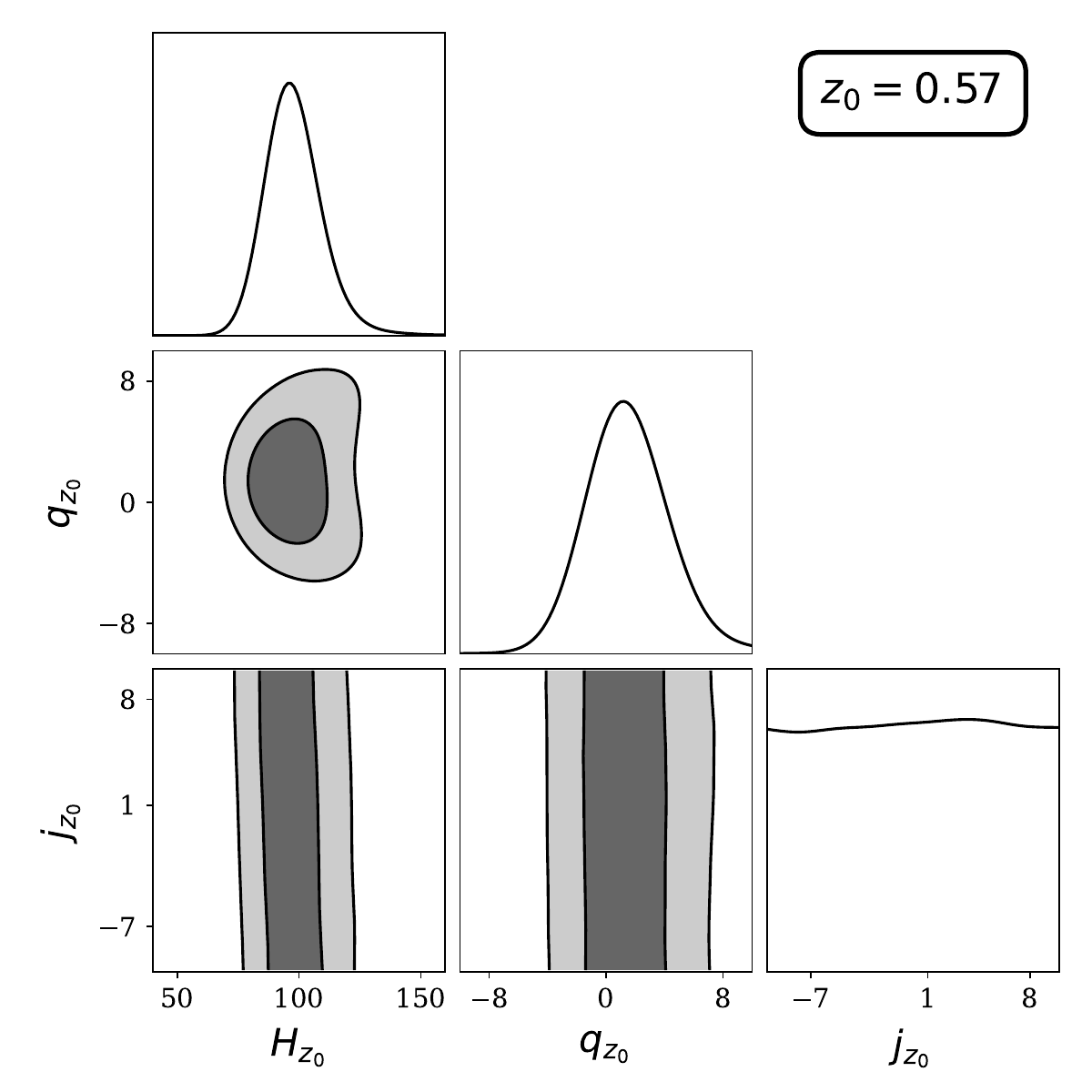}
    \caption{Posterior probability distribution for the cosmographic parameters in the full joint fit of all velocity dispersion groups.}
    \label{fig:posteriorsjoint}
\end{figure}

The posterior distribution of the cosmographic parameters is shown in Fig. \ref{fig:posteriorsjoint}. The parameters appear weakly degenerate and are well constrained within the prior, except for $j_{z_0}$, and the $H_{z_0}$ posterior is very competitive with respect to other observables. Unlike in the cosmographic fit of \citetalias{alvarez25}, where the condition $\text{d}H/\text{d}z>0$ was imposed, we did not need any condition on the redshift derivative of the Hubble parameter, which suggests that the pivotal-redshift expansion yields a more stable reconstruction than the expansion around $z=0$. The fiducial value from the MCMC sampling is $H_{z_0 = 0.57} = 95.1^{+10.9} _{-6.0}$ (stat.) km s$^{-1}$ Mpc$^{-1}$, with a systematic uncertainty of $\pm11.3$ km s$^{-1}$ Mpc$^{-1}$ if all the systematic uncertainty on $H(z = z_0)$ is attributed to $H_{z_0}$. The deceleration parameter is constrained with lower precision than in other recent determinations \citep{fazzari25, sales26}, but the data are still informative with respect to the prior, yielding $q_{z_0 = 0.57} = 1.0^{+2.7}_{-1.8}$ (stat.). The posterior of the jerk parameter is not informative within the adopted prior. Further tests showed that a Gaussian shape is only obtained when the prior is widened to $\sim \pm 250$, that is, to extreme values; in that case, the $1\sigma$ credible interval spans from $-67$ to $51$, and the maximum a posteriori (MAP) estimate becomes unstable, varying from $\sim 0$ to $\sim 5$ between different runs of the MCMC sampler.

The cubic term of the expansion therefore adds no information on the shape of $H(z)$ within our redshift range. We nevertheless keep it in the baseline, since omitting it is not neutral: truncating eq.~(\ref{eq:Dtzz0}) at second order implicitly imposes $j_{z_0}=6+6q_{z_0}+3q_{z_0}^2$, i.e. $j_{z_0}\approx6$ for the $q_{z_0}\approx0$ expected in $\Lambda$CDM at $z_0\approx0.57$ (and truncating it at first order fixes $q_{z_0}=-2$, since a linear $t-z$ relation implies $H\propto(1+z)^{-1}$). As $j$ is the lowest-order kinematic parameter that distinguishes a cosmological constant ($j=1$ in flat $\Lambda$CDM, for any $\Omega_\text{m}$) from dynamical dark energy, leaving it free propagates our ignorance of the dynamics of the expansion into the $H(z)$ credible interval and its covariance. This keeps our reconstruction model-independent and usable to test dynamical cosmological models without assuming any of them, and it maintains consistency with \citetalias{alvarez25} and \cite{fazzari25}.

\begin{figure}[H]
	\includegraphics[width=\columnwidth]{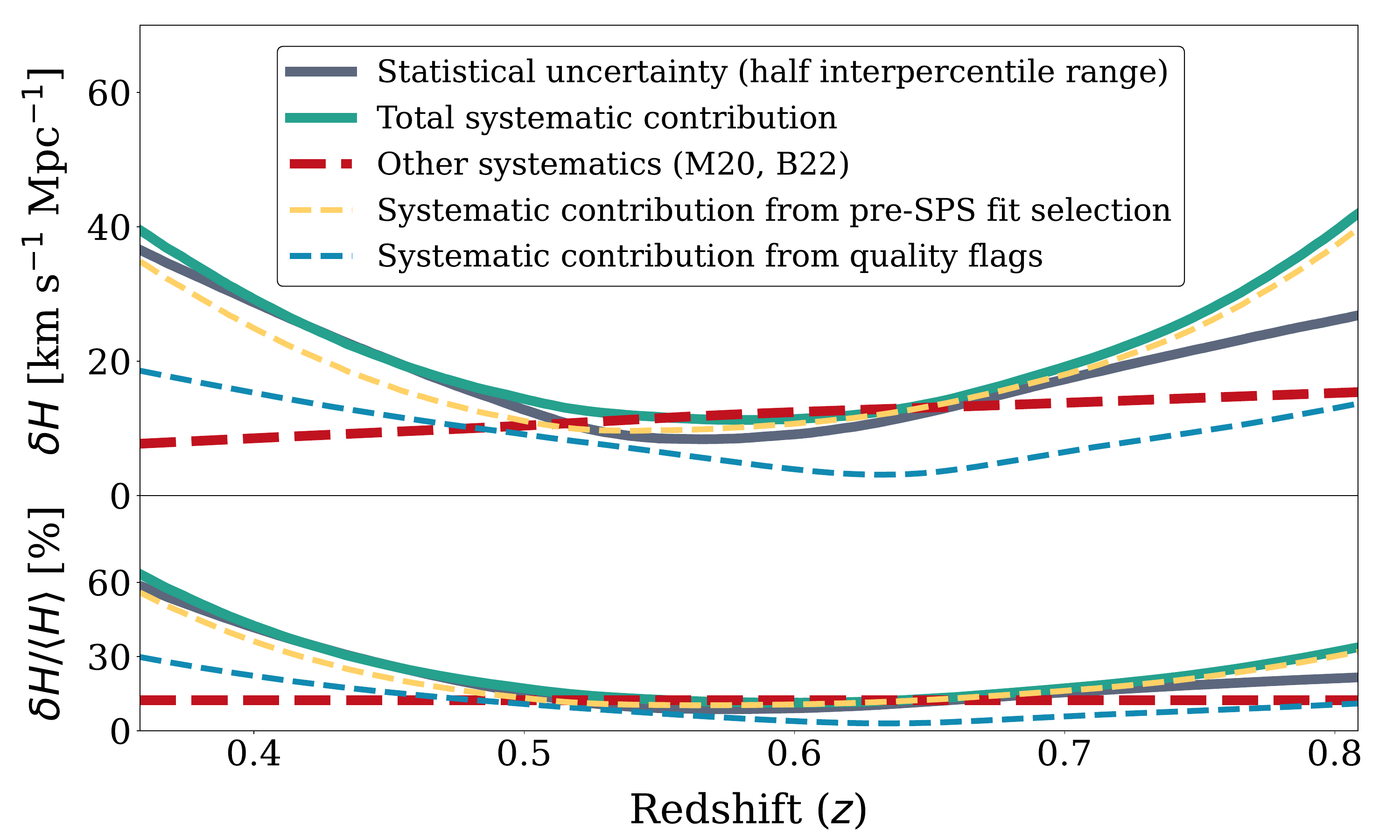}
    \caption{Comparison of statistical (solid dark grey) and systematic (solid green) uncertainties, $\delta H$, in the $H(z)$ sampling. The two separate contributions to the systematic uncertainty, the pre-SPS-fit selection (spectral S/N level and archaeological coherence cut; yellow) and the post-SPS-fit quality flags (blue), are shown as dashed lines. The red dashed line shows our estimate of the other systematics that we could not assess directly (SPS model, SFH and set of Lick indices; see Sec.~\ref{sec:jointcos}). The upper panel shows the absolute uncertainties and the lower panel the relative ones, $\Delta H \equiv \delta H/\langle H \rangle$ (in per cent), where $\langle H \rangle$ is the median of the sampling, also shown in Fig. \ref{fig:Hzjoint} as the thick solid black line.}
    \label{fig:statsyst}
\end{figure}

\begin{figure*}
    \centering
    \includegraphics[width=\textwidth]{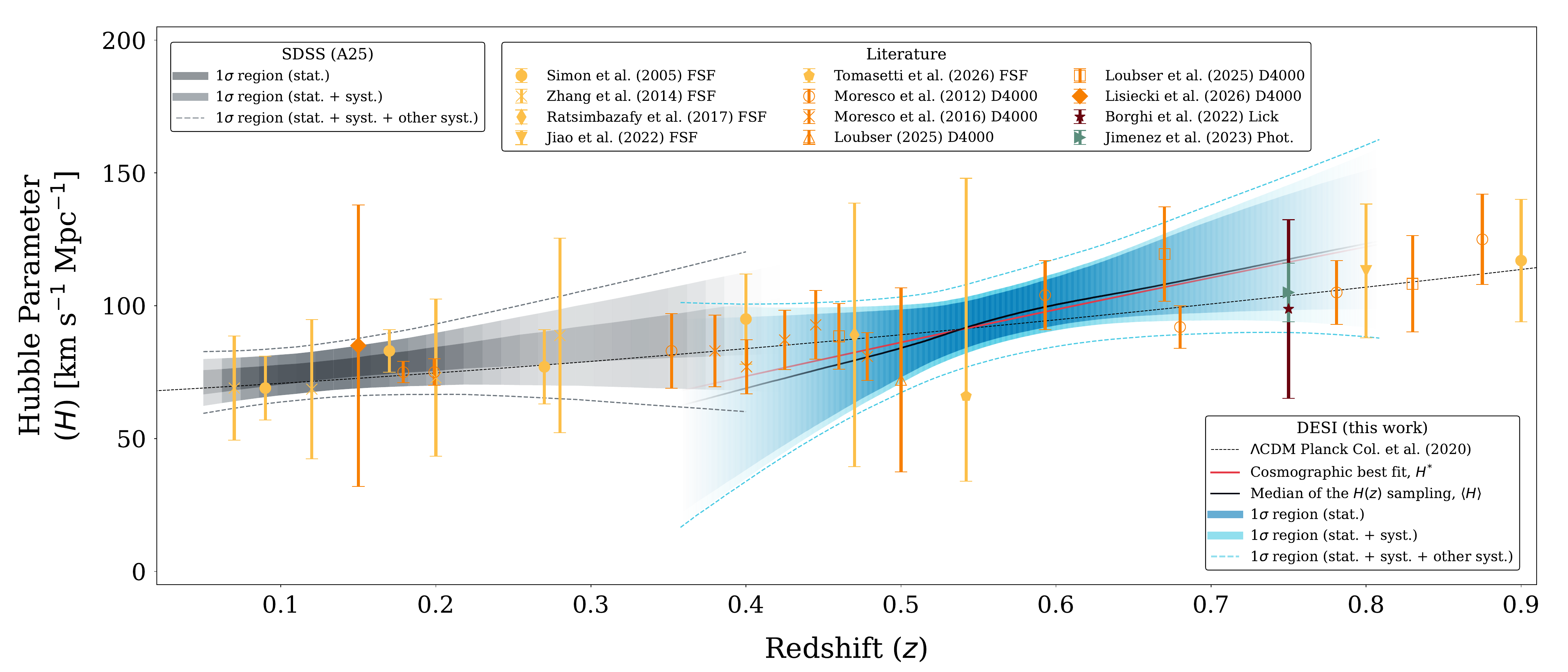}
    \caption{Hubble diagram populated with the finite-difference CC measurements available in the literature at $z<0.9$ (yellow and orange symbols), to which we add the credible interval for $H(z)$ from the joint cosmographic analysis of our $t-z$ relations. In shades of blue, we show the $1\sigma$ region from the statistical uncertainty (darker) and from its combination with the systematic uncertainty arising from our data treatment (lighter). Away from the pivotal redshift, $z_0 \simeq 0.57$, the bands fade to represent the number of stacks contributing to the fit at each redshift (see Fig. \ref{fig:tzmain}). The light-blue dashed lines mark an estimate of the limits of the credible interval when the systematic effects not directly assessed are also included. The solid black line represents the median of the $H(z)$ sampling, $\langle H\rangle$, while the red line is the cosmographic solution, $H(z; z_0; H_{z_0}^*,q_{z_0}^*, j_{z_0}^*)$, for the MAP combination of the parameters. In shades of grey, we show the analogous reconstruction obtained in \protect\citetalias{alvarez25} from SDSS stacks (cosmographic expansion around $z=0$), following the same conventions. The dotted line shows the $\Lambda$CDM $H(z)$ for the parameters of \protect\cite{planck18}.}
    \label{fig:Hzjoint}
\end{figure*}

With the posterior distribution of the cosmographic parameters, we construct the overall sampling of $H(z)$. Before presenting the credible interval, we briefly discuss the systematic uncertainty budget. As shown in Fig.~\ref{fig:statsyst}, the systematics associated with the methodology described in Sec.~\ref{sec:systematics} (bold green line) are of the same order as the statistical uncertainty (bold grey line). The latter is computed, at each redshift, as half the range between the 16th and 84th percentiles of the $H$ values obtained from the sampling of the cosmographic parameters shown in Fig.~\ref{fig:posteriorsjoint}. We analyse the pre- and post-SPS-fit selection effects separately (yellow and blue dashed lines, respectively; see Sec.~\ref{sec:systematics}), and find that the former have a slightly larger impact than the latter (the quality flags). Among the pre-SPS-fit systematics, the contribution from the choice of the spectral S/N level of the stacks is marginally larger than that of the archaeological coherence cut, by a ratio of roughly $3:2$. The upper panel shows the uncertainty, $\delta H$, in absolute units (km\,s$^{-1}$\,Mpc$^{-1}$), while the lower panel shows the relative uncertainty, $\Delta H \equiv \delta H/\langle H \rangle$, where $\langle H \rangle$ is the median of the sampling of the Hubble function at each redshift (bold black line in Fig.~\ref{fig:Hzjoint}).

Figure~\ref{fig:statsyst} also includes a bold red dashed line representing the typical impact of the \textit{other} systematic effects listed in Sec.~\ref{sec:systematics}, which we could not assess directly. This contribution is estimated from the results of \cite{moresco20}, who performed a thorough analysis of the effect of using different, but comparably reliable, SPS models, and of \cite{borghi22b}, who analysed the effect of different SFH models and of the set of Lick indices included in the SPS fit. \cite{moresco20} reported biases of up to $\sim 9\%$, a value that we adopt to remain conservative relative to the precision of our measurement, whereas \cite{borghi22b} reported biases of $0.6\sigma$ and $0.4\sigma$ (in units of their statistical uncertainty) with respect to their fiducial measurements. We combine these estimates in quadrature, scaling them to our statistical uncertainty. As a result, this contribution is constant in relative units (lower panel), whereas in absolute units (upper panel) it grows with redshift, following the increase of $H$ itself (Fig.~\ref{fig:Hzjoint}). These additional systematics dominate the error budget near $z_0$, where both the statistical uncertainty and the methodological systematics reach their minimum values.

We now turn to the final reconstruction of the Hubble parameter in Fig. \ref{fig:Hzjoint}. The MAP estimate of $H(z)$, $H^*(z) \equiv H(z; z_0; H_{z_0}^*, q_{z_0}^*, j_{z_0}^*)$ (thick red line), where the starred quantities denote MAP values, closely follows the median of the $H(z)$ sampling at each redshift, $\langle H\rangle$ (thick black line), and is only slightly flatter towards low redshift. This indicates that the posterior of $H(z)$ is close to Gaussian over most of the redshift range. In the vicinity of the pivotal redshift, $z_0 \simeq 0.57$, the width of our credible interval for $H(z)$ is similar to that of precision measurements based on the $\text{D}4000$ feature \citep{moresco12, loubser25, loubser25b}. The only CC measurement in this redshift range derived from Lick-index ages \citep{borghi22b} has an uncertainty similar to that of our reconstruction extrapolated to its redshift. It should be noted that, even with a large dataset, the constraining power of this methodology is concentrated around the pivotal redshift.

Figure~\ref{fig:Hzjoint} also shows, in grey, the $H(z)$ reconstruction obtained in \citetalias{alvarez25} from SDSS stacks with the same methodology (Lick indices fitted with the TMJ model), but with a cosmographic expansion around $z=0$ and the condition $\text{d}H/\text{d}z>0$ imposed in the fit. The two reconstructions cover complementary redshift ranges and only overlap at $0.36\lesssim z\lesssim0.4$, where both lie at the edges of their respective datasets and are therefore least constrained. In this range, their $1\sigma$ (stat.+syst.) regions overlap, in line with the overall continuity of the observed indices across the two surveys (appendix \ref{sec:app_fulltz}, Fig.~\ref{fig:indicesSDSSDESI}). Together, the SDSS and DESI analyses thus provide a CC-based reconstruction of $H(z)$ from $z\approx0.05$ to $z\approx0.8$ obtained with a homogeneous methodology; a joint cosmographic analysis of both datasets is left for future work.

Although the continuous bands in Fig. \ref{fig:Hzjoint} provide an intuitive visualisation of the expansion history, a quantitative use of these results requires a rigorous treatment of their statistical and functional dependencies. Unlike traditional CC measurements, which provide independent $H(z)$ values at discrete redshifts, our cosmographic reconstruction is intrinsically coupled through the global parameters $\{H_{z_0}, q_{z_0}, j_{z_0}\}$. To bridge the gap between our continuous joint fit and the need for practical data points, particularly to populate the $0.5 < z < 0.7$ region, where CC data are sparse, we provide an array with the sampling of $H(z)$ and its full covariance matrix, $\mathbf{C}_{H}$. This transforms the information contained in the $t-z$ relations of hundreds of stacks, built from $\sim 350\,000$ ETGs, into a dense set of $H(z)$ estimates that preserves the smooth trend of the expansion. The total covariance matrix is defined as $\mathbf{C}_{H} = \mathbf{C}_{\text{stat}} + \mathbf{C}_{\text{syst}}$. The statistical component is derived directly from the MCMC sampling: each sample, $s$, of the chain generates a curve, $H(z)_s$, and the elements of $\mathbf{C}_{\text{stat}}$ are
\begin{equation}
    C_{\text{stat}, ij} = \frac{1}{N-1} \sum_{s=1}^{N
    } \left[ H(z_i)_s - \bar{H}(z_i) \right] \left[ H(z_j)_s - \bar{H}(z_j) \right],
    \label{eq:covmatrix}
\end{equation}
where $N$ is the length of the chain and $\bar{H}(z_i)$ is the mean value at redshift $z_i$. This matrix captures the functional correlations: any variation in $q_{z_0}$ or $j_{z_0}$ affects all redshifts coherently, leading to strong off-diagonal terms. The systematic component, $\mathbf{C}_{\text{syst}}$, accounts for the methodological uncertainties described in Sec. \ref{sec:systematics}. Given that these systematics affect the global scaling of the $t-z$ relations, we assume a high degree of correlation between redshifts and model it as $C_{\text{syst}, ij} = \sigma_{\text{syst}}(z_i) \sigma_{\text{syst}}(z_j)$.

As additional material, we provide the synthesised $H(z)$, $\mathbf{H}_\text{obs}$, sampled at $45$ redshifts from $0.36$ to $0.80$ in steps of $\Delta z = 0.01$, together with the corresponding covariance matrix. We advise users of this sampling to implement a full matrix-based likelihood, which accounts for the functional correlations inherent in our cosmographic fit, with
\begin{equation}
    \chi^2 = (\mathbf{H}_\text{test} - \mathbf{H}_\text{obs})^T \ \mathbf{C}_{H}^{-1} \ (\mathbf{H}_\text{test} - \mathbf{H}_\text{obs}),
    \label{eq:covuse}
\end{equation}
where $\mathbf{H}_\text{test}$ is the model to be tested. For analyses requiring non-Gaussian posterior information, the MCMC chains are also available, allowing for direct sample-based inference.

We note that the systematic effects that we could not assess directly (i.e. those arising from the choice of SPS model, SFH and set of Lick indices) are not included in the covariance matrix. Nevertheless, to give the reader a visual sense of the overall credible interval in the Hubble diagram, Fig.~\ref{fig:Hzjoint} shows two light-blue dashed lines marking the bounds of the interpercentile range when all the estimated sources of uncertainty, including these, are taken into account. This representation is similar to that of figure 9 in \citetalias{alvarez25}, which we also reproduce in grey in Fig.~\ref{fig:Hzjoint}.

\begin{figure*}
    \centering
    \includegraphics[width=\textwidth]{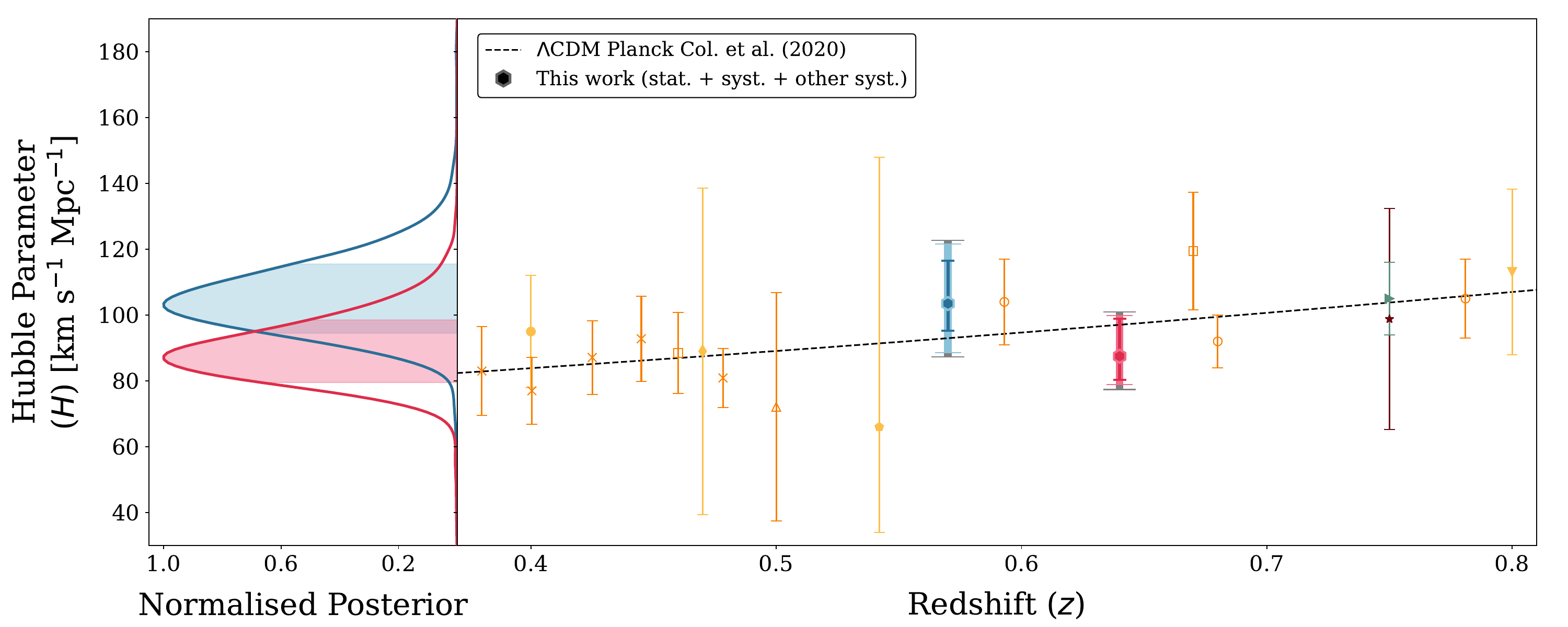}
    \caption{Same literature data as in Fig. \ref{fig:Hzjoint}, restricted to $z > 0.38$. On top, we add our finite-difference measurements of the Hubble parameter for the two super-groups: in blue, the joint fit of the $200-225$, $225-250$ and $250-280$ km s$^{-1}$ groups; in red, the joint fit of the $280-320$ and $320-355$ km s$^{-1}$ groups. The left-hand side of the diagram shows the normalised posteriors of $H$.}
    \label{fig:Hzpuntos}
\end{figure*}

\subsection{Two direct $H(z)$ estimates \label{sec:localcos}}

As anticipated at the beginning of this section, we also derive two direct point-like measurements of the Hubble parameter at the effective redshifts, $z_\text{eff}$ (see Sec. \ref{sec:systematics}). They exploit the unanticipated upward shift of the redshift distribution of the velocity dispersion groups with increasing $\sigma$ (Fig. \ref{fig:tzmain}), a result of the effect of the age quality flags on the $t-z$ relations. This allows us to provide two independent measurements in the way traditionally adopted by the community, in a redshift window ($0.5<z<0.7$) where direct measurements have been scarce. Although the two super-groups represent physically different levels of \textit{quality} as CCs, both are built exclusively from massive ($\sigma > 200 \text{ km s}^{-1}$), spectroscopically selected CC candidates. As discussed in Sec. \ref{sec:systematics}, the redshift range spanned by each super-group in Fig. \ref{fig:tzmain} is comparable to those used in recent works \citep{borghi22b, jiao22, loubser25b}.

In practical terms, we maximise the log-likelihood (\ref{eq:loglike_cosmo}) after replacing the Taylor series in $\{H_{z_0}, q_{z_0}, j_{z_0}\}$ by the linear model
\begin{equation}
    t(z_{i, v};z_\text{eff}; t_{z_\text{eff}, v}; m) =  t_{z_\text{eff}, v}- m  (z_{i, v} - z_\text{eff}),
    \label{eq:newtmodel}
\end{equation}
where $t_{0, v} = t_{z_\text{eff}, v} + m\,z_\text{eff}$ is the present-day average stellar age of each velocity dispersion group in this linear approximation, and the slope, $m$, is common to all the groups of a super-group. We emphasise that the model is anchored at $z_\text{eff}$ ($0.57$ and $0.64$, as defined in Sec. \ref{sec:systematics}) rather than at $z_0$, to ensure the reliability of the finite-difference approximation, and that $H(z_\text{eff})=[(1+z_\text{eff})\,\hat{m}]^{-1}$.

The measurements are shown in Fig. \ref{fig:Hzpuntos}, where they are compared with previous measurements from the literature, and the marginal posteriors of $H$ are shown on the left-hand side of the plot. The shaded central $1 \sigma$ region below each marginal distribution corresponds to the statistical uncertainty (darker error bars in the right panel). The combined statistical and directly assessed systematic uncertainties are represented by the light-colour error bars, and the full estimated uncertainty budget by the grey error bars. The constraints found are
\begin{equation*}
    \begin{aligned}
    H(z \approx 0.57) &= 103.5 ^{+13.0}_{-8.2}\,\text{(stat.)} \pm 28.0\,\text{(all syst.) km s}^{-1} \text{ Mpc}^{-1},\\
    H(z \approx 0.64) &= 87.5 ^{+11.4}_{-7.2}\,\text{(stat.)} \pm 14.3\,\text{(all syst.) km s}^{-1} \text{ Mpc}^{-1}.\\
    \end{aligned}
\end{equation*}

In both cases, the 16th and 84th percentiles are asymmetric with respect to the mode of the posterior distribution, with a longer tail towards high values of $H$, as expected from the non-linear mapping, $H\propto m^{-1}$, of the posterior on the slope, $m \equiv -dt/dz$. The full systematic uncertainty budget, ``(all syst.)'', which combines the contribution derived from our data treatment with our estimate of the indirectly assessed systematics, is of the same order as the statistical uncertainty. For the measurement at lower average velocity dispersion (and effective redshift), the methodological systematics are roughly twice as large as those associated with the choice of SPS model, SFH and set of Lick indices ($\sim25$ vs. $\sim13$ km s$^{-1}$ Mpc$^{-1}$), whereas for the higher effective redshift measurement both contributions are comparable ($\sim10$ km s$^{-1}$ Mpc$^{-1}$). Overall, the relative uncertainties of the direct $H$ estimates with respect to their fiducial values are $\sim29\%$ and $\sim20\%$ for the lower and higher effective redshift measurements, respectively. These comparable error budgets are expected, given that both measurements share the methodology and the sources of systematic uncertainty.

\section{Discussion and conclusions\label{sec:conclusions}}

In this work, we have presented a new, model-independent estimate of the expansion history of the Universe using the cosmic chronometer (CC) approach applied to early-type galaxies (ETGs). We employed the recently released DESI-DR1 spectra of Luminous Red Galaxies (LRGs). The vast scale of the survey provided almost $2$ million LRG spectra that pass our catalogue-level selection, from which we selected a spectroscopically refined subsample of $527\,287$ sources. These spectra were distributed into logarithmically spaced velocity dispersion groups and stacked by increasing redshift, refining the methodology introduced in \citetalias{alvarez25}. This procedure yielded highly stable stacked spectra with a spectral S/N of $150$ to within $1\%$. We emphasise the importance of applying a sensible redshift cut to each velocity dispersion group to avoid the inclusion of young stellar populations, an age regime in which the SPS model used here, TMJ \citep{thomas11}, does not respond with sufficient accuracy.

As part of our analysis, we measured the $25$ standard Lick indices plus $8$ additional spectral features: [OII]$3726$/$3729$ (treated as a doublet due to resolution limits), CaII K, CaII H, D$4000$, D$_\text{n}4000$, H$\beta_0$, [OIII]$5007$ and H$\alpha$. These measurements, performed with the {\fontfamily{qcr}\selectfont pyLick} code \citep{borghi22a} on all the LRGs that pass the catalogue-level selection, are publicly released, together with the DESI-DR1 metadata of each galaxy and the stacked spectra of the baseline (see Data availability). Each stacked spectrum includes the information needed to cross-reference its constituent galaxies, whose individual spectra are publicly available in DESI-DR1.

For the central cosmographic analysis, we implemented a pivotal cosmographic framework, as recently explored by \cite{fazzari25}, which allows for a Taylor expansion of the Hubble parameter around a non-zero pivotal redshift, $z_0$, here the median redshift of our stacks. This provides a more flexible alternative to traditional expansions around $z=0$. Through this approach, we obtained a Hubble parameter estimate at $z_0 \approx 0.57$ of $H = 95.1^{+10.9}_{-6.0} \text{ (stat.)} \pm 11.3 \text{ (syst.) km s}^{-1} \text{ Mpc}^{-1}$. The systematic uncertainty budget accounts for all the methodological choices made after the spectro-photometric selection, namely the spectral S/N level of the stacks, the archaeological coherence cut and the quality flags in the $t-z$ plane, while the uncertainty of the velocity dispersion corrections is included in the statistical error by construction. While the choice of SPS model is a known source of systematic shifts \citep[e.g.][]{moresco20}, exploring multiple models exceeds the scope of this work; however, the literature suggests that this contribution is likely of the same order as our combined internal systematics. Evolutionary effects that could mix galaxy evolution with the cosmological signal over our redshift range, such as progenitor bias, are expected to be negligible for our strictly selected, massive sample below $z\sim0.8$ (Sec.~\ref{sec:systematics}; \citealt{ditrani26}).

The primary strength of this cosmographic approach lies not in a single local measurement, but in the reconstructed expansion history across the redshift range. We provide a set of median $H(z)$ values sampled in steps of $\Delta z = 0.01$ from $z = 0.36$ to $z = 0.80$, the limits of our data. Alternatively, the expansion history can be described with the cosmographic expansion of $H$ (eq.~\ref{eq:Hexplicit}) evaluated at the MAP values of $\{H_{z_0}, q_{z_0}, j_{z_0}\}$ (Sec. \ref{sec:jointcos}). Although our data do not constrain $j_{z_0}$, we keep it free to avoid implicit assumptions on the dynamics of the expansion, so that the reconstruction can be used to test dynamical cosmological models. These measurements are, by construction, not independent, and we therefore provide the full covariance matrix, which includes the systematic contributions, as well as the MCMC chains of the baseline cosmographic fit. Our reconstruction complements the SDSS-based one of \citetalias{alvarez25} at lower redshift, with which it is consistent in the narrow redshift range where both overlap; accordingly, the Lick indices measured on the DESI stacks connect well with those of the SDSS stacks.

To allow for comparison with the broader CC literature, we also provide two independent local measurements derived using the finite-difference approximation, $H(z_\text{eff}) \approx -\Delta z/[(1 + z_\text{eff})\Delta t]$, on two separate super-groups of velocity dispersion. We estimate $H(z \approx 0.57) = 103.5 ^{+13.0}_{-8.2} \text{ (stat.) } \pm 28.0 \text{ (all syst.) km s}^{-1} \text{ Mpc}^{-1}$ and $H(z \approx 0.64) = 87.5 ^{+11.4}_{-7.2} \text{ (stat.) } \pm 14.3 \text{ (all syst.) km s}^{-1} \text{ Mpc}^{-1}$, where the systematic budget includes both the directly assessed contributions and our estimate of those that could not be assessed directly. Both measurements come from physically different CCs (massive and very massive ETGs) and should therefore not be interpreted as the redshift evolution of the same observable. The latter measurement should be regarded as the most robust, as it comes from the \textit{reddest envelope} of passively evolving ETGs, which, according to the CC framework, constitutes the purest sample. Quoting both measurements at data-driven effective redshifts, $z_\text{eff}$, rather than at the median redshift of each super-group, removes the bias associated with departures from linearity within the redshift window (Sec.~\ref{sec:systematics}). These measurements represent an independent contribution to the current observational constraints on the expansion of the Universe.

Finally, this work also highlights the limitations of the Lick-index plus TMJ approach for CC studies at intermediate redshift, which we summarise here for the benefit of future works. First, the TMJ predictions for the Balmer indices rise much more steeply towards young ages than the observed indices, so that ages $\lesssim 2.5$ Gyr cannot be reliably recovered (appendix~\ref{sec:app_TMJ}). This forced us to introduce the archaeological coherence cut, which removes a large fraction of the galaxies of the lowest velocity dispersion groups at $z\gtrsim0.5$ (Table~\ref{tab:archaeologycuttabla}). Second, the age--metallicity degeneracy requires a restrictive metallicity prior: without it, a high-metallicity, low-age solution emerges and the $t-z$ relations lose their coherence (Fig.~\ref{fig:extendedmetallicity}). Third, the flattening of the index--age relations at older ages degrades the precision of the derived ages at fixed spectral S/N, which, together with index oscillations such as the one at $z\approx0.5$ (Fig.~\ref{fig:indicesSDSSDESI}), motivated the quality flags in the $t-z$ plane. These selections only determine which stacks enter the analysis and do not impose any prior on the $t-z$ relations, and their impact is propagated into our systematic budget. Still, they reduce the amount of usable data and the redshift range over which each velocity dispersion group can be exploited, and they stem from the SPS model rather than from the data. Within these limits, TMJ yields stable age trends over most of the parameter space sampled here, and it remains one of the most complete models that directly predict Lick indices for non-solar abundance ratios. Index-based SPS models with better-calibrated predictions at young ages and super-solar metallicities \citep[e.g.][]{knowles23, park25}, together with cross-checks based on full spectral fitting, will be key to fully exploiting forthcoming CC datasets.

\begin{acknowledgements}
This work was partially funded from the projects: “Data Science
methods for MultiMessenger Astrophysics \& Multi-Survey Cosmology” funded by the Italian
Ministry of University and Research, Programmazione triennale 2021/2023 (DM n.2503 dd.
9 December 2019), Programma Congiunto Scuole; EU H2020-MSCA-ITN-2019 n. 860744
BiD4BESt: Big Data applications for black hole Evolution STudies; Italian Research Center
on High Performance Computing Big Data and Quantum Computing (ICSC), project funded
by European Union — NextGenerationEU — and National Recovery and Resilience Plan
(NRRP) — Mission 4 Component 2 within the activities of Spoke 3 (Astrophysics and
Cosmos Observations); European Union — NextGenerationEU under the PRIN MUR 2022
project n. 20224JR28W “Charting unexplored avenues in Dark Matter”; INAF Large Grant 2022 funding scheme with the project “MeerKAT and LOFAR Team up: a Unique Radio
Window on Galaxy/AGN co-Evolution; INAF GO-GTO Normal 2023 funding scheme with
the project “Serendipitous H-ATLAS-fields Observations of Radio Extragalactic Sources
(SHORES)”. MM acknowledges the financial contribution from the grant PRIN-MUR 2022 2022NY2ZRS 001 “Optimizing the extraction of cosmological information from Large Scale Structure analysis in view of the next large spectroscopic surveys” supported by NextGenerationEU, and the financial contribution from the grant ASI n. 2024-10-HH.0 “Attività scientifiche per la missione Euclid – fase E”. We further want to mention the helpful contribution and counsel of Nicola Borghi.\\ \\

\textit{Software.} {\fontfamily{qcr}\selectfont NumPy}  \citep{harris20}, {\fontfamily{qcr}\selectfont SciPy}  \citep{virtanen20}, {\fontfamily{qcr}\selectfont AstroPy}  \citep{astropy18}, {\fontfamily{qcr}\selectfont pyLick}  \citep{borghi22a}, {\fontfamily{qcr}\selectfont emcee}  \citep{FOREMAN13}, {\fontfamily{qcr}\selectfont Matplotlib}  \citep{hunter07}. This research uses services or data provided by the SPectra Analysis and Retrievable Catalog Lab \citep[SPARCL -][]{juneau25}, which is part of the Community Science and Data Center (CSDC) program at NSF National Optical-Infrared Astronomy Research Laboratory. NOIRLab is operated by the Association of Universities for Research in Astronomy (AURA), Inc. under a cooperative agreement with the National Science Foundation.\\ \\

\textit{Data Availability.} The data products and software of this work are publicly available. Owing to their size, the catalogue-level tables of the selected LRGs ({\fontfamily{qcr}\selectfont PARENT0}--{\fontfamily{qcr}\selectfont PARENT4.fits}) are available on Zenodo (\href{https://doi.org/10.5281/zenodo.21535100}{doi:10.5281/zenodo.21535100}). They contain the DESI-DR1 metadata of each galaxy and the $33$ spectral features measured with {\fontfamily{qcr}\selectfont pyLick} on its spectrum, before and after the velocity dispersion correction. The full pipeline is available at \url{https://github.com/caalvarez1999/DESI-DR1-LRGs}. The repository also includes the stacked spectra of the baseline configuration ({\fontfamily{qcr}\selectfont spectra/stacks}), each listing the DESI {\fontfamily{qcr}\selectfont TARGETID}s of its galaxies, and the summary tables of the stacks and their TMJ posteriors for all the configurations explored. It also contains the MCMC chains of the baseline cosmographic fit and the $H(z)$ sampling with its covariance matrix. The individual spectra are publicly available from DESI-DR1 through SPARCL. Both repositories include a README describing their contents.

\end{acknowledgements}

\bibliographystyle{aa} 
\bibliography{main} 

\begin{thebibliography}{126}
\expandafter\ifx\csname natexlab\endcsname\relax\def\natexlab#1{#1}\fi

\bibitem[{{Adame} {et~al.}(2025){Adame}, {Aguilar}, {Ahlen}, {Alam},
  {Alexander}, {Alvarez}, {Alves}, {Anand}, {Andrade}, {Armengaud}, {Avila},
  {Aviles}, {Awan}, {Bahr-Kalus}, {Bailey}, {Baltay}, {Bault}, {Behera},
  {BenZvi}, {Bera}, {Beutler}, {Bianchi}, {Blake}, {Blum}, {Brieden},
  {Brodzeller}, {Brooks}, {Buckley-Geer}, {Burtin}, {Calderon}, {Canning},
  {Carnero Rosell}, {Cereskaite}, {Cervantes-Cota}, {Chabanier}, {Chaussidon},
  {Chaves-Montero}, {Chen}, {Chen}, {Claybaugh}, {Cole}, {Cuceu}, {Davis},
  {Dawson}, {de la Macorra}, {de Mattia}, {Deiosso}, {Dey}, {Dey}, {Ding},
  {Doel}, {Edelstein}, {Eftekharzadeh}, {Eisenstein}, {Elliott}, {Fagrelius},
  {Fanning}, {Ferraro}, {Ereza}, {Findlay}, {Flaugher}, {Font-Ribera},
  {Forero-S{\'a}nchez}, {Forero-Romero}, {Frenk}, {Garcia-Quintero},
  {Gazta{\~n}aga}, {Gil-Mar{\'\i}n}, {Gontcho a Gontcho}, {Gonzalez-Morales},
  {Gonzalez-Perez}, {Gordon}, {Green}, {Gruen}, {Gsponer}, {Gutierrez}, {Guy},
  {Hadzhiyska}, {Hahn}, {Hanif}, {Herrera-Alcantar}, {Honscheid}, {Howlett},
  {Huterer}, {Ir{\v{s}}i{\v{c}}}, {Ishak}, {Juneau}, {Kara{\c{c}}ayl{\i}},
  {Kehoe}, {Kent}, {Kirkby}, {Kremin}, {Krolewski}, {Lai}, {Lan}, {Landriau},
  {Lang}, {Lasker}, {Le Goff}, {Le Guillou}, {Leauthaud}, {Levi}, {Li},
  {Linder}, {Lodha}, {Magneville}, {Manera}, {Margala}, {Martini}, {Maus},
  {McDonald}, {Medina-Varela}, {Meisner}, {Mena-Fern{\'a}ndez}, {Miquel},
  {Moon}, {Moore}, {Moustakas}, {Mueller}, {Mu{\~n}oz-Guti{\'e}rrez}, {Myers},
  {Nadathur}, {Napolitano}, {Neveux}, {Newman}, {Nguyen}, {Nie}, {Niz},
  {Noriega}, {Padmanabhan}, {Paillas}, {Palanque-Delabrouille}, {Pan},
  {Penmetsa}, {Percival}, {Pieri}, {Pinon}, {Poppett}, {Porredon}, {Prada},
  {P{\'e}rez-Fern{\'a}ndez}, {P{\'e}rez-R{\`a}fols}, {Rabinowitz}, {Raichoor},
  {Ram{\'\i}rez-P{\'e}rez}, {Ramirez-Solano}, {Rashkovetskyi}, {Ravoux},
  {Rezaie}, {Rich}, {Rocher}, {Rockosi}, {Roe}, {Rosado-Marin}, {Ross},
  {Rossi}, {Ruggeri}, {Ruhlmann-Kleider}, {Samushia}, {Sanchez}, {Saulder},
  {Schlafly}, {Schlegel}, {Schubnell}, {Seo}, {Shafieloo}, {Sharples},
  {Silber}, {Slosar}, {Smith}, {Sprayberry}, {Tan}, {Tarl{\'e}}, {Taylor},
  {Trusov}, {Ure{\~n}a-L{\'o}pez}, {Vaisakh}, {Valcin}, {Valdes},
  {Vargas-Maga{\~n}a}, {Verde}, {Walther}, {Wang}, {Wang}, {Weaver},
  {Weaverdyck}, {Wechsler}, {Weinberg}, {White}, {Yu}, {Yu}, {Yuan},
  {Y{\`e}che}, {Zaborowski}, {Zarrouk}, {Zhang}, {Zhao}, {Zhao}, {Zhou}, \&
  {Zhuang}}]{ADAME25}
{Adame}, A.~G., {Aguilar}, J., {Ahlen}, S., {et~al.} 2025, \jcap, 2025, 021

\bibitem[{Alvarez {et~al.}(2025)Alvarez, Cueli, Bressan, Boco, Haridasu, Bosi,
  Danese, \& Lapi}]{alvarez25}
Alvarez, C.~A., Cueli, M.~M., Bressan, A., {et~al.} 2025, A\&A, 703, A26

\bibitem[{Baldry {et~al.}(2004)Baldry, Glazebrook, Brinkmann, Ivezić, Lupton,
  Nichol, \& Szalay}]{baldry04}
Baldry, I.~K., Glazebrook, K., Brinkmann, J., {et~al.} 2004, The Astrophysical
  Journal, 600, 681

\bibitem[{{Bargiacchi, G.} {et~al.}(2021){Bargiacchi, G.}, {Risaliti, G.},
  {Benetti, M.}, {Capozziello, S.}, {Lusso, E.}, {Saccardi, A.}, \& {Signorini,
  M.}}]{bargiacchi21}
{Bargiacchi, G.}, {Risaliti, G.}, {Benetti, M.}, {et~al.} 2021, A\&A, 649, A65

\bibitem[{Barone {et~al.}(2018)Barone, D’Eugenio, Colless, Scott, Sande,
  Bland-Hawthorn, Brough, Bryant, Cortese, Croom, Foster, Goodwin,
  Konstantopoulos, Lawrence, Lorente, Medling, Owers, \& Richards}]{barone18}
Barone, T.~M., D’Eugenio, F., Colless, M., {et~al.} 2018, The Astrophysical
  Journal, 856, 64

\bibitem[{Beasley {et~al.}(2000)Beasley, Sharples, Bridges, Hanes, Zepf,
  Ashman, \& Geisler}]{beasley00}
Beasley, M.~A., Sharples, R.~M., Bridges, T.~J., {et~al.} 2000, Monthly Notices
  of the Royal Astronomical Society, 318, 1249

\bibitem[{Bell {et~al.}(2004)Bell, Wolf, Meisenheimer, Rix, Borch, Dye,
  Kleinheinrich, Wisotzki, \& McIntosh}]{bell04}
Bell, E.~F., Wolf, C., Meisenheimer, K., {et~al.} 2004, The Astrophysical
  Journal, 608, 752

\bibitem[{Bernardi {et~al.}(2003)Bernardi, Sheth, Annis, Burles, Eisenstein,
  Finkbeiner, Hogg, Lupton, Schlegel, SubbaRao, Bahcall, Blakeslee, Brinkmann,
  Castander, Connolly, Csabai, Doi, Fukugita, Frieman, Heckman, Hennessy,
  Ivezić, Knapp, Lamb, McKay, Munn, Nichol, Okamura, Schneider, Thakar, \&
  York}]{bernardi03}
Bernardi, M., Sheth, R.~K., Annis, J., {et~al.} 2003, The Astronomical Journal,
  125, 1817

\bibitem[{{Bernardi} {et~al.}(2006){Bernardi}, {Sheth}, {Nichol}, {Miller},
  {Schlegel}, {Frieman}, {Schneider}, {Subbarao}, {York}, \&
  {Brinkmann}}]{bernardi06}
{Bernardi}, M., {Sheth}, R.~K., {Nichol}, R.~C., {et~al.} 2006, \aj, 131, 2018

\bibitem[{Bezanson {et~al.}(2011)Bezanson, van Dokkum, Franx, Brammer,
  Brinchmann, Kriek, Labbé, Quadri, Rix, van~de Sande, Whitaker, \&
  Williams}]{bezanson11}
Bezanson, R., van Dokkum, P.~G., Franx, M., {et~al.} 2011, The Astrophysical
  Journal Letters, 737, L31

\bibitem[{Bolton {et~al.}(2012)Bolton, Schlegel, Aubourg, Bailey, Bhardwaj,
  Brownstein, Burles, Chen, Dawson, Eisenstein, Gunn, Knapp, Loomis, Lupton,
  Maraston, Muna, Myers, Olmstead, Padmanabhan, Pâris, Percival, Petitjean,
  Rockosi, Ross, Schneider, Shu, Strauss, Thomas, Tremonti, Wake, Weaver, \&
  Wood-Vasey}]{bolton12}
Bolton, A.~S., Schlegel, D.~J., Aubourg, E., {et~al.} 2012, The Astronomical
  Journal, 144, 144

\bibitem[{Borghi {et~al.}(2022)Borghi, Moresco, \& Cimatti}]{borghi22b}
Borghi, N., Moresco, M., \& Cimatti, A. 2022, The Astrophysical Journal
  Letters, 928, L4

\bibitem[{{Borghi} {et~al.}(2022){Borghi}, {Moresco}, {Cimatti},
  {et~al.}}]{borghi22a}
{Borghi}, N., {Moresco}, M., {Cimatti}, A., {et~al.} 2022, The Astrophysical
  Journal, 927, 164

\bibitem[{{Bruzual} \& {Charlot}(2003)}]{BRUZUAL03}
{Bruzual}, G. \& {Charlot}, S. 2003, \mnras, 344, 1000

\bibitem[{Capozziello {et~al.}(2011)Capozziello, Lazkoz, \&
  Salzano}]{capozziello12}
Capozziello, S., Lazkoz, R., \& Salzano, V. 2011, Phys. Rev. D, 84, 124061

\bibitem[{Carson \& Nichol(2010)}]{carson10}
Carson, D.~P. \& Nichol, R.~C. 2010, MNRAS, 408, 213

\bibitem[{Cassisi {et~al.}(1997)Cassisi, Castellani, \& Castellani}]{cassisi97}
Cassisi, S., Castellani, M., \& Castellani, V. 1997, A\&A, 317, 108

\bibitem[{Cattoën \& Visser(2007)}]{cattoen07}
Cattoën, C. \& Visser, M. 2007, Classical and Quantum Gravity, 24, 5985

\bibitem[{{Chavez} \& {Bertone}(2011)}]{cahvez11}
{Chavez}, M. \& {Bertone}, E. 2011, \apss, 335, 193

\bibitem[{Clemens {et~al.}(2006)Clemens, Bressan, Nikolic, Alexander, \&
  Annibali}]{clemens06}
Clemens, M.~S., Bressan, A., Nikolic, B., Alexander, P., \& Annibali, F. 2006,
  MNRAS, 702

\bibitem[{{Clemens} {et~al.}(2009){Clemens}, {Bressan}, {Nikolic}, \&
  {Rampazzo}}]{clemens09}
{Clemens}, M.~S., {Bressan}, A., {Nikolic}, B., \& {Rampazzo}, R. 2009, MNRAS,
  392, L35

\bibitem[{Clerici {et~al.}(2024)Clerici, Schnorr-Müller, Trevisan, \&
  Ricci}]{clerici24}
Clerici, K.~S., Schnorr-Müller, A., Trevisan, M., \& Ricci, T.~V. 2024,
  Monthly Notices of the Royal Astronomical Society, 531, 1034

\bibitem[{Collaboration {et~al.}(2018)Collaboration, Price-Whelan, Sipőcz,
  Günther, Lim, Crawford, Conseil, Shupe, Craig, Dencheva, Ginsburg,
  VanderPlas, Bradley, Pérez-Suárez, de~Val-Borro, Contributors), Aldcroft,
  Cruz, Robitaille, Tollerud, Committee), Ardelean, Babej, Bach, Bachetti,
  Bakanov, Bamford, Barentsen, Barmby, Baumbach, Berry, Biscani, Boquien,
  Bostroem, Bouma, Brammer, Bray, Breytenbach, Buddelmeijer, Burke, Calderone,
  Rodríguez, Cara, Cardoso, Cheedella, Copin, Corrales, Crichton, D’Avella,
  Deil, Depagne, Dietrich, Donath, Droettboom, Earl, Erben, Fabbro, Ferreira,
  Finethy, Fox, Garrison, Gibbons, Goldstein, Gommers, Greco, Greenfield,
  Groener, Grollier, Hagen, Hirst, Homeier, Horton, Hosseinzadeh, Hu, Hunkeler,
  Ivezić, Jain, Jenness, Kanarek, Kendrew, Kern, Kerzendorf, Khvalko, King,
  Kirkby, Kulkarni, Kumar, Lee, Lenz, Littlefair, Ma, Macleod, Mastropietro,
  McCully, Montagnac, Morris, Mueller, Mumford, Muna, Murphy, Nelson, Nguyen,
  Ninan, Nöthe, Ogaz, Oh, Parejko, Parley, Pascual, Patil, Patil, Plunkett,
  Prochaska, Rastogi, Janga, Sabater, Sakurikar, Seifert, Sherbert,
  Sherwood-Taylor, Shih, Sick, Silbiger, Singanamalla, Singer, Sladen, Sooley,
  Sornarajah, Streicher, Teuben, Thomas, Tremblay, Turner, Terrón, Kerkwijk,
  de~la Vega, Watkins, Weaver, Whitmore, Woillez, Zabalza, \&
  Contributors)}]{astropy18}
Collaboration, T.~A., Price-Whelan, A.~M., Sipőcz, B.~M., {et~al.} 2018, The
  Astronomical Journal, 156, 123

\bibitem[{{Conroy}(2013)}]{conroy13}
{Conroy}, C. 2013, \araa, 51, 393

\bibitem[{{Conroy} {et~al.}(2014){Conroy}, {Graves}, \& {van
  Dokkum}}]{CONROY14}
{Conroy}, C., {Graves}, G.~J., \& {van Dokkum}, P.~G. 2014, \apj, 780, 33

\bibitem[{Conroy \& Gunn(2010)}]{conroy10}
Conroy, C. \& Gunn, J.~E. 2010, The Astrophysical Journal, 712, 833

\bibitem[{{Cowie} {et~al.}(1996){Cowie}, {Songaila}, {Hu}, \&
  {Cohen}}]{COWIE96}
{Cowie}, L.~L., {Songaila}, A., {Hu}, E.~M., \& {Cohen}, J.~G. 1996, \aj, 112,
  839

\bibitem[{{DESI Collaboration} {et~al.}(2026){DESI Collaboration}, Abdul~Karim,
  Adame, Aguado, Aguilar, Ahlen, Alam, Aldering, Alexander, Alfarsy, Allen,
  Allende~Prieto, Alves, Anand, Andrade, Armengaud, Avila, Aviles, Awan,
  Bailey, Baleato~Lizancos, Ballester, Bault, Bautista, Bean, Behera, BenZvi,
  Beraldo~e Silva, Bermejo-Climent, Beutler, Bianchi, Blake, Blum, Bolton,
  Bonici, Brieden, Brodzeller, Brooks, Buckley-Geer, Burtin, Byström, Canning,
  Carnero~Rosell, Carr, Carrilho, Casas, Castander, Cereskaite, Cervantes-Cota,
  Chaussidon, Chaves-Montero, Chen, Chen, Circosta, Claybaugh, Cole, Cooper,
  Cousinou, Cuceu, Davis, Dawson, de~Belsunce, de~la Cruz, de~la Macorra,
  de~Mattia, Deiosso, Della~Costa, Demina, Demirbozan, DeRose, Dey, Dey, Ding,
  Ding, Doel, Douglass, Dowicz, Ebina, Edelstein, Eisenstein, Elbers, Emas,
  Escoffier, Fagrelius, Fan, Fanning, Favole, Fawcett, Fernández-García,
  Ferraro, Findlay, Font-Ribera, Forero-Romero, Forero-Sánchez, Frenk,
  Gänsicke, Galbany, García-Bellido, Garcia-Quintero, Garrison, Gaztañaga,
  Gil-Marín, Gloudemans, Gnedin, Gontcho A~Gontcho, Gonzalez,
  Gonzalez-Morales, Gonzalez-Perez, Gordon, Graur, Green, Gruen, Gsponer,
  Guandalin, Gutierrez, Guy, Hahn, Han, Han, He, Herrera-Alcantar, Heydenreich,
  Honscheid, Hou, Howlett, Huterer, Iršič, Ishak, Jacques, Jiang, Jimenez,
  Jing, Joachimi, Joudaki, Joyce, Jullo, Juneau, Karaçaylı, Karim, Kehoe,
  Kent, Khederlarian, Kirkby, Kisner, Kitaura, Kizhuprakkat, Kong, Koposov,
  Kremin, Krolewski, Lahav, Lai, Lamman, Lan, Landriau, Lang, Lange, Lasker,
  Le~Goff, Le~Guillou, Leauthaud, Levi, Li, Li, Liu, Lodha, Lokken, Luo, Luo,
  Magneville, Manera, Manser, Margala, Martini, Maus, McCullough, McDonald,
  Medina, Medina-Varela, Meisner, Mena-Fernández, Menegas, Meneses-Rizo,
  Mezcua, Miquel, Montero-Camacho, Moon, Moustakas, Muñoz-Gutiérrez, Mu~noz
  Santos, Myers, Myles, Nadathur, Najita, Napolitano, Newman, Nikakhtar,
  Nikutta, Niz, Noriega, Nugent, Padmanabhan, Paillas, Palanque-Delabrouille,
  Palmese, Pan, Pan, Parkinson, Peacock, Ibanez, Percival, Pérez-Fernández,
  Pérez-Ràfols, Peterson, Piat, Pieri, Pinon, Poppett, Porredon, Prada,
  Pucha, Qin, Rabinowitz, Raichoor, Ramírez-Pérez, Ramirez-Solano,
  Rashkovetskyi, Ravoux, Ried~Guachalla, Riley, Rocher, Rockosi, Rohlf,
  Rosado-Marín, Ross, Ross, Rossi, Ruggeri, Ruhlmann-Kleider, Sabiu, Said,
  Sailer, Saintonge, Salcedo~Hernandez, Samushia, Sanchez, Sanders, Sandford,
  Satyavolu, Saulder, Saydjari, Schlafly, Schlegel, Scholte, Schubnell,
  Semenaite, Seo, Shafieloo, Sharples, Silber, Sinigaglia, Siudek, Slepian,
  Smith, Soumagnac, Sprayberry, Suárez-Pérez, Swanson, Tan, Tarlé, Taylor,
  Thomas, Tojeiro, Turner, Turner, Ureña-López, Vaisakh, Valluri,
  Valogiannis, Vargas-Magaña, Verde, Vielzeuf, Walther, Wang, Wang, Wang,
  Weaver, Weaverdyck, Wechsler, Weinberg, White, Whitford, Wolfson, Yang,
  Yèche, Youles, Yu, Yuan, Zaborowski, Zarrouk, Zhang, Zhao, Zhao, Zheng,
  Zhou, Zhou, Zhou, Zou, Zou, \& Zu}]{abdulkarim26}
{DESI Collaboration}, Abdul~Karim, M., Adame, A.~G., {et~al.} 2026, The
  Astronomical Journal, 171, 285

\bibitem[{{Di Teodoro} {et~al.}(2016){Di Teodoro}, {Fraternali}, \&
  {Miller}}]{diteodoro16}
{Di Teodoro}, E.~M., {Fraternali}, F., \& {Miller}, S.~H. 2016, \aap, 594, A77

\bibitem[{Ditrani {et~al.}(2026)Ditrani, Fossati, Longhetti, Barbera, Iovino,
  Maraston, Thomas, \& Bevacqua}]{ditrani26}
Ditrani, F.~R., Fossati, M., Longhetti, M., {et~al.} 2026, Monthly Notices of
  the Royal Astronomical Society, 548, stag766

\bibitem[{{Dressler} \& {Gunn}(1983)}]{dressler83}
{Dressler}, A. \& {Gunn}, J.~E. 1983, \apj, 270, 7

\bibitem[{{Dressler} \& {Gunn}(1992)}]{dressler92}
{Dressler}, A. \& {Gunn}, J.~E. 1992, \apjs, 78, 1

\bibitem[{Dunsby \& Luongo(2016)}]{dunsby16}
Dunsby, P. K.~S. \& Luongo, O. 2016, International Journal of Geometric Methods
  in Modern Physics, 13, 1630002

\bibitem[{Eftekhari {et~al.}(2019)Eftekhari, Mosleh, Vazdekis, \&
  Tavasoli}]{eftekhari19}
Eftekhari, E., Mosleh, M., Vazdekis, A., \& Tavasoli, S. 2019, Monthly Notices
  of the Royal Astronomical Society, 486, 3788

\bibitem[{{Faber} \& {Jackson}(1976)}]{faber76}
{Faber}, S.~M. \& {Jackson}, R.~E. 1976, \apj, 204, 668

\bibitem[{{Falc{\'o}n-Barroso} {et~al.}(2011){Falc{\'o}n-Barroso},
  {S{\'a}nchez-Bl{\'a}zquez}, {Vazdekis}, {Ricciardelli}, {Cardiel}, {Cenarro},
  {Gorgas}, \& {Peletier}}]{falconbarroso11}
{Falc{\'o}n-Barroso}, J., {S{\'a}nchez-Bl{\'a}zquez}, P., {Vazdekis}, A.,
  {et~al.} 2011, \aap, 532, A95

\bibitem[{Favale {et~al.}(2024)Favale, Dainotti, Gómez-Valent, \&
  Migliaccio}]{favale24}
Favale, A., Dainotti, M.~G., Gómez-Valent, A., \& Migliaccio, M. 2024, Journal
  of High Energy Astrophysics, 44, 323

\bibitem[{Favale {et~al.}(2023)Favale, Gómez-Valent, \& Migliaccio}]{favale23}
Favale, A., Gómez-Valent, A., \& Migliaccio, M. 2023, Monthly Notices of the
  Royal Astronomical Society, 523, 3406

\bibitem[{Favale {et~al.}(2026)Favale, Gómez-Valent, \& Migliaccio}]{favale26}
Favale, A., Gómez-Valent, A., \& Migliaccio, M. 2026, Monthly Notices of the
  Royal Astronomical Society, 546, stag303

\bibitem[{Fazzari {et~al.}(2025)Fazzari, Giarè, \& Valentino}]{fazzari25}
Fazzari, E., Giarè, W., \& Valentino, E.~D. 2025, Cosmographic Footprints of
  Dynamical Dark Energy

\bibitem[{Fitzpatrick \& Graves(2015)}]{fitzpatrick15}
Fitzpatrick, P.~J. \& Graves, G.~J. 2015, Monthly Notices of the Royal
  Astronomical Society, 447, 1383

\bibitem[{Foreman-Mackey {et~al.}(2013)Foreman-Mackey, Hogg, Lang,
  {et~al.}}]{FOREMAN13}
Foreman-Mackey, D., Hogg, D.~W., Lang, D., {et~al.} 2013, Publications of the
  Astronomical Society of the Pacific, 125, 306

\bibitem[{{Gallazzi} {et~al.}(2005){Gallazzi}, {Charlot}, {Brinchmann},
  {White}, \& {Tremonti}}]{GALLAZZI05}
{Gallazzi}, A., {Charlot}, S., {Brinchmann}, J., {White}, S. D.~M., \&
  {Tremonti}, C.~A. 2005, \mnras, 362, 41

\bibitem[{{Girardi} {et~al.}(2000){Girardi}, {Bressan}, {Bertelli}, \&
  {Chiosi}}]{girardi00}
{Girardi}, L., {Bressan}, A., {Bertelli}, G., \& {Chiosi}, C. 2000, \aaps, 141,
  371

\bibitem[{G\'omez-Valent {et~al.}(2024)G\'omez-Valent, Favale, Migliaccio, \&
  Sen}]{gomezvalent24}
G\'omez-Valent, A., Favale, A., Migliaccio, M., \& Sen, A.~A. 2024, Phys. Rev.
  D, 109, 023525

\bibitem[{González {et~al.}(2023)González, Leon, \&
  Fernandez-Anaya}]{gonzalez23}
González, E., Leon, G., \& Fernandez-Anaya, G. 2023, Fractal and Fractional, 7

\bibitem[{Goodman \& Weare(2010)}]{GOODMAN10}
Goodman, J. \& Weare, J. 2010, Communications in Applied Mathematics and
  Computational Science, 5, 65

\bibitem[{Gruber \& Luongo(2014)}]{gruber14}
Gruber, C. \& Luongo, O. 2014, Phys. Rev. D, 89, 103506

\bibitem[{Harris {et~al.}(2020)Harris, Millman, van~der Walt, Gommers,
  Virtanen, Cournapeau, Wieser, Taylor, Berg, Smith, Kern, Picus, Hoyer, van
  Kerkwijk, Brett, Haldane, del Río, Wiebe, Peterson, Gérard-Marchant,
  Sheppard, Reddy, Weckesser, Abbasi, Gohlke, \& Oliphant}]{harris20}
Harris, C.~R., Millman, K.~J., van~der Walt, S.~J., {et~al.} 2020, Nature, 585,
  357

\bibitem[{Hunter(2007)}]{hunter07}
Hunter, J.~D. 2007, Computing in Science \& Engineering, 9, 90

\bibitem[{{Ilbert, O.} {et~al.}(2013){Ilbert, O.}, {McCracken, H. J.}, {Le
  Fèvre, O.}, {Capak, P.}, {Dunlop, J.}, {Karim, A.}, {Renzini, M. A.},
  {Caputi, K.}, {Boissier, S.}, {Arnouts, S.}, {Aussel, H.}, {Comparat, J.},
  {Guo, Q.}, {Hudelot, P.}, {Kartaltepe, J.}, {Kneib, J. P.}, {Krogager, J.
  K.}, {Le Floc’h, E.}, {Lilly, S.}, {Mellier, Y.}, {Milvang-Jensen, B.},
  {Moutard, T.}, {Onodera, M.}, {Richard, J.}, {Salvato, M.}, {Sanders, D. B.},
  {Scoville, N.}, {Silverman, J. D.}, {Taniguchi, Y.}, {Tasca, L.}, {Thomas,
  R.}, {Toft, S.}, {Tresse, L.}, {Vergani, D.}, {Wolk, M.}, \& {Zirm,
  A.}}]{ilbert13}
{Ilbert, O.}, {McCracken, H. J.}, {Le Fèvre, O.}, {et~al.} 2013, A\&A, 556,
  A55

\bibitem[{Jiao {et~al.}(2023)Jiao, Borghi, Moresco, \& Zhang}]{jiao22}
Jiao, K., Borghi, N., Moresco, M., \& Zhang, T.-J. 2023, The Astrophysical
  Journal Supplement Series, 265, 48

\bibitem[{Jimenez \& Loeb(2002)}]{jimenez02}
Jimenez, R. \& Loeb, A. 2002, ApJ, 573, 37

\bibitem[{Jimenez {et~al.}(2023)Jimenez, Moresco, Verde, \&
  Wandelt}]{jimenez23}
Jimenez, R., Moresco, M., Verde, L., \& Wandelt, B.~D. 2023, Journal of
  Cosmology and Astroparticle Physics, 2023, 047

\bibitem[{Johansson {et~al.}(2012)Johansson, Thomas, \& Maraston}]{johansson12}
Johansson, J., Thomas, D., \& Maraston, C. 2012, MNRAS, 421, 1908

\bibitem[{Juneau {et~al.}(2025)Juneau, Jacques, Pothier, Bolton, Weaver, Pucha,
  McManus, Nikutta, \& Olsen}]{juneau25}
Juneau, S., Jacques, A., Pothier, S., {et~al.} 2025, SPARCL: SPectra Analysis
  and Retrievable Catalog Lab

\bibitem[{{Kauffmann} {et~al.}(2003){Kauffmann}, {Heckman}, {White}, {Charlot},
  {Tremonti}, {Brinchmann}, {Bruzual}, {Peng}, {Seibert}, {Bernardi},
  {Blanton}, {Brinkmann}, {Castander}, {Cs{\'a}bai}, {Fukugita}, {Ivezic},
  {Munn}, {Nichol}, {Padmanabhan}, {Thakar}, {Weinberg}, \&
  {York}}]{KAUFFMANN03}
{Kauffmann}, G., {Heckman}, T.~M., {White}, S. D.~M., {et~al.} 2003, \mnras,
  341, 33

\bibitem[{{Knowles} {et~al.}(2023){Knowles}, {Sansom}, {Vazdekis}, \& {Allende
  Prieto}}]{knowles23}
{Knowles}, A.~T., {Sansom}, A.~E., {Vazdekis}, A., \& {Allende Prieto}, C.
  2023, \mnras, 523, 3450

\bibitem[{La~Barbera {et~al.}(2013)La~Barbera, Ferreras, Vazdekis, de~la Rosa,
  de~Carvalho, Trevisan, Falcón-Barroso, \& Ricciardelli}]{labarbera13}
La~Barbera, F., Ferreras, I., Vazdekis, A., {et~al.} 2013, MNRAS, 433, 3017

\bibitem[{Leaf \& Melia(2017)}]{leaf17}
Leaf, K. \& Melia, F. 2017, Monthly Notices of the Royal Astronomical Society,
  470, 2320

\bibitem[{Lisiecki {et~al.}(2026)Lisiecki, Principi~Cavaterra, Pollo,
  {Spiniello, Chiara}, {Hunt, Laura}, {Rosen, Charlie}, {Biesiada, Marek},
  {Donevski, Darko}, {Matera, Patryk}, {Lorenzon, Giuliano}, {Małek,
  Katarzyna}, \& {Mills, John}}]{lisiecki26}
Lisiecki, K., Principi~Cavaterra, N., Pollo, A., {et~al.} 2026, A\&A, 714, A268

\bibitem[{Loubser(2025)}]{loubser25b}
Loubser, S.~I. 2025, Monthly Notices of the Royal Astronomical Society, 544,
  3064

\bibitem[{{Loubser} {et~al.}(2025){Loubser}, {Alabi}, {Hilton}, {Ma}, {Tang},
  {Hatamkhani}, {Cress}, {Skelton}, \& {Nkosi}}]{loubser25}
{Loubser}, S.~I., {Alabi}, A.~B., {Hilton}, M., {et~al.} 2025, \mnras, 540,
  3135

\bibitem[{Lozano~Torres(2023)}]{lozanotorres23}
Lozano~Torres, J.~A. 2023, Astronomy, 2, 300

\bibitem[{Maraston(2005)}]{maraston05}
Maraston, C. 2005, MNRAS, 362, 799

\bibitem[{Maraston {et~al.}(2013)Maraston, Pforr, Henriques, Thomas, Wake,
  Brownstein, Capozzi, Tinker, Bundy, Skibba, Beifiori, Nichol, Edmondson,
  Schneider, Chen, Masters, Steele, Bolton, York, Weaver, Higgs, Bizyaev,
  Brewington, Malanushenko, Malanushenko, Snedden, Oravetz, Pan, Shelden, \&
  Simmons}]{maraston13}
Maraston, C., Pforr, J., Henriques, B.~M., {et~al.} 2013, MNRAS, 435, 2764

\bibitem[{Maraston \& Strömbäck(2011)}]{maraston11}
Maraston, C. \& Strömbäck, G. 2011, Monthly Notices of the Royal Astronomical
  Society, 418, 2785

\bibitem[{Masters {et~al.}(2011)Masters, Maraston, Nichol, Thomas, Beifiori,
  Bundy, Edmondson, Higgs, Leauthaud, Mandelbaum, Pforr, Ross, Ross, Schneider,
  Skibba, Tinker, Tojeiro, Wake, Brinkmann, \& Weaver}]{masters11}
Masters, K.~L., Maraston, C., Nichol, R.~C., {et~al.} 2011, MNRAS, 418, 1055

\bibitem[{{Matteucci} \& {Greggio}(1986)}]{matteucci86}
{Matteucci}, F. \& {Greggio}, L. 1986, \aap, 154, 279

\bibitem[{McLure {et~al.}(2018)McLure, Pentericci, Cimatti, Dunlop, Elbaz,
  Fontana, Nandra, Amorin, Bolzonella, Bongiorno, Carnall, Castellano,
  Cirasuolo, Cucciati, Cullen, DeBarros, Finkelstein, Fontanot, Franzetti,
  Fumana, Gargiulo, Garilli, Guaita, Hartley, Iovino, Jarvis, Juneau, Karman,
  Maccagni, Marchi, Mármol-Queraltó, Pompei, Pozzetti, Scodeggio, Sommariva,
  Talia, Almaini, Balestra, Bardelli, Bell, Bourne, Bowler, Brusa, Buitrago,
  Caputi, Cassata, Charlot, Citro, Cresci, Cristiani, Curtis-Lake, Dickinson,
  Fazio, Ferguson, Fiore, Franco, Fynbo, Galametz, Georgakakis, Giavalisco,
  Grazian, Hathi, Jung, Kim, Koekemoer, Khusanova, LeFèvre, Lotz, Mannucci,
  Maltby, Matsuoka, McLeod, Mendez-Hernandez, Mendez-Abreu, Mignoli, Moresco,
  Mortlock, Nonino, Pannella, Papovich, Popesso, Rosario, Salvato, Santini,
  Schaerer, Schreiber, Stark, Tasca, Thomas, Treu, Vanzella, Wild, Williams,
  Zamorani, \& Zucca}]{mclure18}
McLure, R.~J., Pentericci, L., Cimatti, A., {et~al.} 2018, MNRAS, 479, 25

\bibitem[{Moresco(2015)}]{moresco15}
Moresco, M. 2015, Monthly Notices of the Royal Astronomical Society: Letters,
  450, L16

\bibitem[{Moresco(2024)}]{moresco24}
Moresco, M. 2024, Addressing the Hubble Tension with Cosmic Chronometers, ed.
  E.~Di~Valentino \& D.~Brout (Singapore: Springer Nature Singapore), 277--293

\bibitem[{Moresco(2026)}]{moresco26}
Moresco, M. 2026, in Encyclopedia of Astrophysics (First Edition), first
  edition edn., ed. I.~Mandel (Oxford: Elsevier), 489--507

\bibitem[{{Moresco} {et~al.}(2022){Moresco}, {Amati}, {Amendola}, {Birrer},
  {Blakeslee}, {Cantiello}, {Cimatti}, {Darling}, {Della Valle}, {Fishbach},
  {Grillo}, {Hamaus}, {Holz}, {Izzo}, {Jimenez}, {Lusso}, {Meneghetti},
  {Piedipalumbo}, {Pisani}, {Pourtsidou}, {Pozzetti}, {Quartin}, {Risaliti},
  {Rosati}, \& {Verde}}]{moresco22}
{Moresco}, M., {Amati}, L., {Amendola}, L., {et~al.} 2022, Living Reviews in
  Relativity, 25, 6

\bibitem[{Moresco {et~al.}(2012)Moresco, Cimatti, Jimenez, Pozzetti, Zamorani,
  Bolzonella, Dunlop, Lamareille, Mignoli, Pearce, Rosati, Stern, Verde, Zucca,
  Carollo, Contini, Kneib, Fèvre, Lilly, Mainieri, Renzini, Scodeggio,
  Balestra, Gobat, McLure, Bardelli, Bongiorno, Caputi, Cucciati, de~la Torre,
  de~Ravel, Franzetti, Garilli, Iovino, Kampczyk, Knobel, Kovač, Borgne, Brun,
  Maier, Pelló, Peng, Perez-Montero, Presotto, Silverman, Tanaka, Tasca,
  Tresse, Vergani, Almaini, Barnes, Bordoloi, Bradshaw, Cappi, Chuter,
  Cirasuolo, Coppa, Diener, Foucaud, Hartley, Kamionkowski, Koekemoer,
  López-Sanjuan, McCracken, Nair, Oesch, Stanford, \& Welikala}]{moresco12}
Moresco, M., Cimatti, A., Jimenez, R., {et~al.} 2012, JCAP, 2012, 006

\bibitem[{Moresco {et~al.}(2011)Moresco, Cimatti, Jimenez, Pozzetti, Zamorani,
  Bolzonella, Dunlop, Lamareille, Mignoli, Pearce, Rosati, Stern, Verde, Zucca,
  Carollo, Contini, Kneib, Fèvre, Lilly, Mainieri, Renzini, Scodeggio,
  Balestra, Gobat, McLure, Bardelli, Bongiorno, Caputi, Cucciati, de~la Torre,
  de~Ravel, Franzetti, Garilli, Iovino, Kampczyk, Knobel, Kovač, Borgne, Brun,
  Maier, Pelló, Peng, Perez-Montero, Presotto, Silverman, Tanaka, Tasca,
  Tresse, Vergani, Almaini, Barnes, Bordoloi, Bradshaw, Cappi, Chuter,
  Cirasuolo, Coppa, Diener, Foucaud, Hartley, Kamionkowski, Koekemoer,
  López-Sanjuan, McCracken, Nair, Oesch, Stanford, \& Welikala}]{moresco11}
Moresco, M., Cimatti, A., Jimenez, R., {et~al.} 2011, JCAP, 2011, 045

\bibitem[{{Moresco} {et~al.}(2020){Moresco}, {Jimenez}, {Verde}, {Cimatti}, \&
  {Pozzetti}}]{moresco20}
{Moresco}, M., {Jimenez}, R., {Verde}, L., {Cimatti}, A., \& {Pozzetti}, L.
  2020, \apj, 898, 82

\bibitem[{Moresco {et~al.}(2018)Moresco, Jimenez, Verde, Pozzetti, Cimatti, \&
  Citro}]{moresco18}
Moresco, M., Jimenez, R., Verde, L., {et~al.} 2018, ApJ, 868

\bibitem[{Moresco {et~al.}(2016)Moresco, Pozzetti, Cimatti, Jimenez, Maraston,
  Verde, Thomas, Citro, Tojeiro, \& Wikinson}]{moresco16a}
Moresco, M., Pozzetti, L., Cimatti, A., {et~al.} 2016, JCAP, 2016, 014

\bibitem[{{Moustakas} {et~al.}(2023){Moustakas}, {Buhler}, {Scholte}, {Dey}, \&
  {Khederlarian}}]{moustakas23}
{Moustakas}, J., {Buhler}, J., {Scholte}, D., {Dey}, B., \& {Khederlarian}, A.
  2023, {FastSpecFit: Fast spectral synthesis and emission-line fitting of DESI
  spectra}, Astrophysics Source Code Library, record ascl:2308.005

\bibitem[{Nomoto {et~al.}(2006)Nomoto, Tominaga, Umeda, Kobayashi, \&
  Maeda}]{nomoto06}
Nomoto, K., Tominaga, N., Umeda, H., Kobayashi, C., \& Maeda, K. 2006, Nuclear
  Physics A, 777, 424, special Isseu on Nuclear Astrophysics

\bibitem[{Park {et~al.}(2025)Park, Conroy, Johnson, Leja, Dotter, \&
  Cargile}]{park25}
Park, M., Conroy, C., Johnson, B.~D., {et~al.} 2025, The Astrophysical Journal,
  994, 165

\bibitem[{Peronaci {et~al.}(2026)Peronaci, Martinelli, \&
  Nesseris}]{peronaci26}
Peronaci, M., Martinelli, M., \& Nesseris, S. 2026, Journal of Cosmology and
  Astroparticle Physics, 2026, 063

\bibitem[{{Planck Collaboration} {et~al.}(2020){Planck Collaboration},
  {Aghanim, N.}, {Akrami, Y.}, {Ashdown, M.}, {Aumont, J.}, {Baccigalupi, C.},
  {Ballardini, M.}, {Banday, A. J.}, {Barreiro, R. B.}, {Bartolo, N.}, {Basak,
  S.}, {Battye, R.}, {Benabed, K.}, {Bernard, J.-P.}, {Bersanelli, M.},
  {Bielewicz, P.}, {Bock, J. J.}, {Bond, J. R.}, {Borrill, J.}, {Bouchet, F.
  R.}, {Boulanger, F.}, {Bucher, M.}, {Burigana, C.}, {Butler, R. C.},
  {Calabrese, E.}, {Cardoso, J.-F.}, {Carron, J.}, {Challinor, A.}, {Chiang, H.
  C.}, {Chluba, J.}, {Colombo, L. P. L.}, {Combet, C.}, {Contreras, D.},
  {Crill, B. P.}, {Cuttaia, F.}, {de Bernardis, P.}, {de Zotti, G.},
  {Delabrouille, J.}, {Delouis, J.-M.}, {Di Valentino, E.}, {Diego, J. M.},
  {Doré, O.}, {Douspis, M.}, {Ducout, A.}, {Dupac, X.}, {Dusini, S.},
  {Efstathiou, G.}, {Elsner, F.}, {Enßlin, T. A.}, {Eriksen, H. K.}, {Fantaye,
  Y.}, {Farhang, M.}, {Fergusson, J.}, {Fernandez-Cobos, R.}, {Finelli, F.},
  {Forastieri, F.}, {Frailis, M.}, {Fraisse, A. A.}, {Franceschi, E.}, {Frolov,
  A.}, {Galeotta, S.}, {Galli, S.}, {Ganga, K.}, {Génova-Santos, R. T.},
  {Gerbino, M.}, {Ghosh, T.}, {González-Nuevo, J.}, {Górski, K. M.},
  {Gratton, S.}, {Gruppuso, A.}, {Gudmundsson, J. E.}, {Hamann, J.}, {Handley,
  W.}, {Hansen, F. K.}, {Herranz, D.}, {Hildebrandt, S. R.}, {Hivon, E.},
  {Huang, Z.}, {Jaffe, A. H.}, {Jones, W. C.}, {Karakci, A.}, {Keihänen, E.},
  {Keskitalo, R.}, {Kiiveri, K.}, {Kim, J.}, {Kisner, T. S.}, {Knox, L.},
  {Krachmalnicoff, N.}, {Kunz, M.}, {Kurki-Suonio, H.}, {Lagache, G.},
  {Lamarre, J.-M.}, {Lasenby, A.}, {Lattanzi, M.}, {Lawrence, C. R.}, {Le
  Jeune, M.}, {Lemos, P.}, {Lesgourgues, J.}, {Levrier, F.}, {Lewis, A.},
  {Liguori, M.}, {Lilje, P. B.}, {Lilley, M.}, {Lindholm, V.}, {López-Caniego,
  M.}, {Lubin, P. M.}, {Ma, Y.-Z.}, {Macías-Pérez, J. F.}, {Maggio, G.},
  {Maino, D.}, {Mandolesi, N.}, {Mangilli, A.}, {Marcos-Caballero, A.}, {Maris,
  M.}, {Martin, P. G.}, {Martinelli, M.}, {Martínez-González, E.},
  {Matarrese, S.}, {Mauri, N.}, {McEwen, J. D.}, {Meinhold, P. R.},
  {Melchiorri, A.}, {Mennella, A.}, {Migliaccio, M.}, {Millea, M.}, {Mitra,
  S.}, {Miville-Deschênes, M.-A.}, {Molinari, D.}, {Montier, L.}, {Morgante,
  G.}, {Moss, A.}, {Natoli, P.}, {Nørgaard-Nielsen, H. U.}, {Pagano, L.},
  {Paoletti, D.}, {Partridge, B.}, {Patanchon, G.}, {Peiris, H. V.}, {Perrotta,
  F.}, {Pettorino, V.}, {Piacentini, F.}, {Polastri, L.}, {Polenta, G.},
  {Puget, J.-L.}, {Rachen, J. P.}, {Reinecke, M.}, {Remazeilles, M.}, {Renzi,
  A.}, {Rocha, G.}, {Rosset, C.}, {Roudier, G.}, {Rubiño-Martín, J. A.},
  {Ruiz-Granados, B.}, {Salvati, L.}, {Sandri, M.}, {Savelainen, M.}, {Scott,
  D.}, {Shellard, E. P. S.}, {Sirignano, C.}, {Sirri, G.}, {Spencer, L. D.},
  {Sunyaev, R.}, {Suur-Uski, A.-S.}, {Tauber, J. A.}, {Tavagnacco, D.}, {Tenti,
  M.}, {Toffolatti, L.}, {Tomasi, M.}, {Trombetti, T.}, {Valenziano, L.},
  {Valiviita, J.}, {Van Tent, B.}, {Vibert, L.}, {Vielva, P.}, {Villa, F.},
  {Vittorio, N.}, {Wandelt, B. D.}, {Wehus, I. K.}, {White, M.}, {White, S. D.
  M.}, {Zacchei, A.}, \& {Zonca, A.}}]{planck18}
{Planck Collaboration}, {Aghanim, N.}, {Akrami, Y.}, {et~al.} 2020, A\&A, 641,
  A6

\bibitem[{Ratsimbazafy {et~al.}(2017)Ratsimbazafy, Loubser, Crawford, Cress,
  Bassett, Nichol, \& Väisänen}]{ratsimbazafy17}
Ratsimbazafy, A.~L., Loubser, S.~I., Crawford, S.~M., {et~al.} 2017, MNRAS,
  467, 3239

\bibitem[{Renzini(2006)}]{renzini06}
Renzini, A. 2006, \araa, 44, 141

\bibitem[{{Riess} {et~al.}(2022){Riess}, {Yuan}, {Macri}, {Scolnic}, {Brout},
  {Casertano}, {Jones}, {Murakami}, {Anand}, {Breuval}, {Brink}, {Filippenko},
  {Hoffmann}, {Jha}, {D'arcy Kenworthy}, {Mackenty}, {Stahl}, \&
  {Zheng}}]{RIESS22}
{Riess}, A.~G., {Yuan}, W., {Macri}, L.~M., {et~al.} 2022, \apjl, 934, L7

\bibitem[{{Rose}(1985)}]{rose85}
{Rose}, J.~A. 1985, \aj, 90, 1927

\bibitem[{Ruchika {et~al.}(2025)Ruchika, Mukherjee, \& Favale}]{ruchika26}
Ruchika, Mukherjee, P., \& Favale, A. 2025 [\eprint[arXiv]{2510.03742}]

\bibitem[{{Sales} {et~al.}(2025){Sales}, {de Farias}, {Queiroz}, {Santos},
  {Batista}, {Oliveira}, {Santana}, {Wuensche}, {Villela}, \&
  {Vieira}}]{sales26}
{Sales}, L.~L., {de Farias}, K. E.~L., {Queiroz}, A.~R., {et~al.} 2025, arXiv
  e-prints, arXiv:2507.06975

\bibitem[{{S{\'a}nchez-Bl{\'a}zquez} {et~al.}(2006){S{\'a}nchez-Bl{\'a}zquez},
  {Peletier}, {Jim{\'e}nez-Vicente}, {Cardiel}, {Cenarro},
  {Falc{\'o}n-Barroso}, {Gorgas}, {Selam}, \& {Vazdekis}}]{sanchezblazquez06}
{S{\'a}nchez-Bl{\'a}zquez}, P., {Peletier}, R.~F., {Jim{\'e}nez-Vicente}, J.,
  {et~al.} 2006, \mnras, 371, 703

\bibitem[{{S\'anchez-Bl\'azquez, P.} {et~al.}(2009){S\'anchez-Bl\'azquez, P.},
  {Jablonka, P.}, {Noll, S.}, {Poggianti, B. M.}, {Moustakas, J.},
  {Milvang-Jensen, B.}, {Halliday, C.}, {Arag\'on-Salamanca, A.}, {Saglia, R.
  P.}, {Desai, V.}, {De Lucia, G.}, {Clowe, D. I.}, {Pell\'o, R.}, {Rudnick,
  G.}, {Simard, L.}, {White, S. D. M.}, \& {Zaritsky, D.}}]{sanchezblazquez09}
{S\'anchez-Bl\'azquez, P.}, {Jablonka, P.}, {Noll, S.}, {et~al.} 2009, A\&A,
  499, 47

\bibitem[{Shen {et~al.}(2003)Shen, Mo, White, Blanton, Kauffmann, Voges,
  Brinkmann, \& Csabai}]{shen03}
Shen, S., Mo, H.~J., White, S. D.~M., {et~al.} 2003, Monthly Notices of the
  Royal Astronomical Society, 343, 978

\bibitem[{Sheth {et~al.}(2003)Sheth, Bernardi, Schechter, Burles, Eisenstein,
  Finkbeiner, Frieman, Lupton, Schlegel, Subbarao, Shimasaku, Bahcall,
  Brinkmann, \& Ivezić}]{sheth03}
Sheth, R.~K., Bernardi, M., Schechter, P.~L., {et~al.} 2003, The Astrophysical
  Journal, 594, 225

\bibitem[{Shimasaku {et~al.}(2001)Shimasaku, Fukugita, Doi, Hamabe, Ichikawa,
  Okamura, Sekiguchi, Yasuda, Brinkmann, Csabai, Ichikawa, Ivezić, Kunszt,
  Schneider, Szokoly, Watanabe, \& York}]{shimasaku01}
Shimasaku, K., Fukugita, M., Doi, M., {et~al.} 2001, ApJ, 122, 1238

\bibitem[{Simon {et~al.}(2005)Simon, Verde, \& Jimenez}]{simon05}
Simon, J., Verde, L., \& Jimenez, R. 2005, Phys. Rev. D, 71, 123001

\bibitem[{Stern {et~al.}(2010)Stern, Jimenez, Verde, Kamionkowski, \&
  Stanford}]{stern10}
Stern, D., Jimenez, R., Verde, L., Kamionkowski, M., \& Stanford, S.~A. 2010,
  JCAP, 2010, 008

\bibitem[{Strateva {et~al.}(2001)Strateva, Željko Ivezić, Knapp, Narayanan,
  Strauss, Gunn, Lupton, Schlegel, Bahcall, Brinkmann, Brunner, Budavári,
  Csabai, Castander, Doi, Fukugita, Győry, Hamabe, Hennessy, Ichikawa, Kunszt,
  Lamb, McKay, Okamura, Racusin, Sekiguchi, Schneider, Shimasaku, \&
  York}]{strateva01}
Strateva, I., Željko Ivezić, Knapp, G.~R., {et~al.} 2001, AJ, 122, 1861

\bibitem[{Sultana {et~al.}(2022)Sultana, Yennapureddy, Melia, \&
  Kazanas}]{sultana22}
Sultana, J., Yennapureddy, M.~K., Melia, F., \& Kazanas, D. 2022, Monthly
  Notices of the Royal Astronomical Society, 514, 5827

\bibitem[{Taylor {et~al.}(2010)Taylor, Franx, Brinchmann, van~der Wel, \& van
  Dokkum}]{taylor10}
Taylor, E.~N., Franx, M., Brinchmann, J., van~der Wel, A., \& van Dokkum, P.~G.
  2010, The Astrophysical Journal, 722, 1

\bibitem[{Terlevich \& Forbes(2002)}]{terlevich02}
Terlevich, A. \& Forbes, D.~A. 2002, Monthly Notices of the Royal Astronomical
  Society, 330, 547

\bibitem[{{Thomas} {et~al.}(2003){Thomas}, {Maraston}, \& {Bender}}]{thomas03}
{Thomas}, D., {Maraston}, C., \& {Bender}, R. 2003, MNRAS, 339, 897

\bibitem[{Thomas {et~al.}(2005)Thomas, {Maraston}, {Bender}, \& {Mendes de
  Oliveira}}]{THOMAS05}
Thomas, D., {Maraston}, C., {Bender}, R., \& {Mendes de Oliveira}, C. 2005,
  \apj, 621, 673

\bibitem[{Thomas {et~al.}(2011)Thomas, Maraston, \& Johansson}]{thomas11}
Thomas, D., Maraston, C., \& Johansson, J. 2011, MNRAS, 412, 2183–2198

\bibitem[{Thomas {et~al.}(2010)Thomas, Maraston, Schawinski, Sarzi, \&
  Silk}]{thomas10}
Thomas, D., Maraston, C., Schawinski, K., Sarzi, M., \& Silk, J. 2010, MNRAS,
  404, 1775

\bibitem[{Thomas {et~al.}(2013)Thomas, Steele, Maraston, Johansson, Beifiori,
  Pforr, Strömbäck, Tremonti, Wake, Bizyaev, Bolton, Brewington, Brownstein,
  Comparat, Kneib, Malanushenko, Malanushenko, Oravetz, Pan, Parejko,
  Schneider, Shelden, Simmons, Snedden, Tanaka, Weaver, \& Yan}]{thomas13}
Thomas, D., Steele, O., Maraston, C., {et~al.} 2013, Monthly Notices of the
  Royal Astronomical Society, 431, 1383

\bibitem[{Tojeiro {et~al.}(2012)Tojeiro, Percival, Wake, Maraston, Skibba,
  Zehavi, Ross, Brinkmann, Conroy, Guo, Manera, Masters, Pforr, Samushia,
  Schneider, Thomas, Weaver, Bizyaev, Brewington, Malanushenko, Malanushenko,
  Oravetz, Pan, Shelden, Simmons, \& Snedden}]{trojeiro12}
Tojeiro, R., Percival, W.~J., Wake, D.~A., {et~al.} 2012, Monthly Notices of
  the Royal Astronomical Society, 424, 136

\bibitem[{Tomasetti {et~al.}(2023)Tomasetti, Moresco, Borghi, Jiao, Cimatti,
  Pozzetti, Carnall, McLure, \& Pentericci}]{tomasetti23}
Tomasetti, E., Moresco, M., Borghi, N., {et~al.} 2023, A\&A, 679, A96

\bibitem[{{Tomasetti} {et~al.}(2026){Tomasetti}, {Moresco}, {Granata},
  {D'Addona}, {Bergamini}, {Grillo}, {Mercurio}, {Rosati}, {Cimatti},
  {Tortorelli}, {Schuldt}, \& {Meneghetti}}]{tomasetti26}
{Tomasetti}, E., {Moresco}, M., {Granata}, G., {et~al.} 2026, \aap, 708, A145

\bibitem[{{Trager} {et~al.}(2000){Trager}, {Faber}, {Worthey}, \&
  {Gonz{\'a}lez}}]{trager00}
{Trager}, S.~C., {Faber}, S.~M., {Worthey}, G., \& {Gonz{\'a}lez}, J.~J. 2000,
  \aj, 119, 1645

\bibitem[{Trager {et~al.}(1998)Trager, Worthey, Faber, Burstein, \&
  González}]{trager98}
Trager, S.~C., Worthey, G., Faber, S.~M., Burstein, D., \& González, J.~J.
  1998, ApJ, 116, 1

\bibitem[{Trujillo {et~al.}(2011)Trujillo, Ferreras, \& de~la
  Rosa}]{trujillo11}
Trujillo, I., Ferreras, I., \& de~la Rosa, I.~G. 2011, Monthly Notices of the
  Royal Astronomical Society, 415, 3903

\bibitem[{van Dokkum {et~al.}(2000)van Dokkum, Franx, Fabricant, Illingworth,
  \& Kelson}]{vandokkum00}
van Dokkum, P.~G., Franx, M., Fabricant, D., Illingworth, G.~D., \& Kelson,
  D.~D. 2000, The Astrophysical Journal, 541, 95

\bibitem[{{Vazdekis} {et~al.}(2015){Vazdekis}, {Coelho}, {Cassisi},
  {Ricciardelli}, {Falc{\'o}n-Barroso}, {S{\'a}nchez-Bl{\'a}zquez}, {La
  Barbera}, {Beasley}, \& {Pietrinferni}}]{vazdekis15}
{Vazdekis}, A., {Coelho}, P., {Cassisi}, S., {et~al.} 2015, \mnras, 449, 1177

\bibitem[{Veale {et~al.}(2017)Veale, Ma, Greene, Thomas, Blakeslee, McConnell,
  Walsh, \& Ito}]{veale17}
Veale, M., Ma, C.-P., Greene, J.~E., {et~al.} 2017, Monthly Notices of the
  Royal Astronomical Society, 471, 1428

\bibitem[{Verde {et~al.}(2017)Verde, Bernal, Heavens, \& Jimenez}]{verde17}
Verde, L., Bernal, J.~L., Heavens, A.~F., \& Jimenez, R. 2017, Monthly Notices
  of the Royal Astronomical Society, 467, 731

\bibitem[{Virtanen {et~al.}(2020)Virtanen, Gommers, Oliphant, Haberland, Reddy,
  Cournapeau, Burovski, Peterson, Weckesser, Bright, {van der Walt}, Brett,
  Wilson, Millman, Mayorov, Nelson, Jones, Kern, Larson, Carey, Polat, Feng,
  Moore, {VanderPlas}, Laxalde, Perktold, Cimrman, Henriksen, Quintero, Harris,
  Archibald, Ribeiro, Pedregosa, {van Mulbregt}, \& {SciPy 1.0
  Contributors}}]{virtanen20}
Virtanen, P., Gommers, R., Oliphant, T.~E., {et~al.} 2020, Nature Methods, 17,
  261

\bibitem[{Visser(2004)}]{visser04}
Visser, M. 2004, Classical and Quantum Gravity, 21, 2603

\bibitem[{White {et~al.}(2011)White, Blanton, Bolton, Schlegel, Tinker,
  Berlind, da~Costa, Kazin, Lin, Maia, McBride, Padmanabhan, Parejko, Percival,
  Prada, Ramos, Sheldon, de~Simoni, Skibba, Thomas, Wake, Zehavi, Zheng,
  Nichol, Schneider, Strauss, Weaver, \& Weinberg}]{white11}
White, M., Blanton, M., Bolton, A., {et~al.} 2011, ApJ, 728, 126

\bibitem[{{Worthey}(1994)}]{worthey94a}
{Worthey}, G. 1994, \apjs, 95, 107

\bibitem[{{Worthey} {et~al.}(1992){Worthey}, {Faber}, \&
  {Gonzalez}}]{worthey92}
{Worthey}, G., {Faber}, S.~M., \& {Gonzalez}, J.~J. 1992, \apj, 398, 69

\bibitem[{{Worthey} {et~al.}(1994){Worthey}, {Faber}, {Gonzalez}, \&
  {Burstein}}]{worthey94b}
{Worthey}, G., {Faber}, S.~M., {Gonzalez}, J.~J., \& {Burstein}, D. 1994, ApJS,
  94, 687

\bibitem[{Worthey \& Ottaviani(1997)}]{worthey97}
Worthey, G. \& Ottaviani, D.~L. 1997, ApJS, 111, 377

\bibitem[{Zhang {et~al.}(2014)Zhang, Han, Shuo, Siqi, Tong-Jie, \&
  Yan-Chun}]{zhang14}
Zhang, C., Han, Z., Shuo, Y., {et~al.} 2014, RAA, 14, 1221

\bibitem[{Zheng {et~al.}(2016)Zheng, Ding, Biesiada, Cao, \& Zhu}]{zheng16}
Zheng, X., Ding, X., Biesiada, M., Cao, S., \& Zhu, Z.-H. 2016, The
  Astrophysical Journal, 825, 17

\bibitem[{Zhou {et~al.}(2023)Zhou, Dey, Newman, Eisenstein, Dawson, Bailey,
  Berti, Guy, Lan, Zou, Aguilar, Ahlen, Alam, Brooks, de~la Macorra, Dey,
  Dhungana, Fanning, Font-Ribera, Gontcho, Honscheid, Ishak, Kisner, Kovács,
  Kremin, Landriau, Levi, Magneville, Manera, Martini, Meisner, Miquel,
  Moustakas, Myers, Nie, Palanque-Delabrouille, Percival, Poppett, Prada,
  Raichoor, Ross, Schlafly, Schlegel, Schubnell, Tarlé, Weaver, Wechsler,
  Yéche, \& Zhou}]{zhou23}
Zhou, R., Dey, B., Newman, J.~A., {et~al.} 2023, The Astronomical Journal, 165,
  58

\end{thebibliography}
\begin{appendix}

\section{Full derivation of the pivotal-redshift cosmographic expansion\label{app:theoretical_cosmography}}

In this appendix we derive the closed-form expression for $G(x;q_{z_0},j_{z_0})$ quoted in eq. (\ref{eq:Dtzz0}) by finding the relation between the $b_m$ (eq. \ref{eq:Dtbm}) and the $a_n$ (eq. \ref{eq:Han}). We begin by evaluating eq. (\ref{eq:qyj}) at $z=z_0$. This gives us two algebraic relations, one from $q_{z_0}$ and one from $j_{z_0}$, which we solve for $H'(z_0)$ and $H''(z_0)$ respectively:
\begin{equation}\begin{aligned}
    H'(z_0) &= \frac{(1+q_{z_0})\,H_{z_0}}{1+z_0} ,\\
    H''(z_0) &= \frac{\left(j_{z_0}-q_{z_0}^2\right)H_{z_0}}{(1+z_0)^2} .
\end{aligned}\end{equation}
These are exactly the two ingredients needed to build the Taylor expansion of $H(z)$ around $z_0$ introduced in eq. (\ref{eq:Han}). Substituting the two derivatives above,
\begin{equation}
    \begin{aligned}
        a_0 &= H_{z_0},\\ a_1 &= \frac{(1+q_{z_0})\,H_{z_0}}{1+z_0}, \\ a_2 &= \frac{\left(j_{z_0}-q_{z_0}^2\right)H_{z_0}}{2(1+z_0)^2} .
    \end{aligned}
\end{equation}
At this point $a_1$ and $a_2$ are still expressed with explicit powers of $(1+z_0)$ in the denominator, which is less convenient than the form adopted in Sec. \ref{sec:CC}. We therefore rescale the expansion using $x\equiv(z-z_0)/(1+z_0)$, so that every power $(z-z_0)^n$ becomes $(1+z_0)^n x^n$. Substituting into eq. (\ref{eq:Han}) and factoring out the leading term $H_{z_0}=a_0$,
\begin{equation}
    H(z) = \sum_{n=0}^{\infty} a_n (1+z_0)^n x^n = H_{z_0}\sum_{n=0}^{\infty}\tilde a_n\, x^n ,
\end{equation}
with
\begin{equation}
     \tilde a_n \equiv \frac{a_n(1+z_0)^n}{H_{z_0}},
\end{equation}
so that $\tilde a_1 = 1+q_{z_0}$ and $\tilde a_2 = (j_{z_0}-q_{z_0}^2)/2$, consistent with the explicit form of $H(x)$ already given in eq. (\ref{eq:Hexplicit}). We now turn to the integrand of eq. (\ref{eq:DtdezdeHz}), and derive its Taylor coefficients $b_m$, eq. (\ref{eq:fbm}), from the $\tilde a_n$ found above. The first step is to write $1+z$ itself in terms of $x$, so that we can write the inverse of the integrand in eq. (\ref{eq:DtdezdeHz}), $P(x) \equiv 1/f(x)$, as
\begin{equation}
    \begin{aligned}
    P(x) = \ &(1+z_0)(1+x)\,H_{z_0}\sum_n \tilde a_n x^n \\
    = \ &H_{z_0}(1+z_0)\Big[1+(1+\tilde a_1)x +(\tilde a_1+\tilde a_2)x^2+\mathcal{O}(x^3)\Big] .
    \end{aligned}
    \label{eq:Px}
\end{equation}
We only keep terms up to $\mathcal{O}(x^2)$, since this is all that is needed to reach the cubic order in $\Delta t$ once $f(z)$ is integrated. Denoting $u_0\equiv H_{z_0}(1+z_0)$, $c_1\equiv 1+\tilde a_1$ and $c_2\equiv\tilde a_1+\tilde a_2$, so that $P(x)=u_0(1+c_1x+c_2x^2+\dots)$, we obtain the Taylor coefficients of $f(x)$ directly by differentiating its inverse, $P$, and evaluating at $x=0$:
\begin{equation}
    \begin{aligned}
        f(0) &= \frac{1}{u_0}, \\
        \frac{\text{d}f}{\text{d}x}\bigg|_0 &= -\frac{P'(0)}{P^2(0)} = -\frac{c_1}{u_0},\\
        \frac{1}{2}\frac{\text{d}^2f}{\text{d}x^2}\bigg|_0 &= -\frac{P''(0)}{2P^2(0)}+\frac{[P'(0)]^2}{P^3(0)} = \frac{c_1^2-c_2}{u_0},
    \end{aligned}
\end{equation}
where we used $P'(0)=u_0c_1$ and $P''(0)=2u_0c_2$, both read off directly from eq. (\ref{eq:Px}). Identifying $\tilde b_m\equiv b_m(1+z_0)^m$, defined via $f(z)=\sum_m\tilde b_m x^m$, these three derivatives give
\begin{equation}\begin{aligned}
    \tilde b_0 &= \frac{1}{u_0},\\
    \tilde b_1 &= -\frac{c_1}{u_0},\\
    \tilde b_2 &= \frac{c_1^2-c_2}{u_0} .
\end{aligned}\end{equation}
Substituting $c_1=2+q_{z_0}$ and $c_2=1+q_{z_0}+(j_{z_0}-q_{z_0}^2)/2$, obtained from $\tilde a_1=1+q_{z_0}$ and $\tilde a_2=(j_{z_0}-q_{z_0}^2)/2$ found above,
\begin{equation}\begin{aligned}
    \tilde b_0 &= \frac{1}{u_0}, \\
    \tilde b_1 &= -\frac{2+q_{z_0}}{u_0} ,\\
    \tilde b_2 &= \frac{3+3q_{z_0}+\frac{3}{2}q_{z_0}^2-\frac{1}{2}j_{z_0}}{u_0} .
\end{aligned}\end{equation}

Finally, we integrate $f(z)=P^{-1}(z)$, after performing a variable change $z \to x$, to find $\Delta t(z, z_0) \equiv t(z_0) - t(z)$,
\begin{equation}
\begin{aligned}
    \Delta t(z,z_0) &= \int_{z_0}^z f(z')\,\text{d}z' \\&= (1+z_0)\int_0^x \sum_m \tilde b_m x'^m\,\text{d}x'\\
    &= (1+z_0)\sum_{m}\frac{\tilde{b}_m}{m+1}x^{m+1}.
\end{aligned}
\end{equation} Since $u_0=H_{z_0}(1+z_0)$, the factor $(1+z_0)$ in the previous equation cancels out with that in the $\tilde{b}_m$, and multiplying $\Delta t$ by $H_{z_0}$ recovers eq. (\ref{eq:Dtzz0}):

\begin{equation}
    \begin{aligned}
    G(x;q_{z_0},j_{z_0}) = \ &x - \left(1+\frac{q_{z_0}}{2}\right)x^2 \\
    &+\left(1+q_{z_0}+\frac{1}{2}q_{z_0}^2-\frac{1}{6}j_{z_0}\right)x^3+\mathcal{O}(x^4) .
    \end{aligned}
\end{equation}

\section{Correction of the velocity dispersion effect}
\label{sec:app_VD}

In this appendix, we present the velocity dispersion corrections used in the data treatment. The correction functions of \citetalias{alvarez25} included the degradation of the instrumental resolution to $\sigma_\text{RMS} = 1.50$ \AA, so that the observed indices, $I_{v, \sigma_\text{IR} = 1.5\text{ \AA}}$, were translated to zero velocity dispersion and to the MILES instrumental resolution ($1.06$ \AA). This is convenient for SDSS spectra, but DESI spectra have a better, although less stable, resolution than the MILES library (from $\sigma_\text{RMS} \approx 0.5$ to $\approx 0.9$ \AA, mainly depending on the arm of the spectrograph), so it is enough to degrade all the observed spectra to $\sigma_\text{IR} = 1.06$ \AA.

\begin{figure*}
    \centering
    \includegraphics[width=\textwidth]{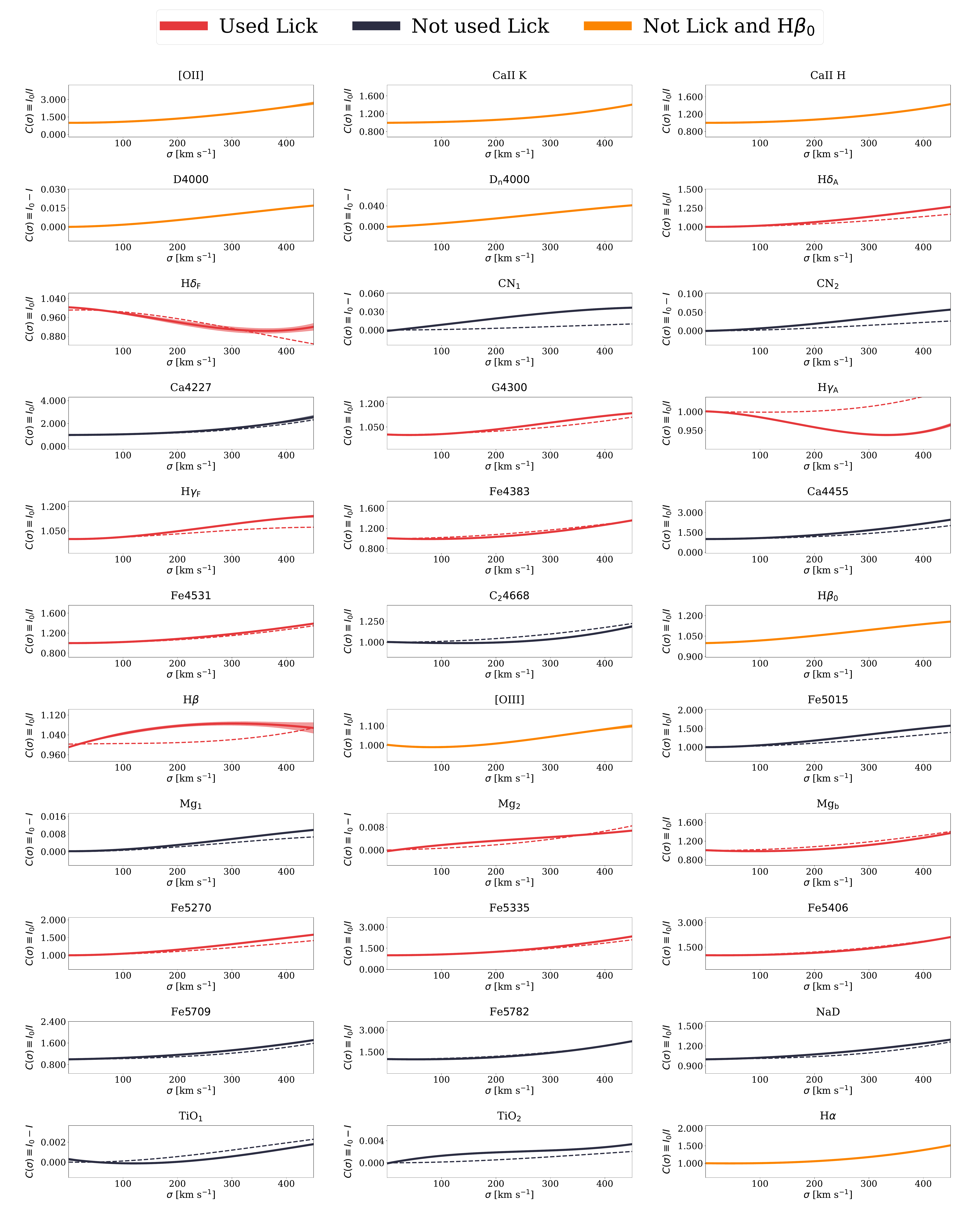}
    \caption{Velocity dispersion correction functions, $C_I(\sigma)$, for the $33$ measured features: Lick indices included in the SPS fit (red), Lick indices not included (dark blue) and the $8$ additional features (orange). The shaded regions show the $1\sigma$ dispersion of the MILES stars. Dashed lines show the corrections of \protect\cite{carson10}.}
    \label{fig:vdcorrs}
\end{figure*}

To make these corrections useful for future works and for spectra with a resolution better than that of MILES, we do not include the effect of the instrumental resolution in them, in the fashion of \citet{carson10} and \citet{clerici24}. Since the MILES v9.1 spectra \citep{falconbarroso11} continuously cover the range from $3525$ \AA \ to $7500$ \AA, we computed correction functions for all the $33$ measured features, although some of them are expected to be negligible ($\sim 1$ for atomic indices, $\sim 0$ for molecular and other indices) or nearly so (e.g. emission features not observed in the spectra of passive galaxies, or features insensitive to velocity dispersion, such as D$_\text{(n)}4000$). In practice, we repeatedly smooth the MILES spectra to increasing velocity dispersions (from $0$ to $450$ km s$^{-1}$), measure the indices at each step, and compute their mean and standard deviation over the MILES stars. We then fit the usual third-order polynomial,\begin{equation}
    I_0\ \square\ I(\sigma) \equiv C_I(\sigma) = a_0 + a_1\sigma + a_2 \sigma^2 + a_3\sigma^3
    \label{eq:Csigma},
\end{equation}where $\square$ represents a division for atomic indices and a subtraction for molecular indices, with the {\fontfamily{qcr}\selectfont scipy.curve\_fit} routine. The corrected index is therefore $I_0 = C_I(\sigma)\,I(\sigma)$ for atomic indices and $I_0 = I(\sigma) + C_I(\sigma)$ for molecular ones, where $\sigma$ is the velocity dispersion of the galaxy or stack.

\begin{table}[H]
    \centering
    \captionsetup[table]{skip=10pt}
    \caption{Coefficients of the correction function for each spectral feature.}
    \label{tab:Cfunctions}

    \resizebox{0.45\textwidth}{!}{%
    \tiny
    \begin{tabular}{|l|cccc|}
    \hline
    Index & $a_0$ & $a_1$ $(\times 10^{-3})$ & $a_2$ $(\times 10^{-6})$ & $a_3$ $(\times 10^{-9})$ \\ \hline
    O$\text{II}      $ & 0.998 & -0.064 & 10.154 & -3.786 \\
    CaII$\text{ K}   $ & 0.998 &  0.103 &  0.610 &  2.598 \\
    CaII$\text{ H}   $ & 0.999 & -0.001 &  1.599 &  1.167 \\
    D$4000           $ & 0.000 &  0.005 &  0.139 & -0.147 \\
    D$_\text{n}4000  $ &-0.001 &  0.051 &  0.200 & -0.238 \\
    H$\delta_\text{A}$ & 1.000 & -0.026 &  1.970 & -1.324 \\
    H$\delta_\text{F}$ & 1.003 & -0.127 & -1.595 &  3.262 \\
    CN$_1            $ &-0.001 &  0.095 &  0.089 & -0.245 \\
    CN$_2            $ & 0.000 &  0.033 &  0.402 & -0.426 \\
    Ca$4227          $ & 0.995 &  0.249 &  2.644 & 10.311 \\
    G$4300           $ & 1.002 & -0.154 &  2.049 & -2.290 \\
    H$\gamma_\text{A}$ & 1.001 & -0.042 & -1.458 &  3.056 \\
    H$\gamma_\text{F}$ & 1.000 & -0.033 &  1.845 & -2.392 \\
    Fe$4383          $ & 1.007 & -0.446 &  3.059 & -0.722 \\
    Ca$4455          $ & 0.999 & -0.037 &  7.939 & -1.469 \\
    Fe$4531          $ & 1.000 & -0.007 &  2.286 & -0.754 \\
    C$_2 4668        $ & 1.003 & -0.174 &  0.223 &  2.401 \\
    H$\beta_0        $ & 0.999 &  0.086 &  1.182 & -1.323 \\
    H$\beta          $ & 0.991 &  0.676 & -1.346 &  0.501 \\
    O$\text{III}     $ & 1.002 & -0.310 &  2.144 & -2.165 \\
    Fe$5015          $ & 0.999 &  0.017 &  5.589 & -6.137 \\
    Mg$_1            $ & 0.000 &  0.000 &  0.093 & -0.098 \\
    Mg$_2            $ & 0.000 &  0.026 & -0.053 &  0.067 \\
    Mg$_\text{b}     $ & 1.007 & -0.496 &  2.785 &  0.275 \\
    Fe$5270          $ & 0.999 &  0.040 &  4.521 & -3.836 \\
    Fe$5335          $ & 0.999 &  0.029 &  5.586 &  2.173 \\
    Fe$5406          $ & 1.002 & -0.216 &  3.796 &  4.855 \\
    Fe$5709          $ & 0.996 &  0.304 &  2.449 &  0.923 \\
    Fe$5782          $ & 1.007 & -0.622 &  5.548 &  4.299 \\
    NaD$             $ & 0.998 &  0.116 &  1.331 & -0.280 \\
    TiO$_1           $ & 0.000 & -0.007 &  0.035 & -0.026 \\
    TiO$_2           $ & 0.000 &  0.019 & -0.064 &  0.085 \\
    H$\alpha         $ & 1.002 & -0.181 &  1.862 &  2.375 \\\hline
    \end{tabular}}
\end{table}

The correction functions are shown in Fig. \ref{fig:vdcorrs}, with different colours for the Lick indices used in the SPS fit (red), those not used (dark blue) and the $8$ additional spectral features (orange), and the coefficients of the polynomial of eq. (\ref{eq:Csigma}) are given in Table \ref{tab:Cfunctions}.

\section{Pathologies of the SPS analysis\label{sec:app_TMJ}}

In this appendix, we present some features of the TMJ models that were taken into account before analysing the data, in order to avoid spurious biases. First, we show that some model index strengths at low ages depart from the trend followed by the galaxy data when redshifts are translated into ages with a typical $\Lambda$CDM cosmology and an age scaling relation. Then, we perform a test SPS fit with the full metallicity range allowed by the model, which reaches almost five times the solar metallicity, and show how the age--metallicity degeneracy naturally arises \citep{worthey94a, cahvez11, barone18, borghi22a} when extremely high metallicities are allowed.

\begin{figure}[H]
	\includegraphics[width=\columnwidth]{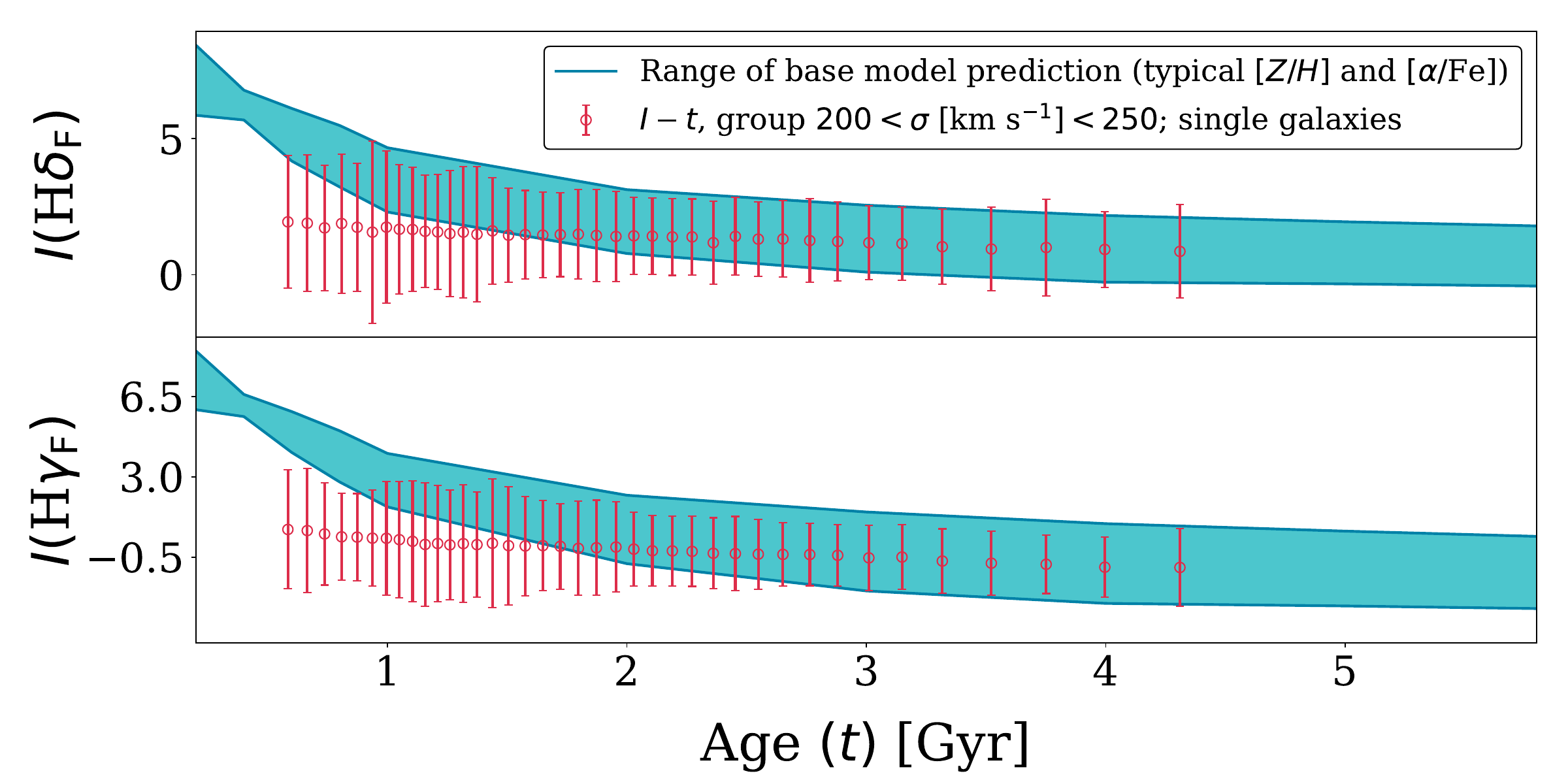}
    \caption{Index strength--age relations for H$\delta_\text{F}$ and H$\gamma_\text{F}$. The blue band shows the region of the parameter space sampled by the base model for the typical metallicities and $\alpha$-enhancements of ETGs. The data points are moving means over $5000$ galaxies each, taken from the galaxies with $200 \leq \sigma\ [\text{km s}^{-1}] < 250$.}
    \label{fig:balmersteep}
\end{figure}

Figure \ref{fig:balmersteep} shows the steep $I-t$ relation of the model at low ages ($t\lesssim 2$ Gyr) for two Balmer lines, H$\delta$ and H$\gamma$. The blue band shows the region of the parameter space sampled by the base grid of the model for the metallicities ($[Z/H]>-0.1$) and $\alpha$-element enhancements ($[\alpha/\text{Fe}]> 0.0$) typical of ETGs. The data points shown correspond to a moving average of $5000$ galaxies from the spectroscopically selected sample without imposing the archaeological coherence cut in redshift (see Sec. \ref{sec:archaeo}), sampling in this way high redshifts (low ages), to show the disagreement with the model. The expected age of each galaxy is obtained from its redshift and velocity dispersion by combining the \citetalias{alvarez25} age--velocity dispersion relation, $t(\sigma)$, with a flat $\Lambda$CDM cosmology.

\begin{figure}
	\includegraphics[width=\columnwidth]{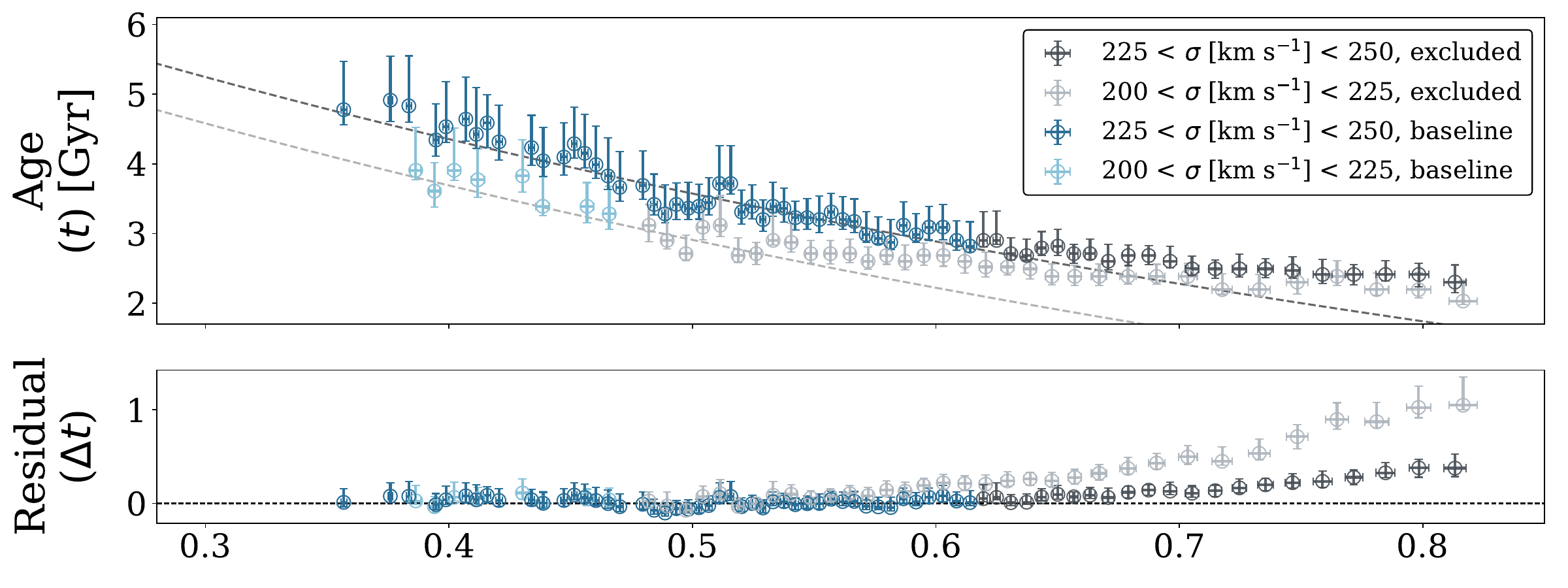}
    \caption{Blind test of the archaeological coherence cut for the two lowest velocity dispersion groups ($200<\sigma \ [\text{km s}^{-1}]<225$ and $225<\sigma \ [\text{km s}^{-1}]<250$). Blue tones show the baseline for these two groups, and grey tones the TMJ ages of the dataset without the cut. Upper panel: $t-z$ relations; lower panel: departure from the ages predicted with a $\Lambda$CDM cosmology.}
    \label{fig:quepasaahi}
\end{figure}

The observed indices follow a much softer trend than the model ones, which grow much faster towards lower ages (from right to left in the figure, i.e. towards higher redshifts). A cut at $2.5$ Gyr is conservative, but still retains a good number of sources in the final sample. This effect is only relevant for the lowest velocity dispersion groups (see Table \ref{tab:archaeologycuttabla}), since for the higher velocity dispersion groups the redshift limit imposed by the observed wavelength window, which must include all the indices used in the SPS fit, is reached before their ages fall below $\sim3$ Gyr. This is illustrated in Fig. \ref{fig:quepasaahi}, which shows a blind test without the archaeological coherence cut for the two lowest velocity dispersion groups, $200<\sigma \ [\text{km s}^{-1}]<225$ and $225<\sigma \ [\text{km s}^{-1}]<250$, reproducing their $t-z$ relations of Fig. \ref{fig:tzmain}. The baseline is shown with the same colours as in Fig. \ref{fig:tzmain}, and the blind test in grey.

The upper panel shows that the ages of the blind test converge to $\sim2.5$ Gyr at high redshift, and that the archaeological coherence cut does not introduce any cosmological bias within the redshift window. Indeed, since the galaxies are stacked in redshift order (Sec. \ref{sec:archaeo}), the stacks of the blind and cut samples are identical up to the redshift cut, except perhaps for the single stack at the limit, which in any case was typically excluded from the baseline because it did not reach the required spectral S/N (low-S/N points in Fig. \ref{fig:snnz}). The lower panel shows the departure from the ages predicted with a $\Lambda$CDM cosmology. This departure is larger for the lowest velocity dispersion group, which supports the idea that the bias arises because the TMJ models do not reproduce the observed indices of galaxies younger than $\sim2.5$ Gyr, at least for the set of indices used in the SPS fit.

This effect could be avoided if all ages were globally shifted upwards, an effect discussed by \cite{borghi22b} as a consequence of the set of Lick indices used in the fit. Such a shift would not modify the cosmological evolution of the ages as long as the average metallicity of each velocity dispersion group remains constant, which is theoretically sound: good CCs drawn from a statistically representative population at different cosmic times should show no significant metallicity evolution with redshift. Different ways of applying SPS models (e.g. FSF) also produce global age offsets while leaving the differential age relations unchanged; for instance, \cite{jiao22} found global variations of $\sim 0.6$ Gyr, similar to those obtained by simply changing the set of Lick indices \citep{borghi22b}.

\begin{figure}
	\includegraphics[width=\columnwidth]{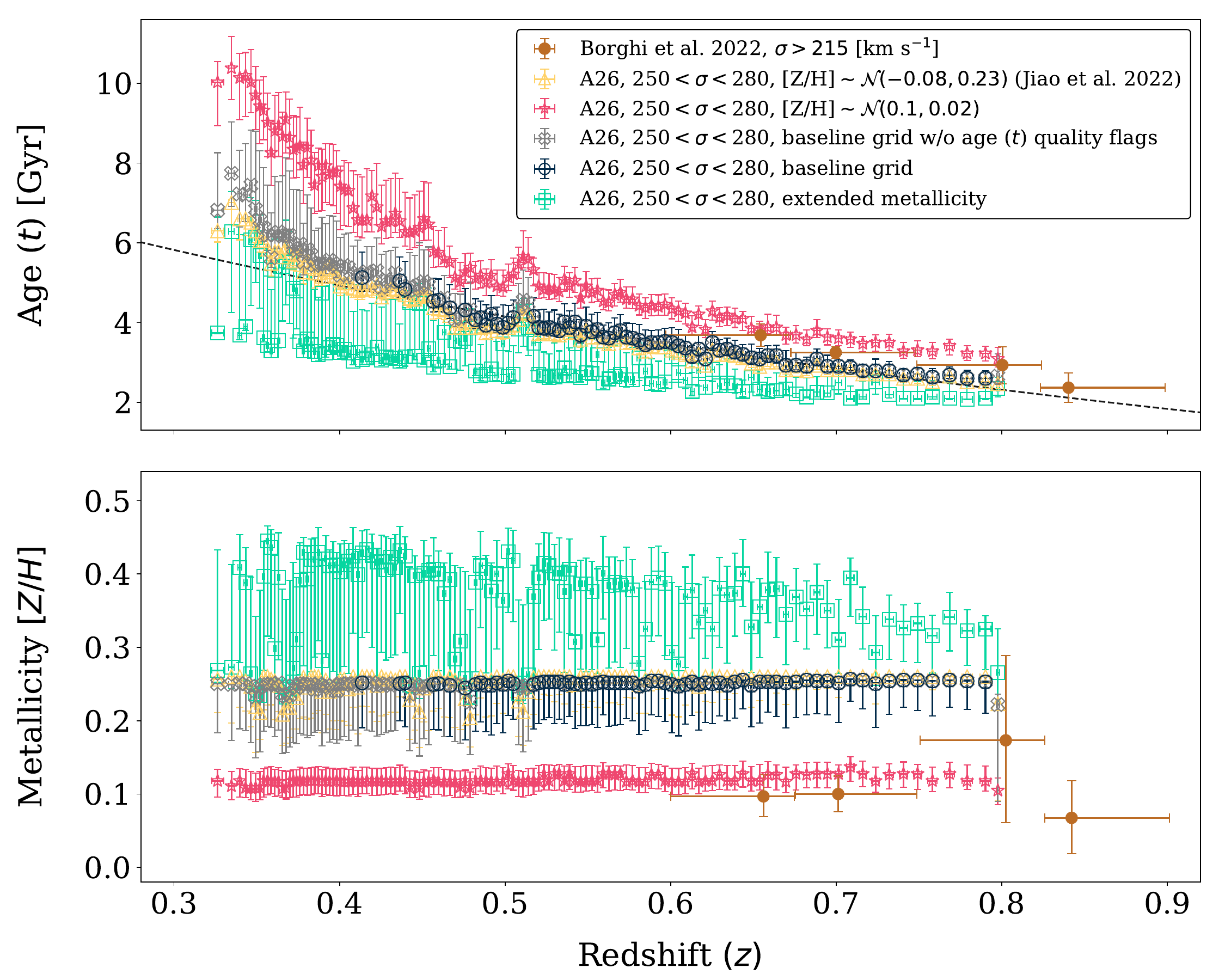}
    \caption{$t-z$ and $\text{[Z/H]}-z$ diagrams for the velocity dispersion group $250 < \sigma\text{ [km s}^{-1}] < 280$. The dark blue trend is the same as in Fig. \ref{fig:tzmain}, while the green trend shows the full data (no quality flags applied) for a free metallicity parameter, $-0.33<\text{[Z/H]}<0.67$. The high velocity dispersion data of \protect\cite{borghi22a} are shown for comparison.}
    \label{fig:extendedmetallicity}
\end{figure}

These changes in global age are also related to changes in the estimated metallicity, owing to the well-known age--metallicity degeneracy \citep{renzini06, conroy13}. In \citetalias{alvarez25}, we restricted the metallicity prior of the TMJ model to the limit of the \cite{knowles23} model, so that metallicities above $[\text{Z}/\text{H}] = 0.26$ ($\sim 1.8\,Z_\odot$) were not allowed. When this restriction is lifted, the smooth $t-z$ relations of Fig. \ref{fig:tzmain} are lost, as shown in Fig. \ref{fig:extendedmetallicity}, where the unrestricted case is represented by green data points. Some of the stacks still agree with the $[\text{Z}/\text{H}] < 0.26$ baseline, which suggests that our restriction selects the more sensible of the two modes produced by the age--metallicity degeneracy.

In addition to the baseline grid with (dark blue) and without (grey) age ($t$) quality flags, and the full parameter space SPS fit with the TMJ model (green), we include two further fits in which Gaussian priors are imposed on the metallicity, together with the estimates for the high velocity dispersion group of \cite{borghi22a} (brown dots). In yellow, we show the case in which the Gaussian prior is centred on the average metallicity obtained by applying the FSF method to the \cite{borghi22a} sample of CCs \citep{jiao22}. We find that a prior width based on the average dispersion reported by \cite{jiao22} is not sufficient to remove the skewness of the metallicity posteriors of our galaxy stacks. The second Gaussian prior, centred at $\text{[Z/H]}=0.1$ and with a narrow width, was chosen to obtain a metallicity posterior consistent with the value observed by \cite{borghi22a} for the high-velocity-dispersion group. Overall, this shifts the estimated ages upwards by $\sim 0.5$ Gyr at high redshift, whereas at low redshift the discrepancy increases to $\sim 2$ Gyr. This reinforces the idea that the TMJ model responds differently in different mass regimes (see also the discussion of Fig. \ref{fig:fulltz}).

We decided not to include a dedicated estimate of the bias that the restriction of the metallicity parameter space introduces in the overall $H(z)$ reconstruction, since it is not clear to us which of the estimates, if any, should be discarded outright. The least arbitrary estimate, shown by the green trend, has no restrictions on the metallicity parameter space and its posteriors close freely within the prior range. It indicates that some points should be discarded because of the irregular and unphysical evolution of their metallicities. We believe that the low-metallicity mode produced by the age--metallicity degeneracy is the \textit{preferred} one, but there is no clear way to impose it other than the baseline restriction, $[\text{Z}/\text{H}] < 0.26$.

\section{Full $t-z$ relations\label{sec:app_fulltz}}

\begin{figure}[H]
	\includegraphics[width=\columnwidth]{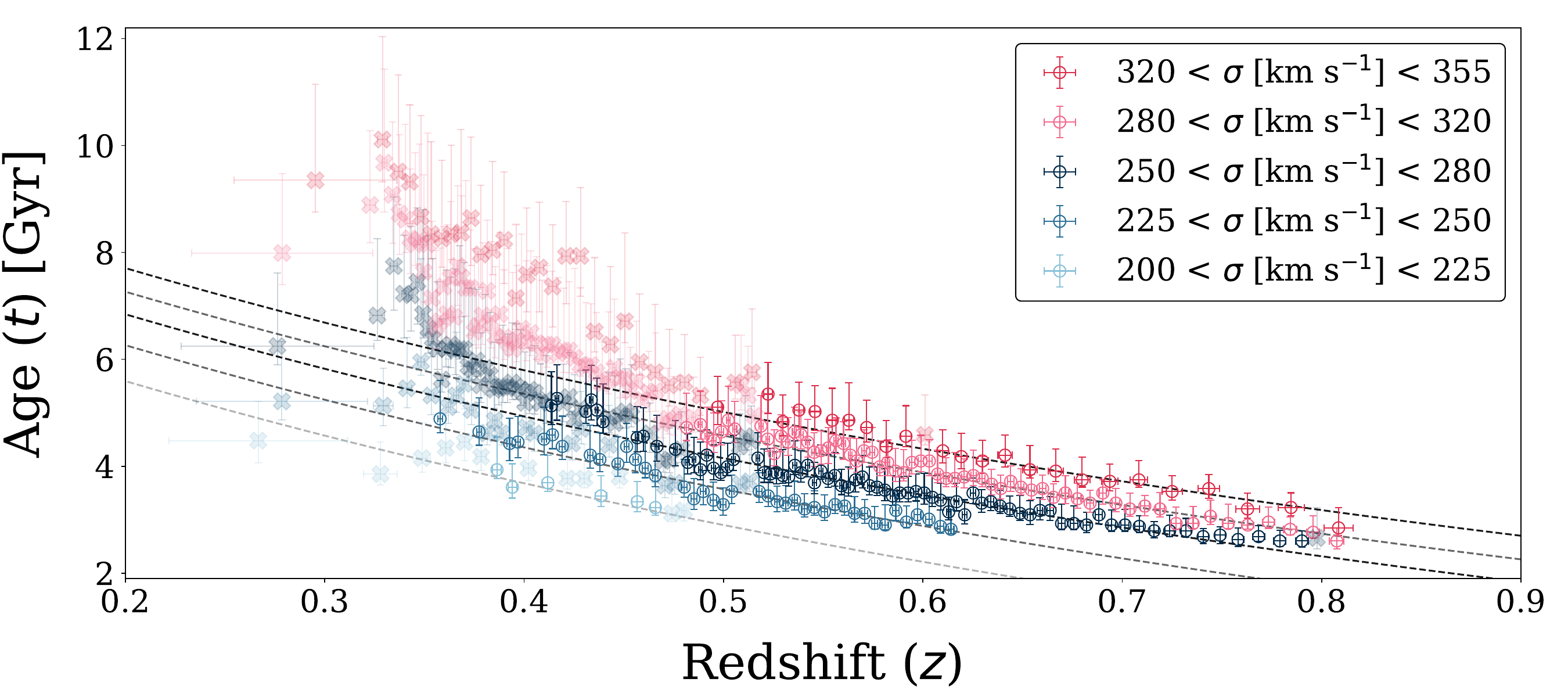}
    \caption{Full $t-z$ diagram, including the stacks discarded by the quality flags (Sec. \ref{sec:QFs}), which are shown as crosses in a lighter colour for each velocity dispersion group.}
    \label{fig:fulltz}
\end{figure}

After the SPS fit, we applied quality flags in redshift, so that only stacks of physically similar sources are retained, and in age, including a cut in the S/N of the ages that avoids local, redshift-concentrated biases such as those presented in this appendix. Figure \ref{fig:fulltz} shows the full $t-z$ relations, including the discarded stacks, which are shown as crosses in a lighter colour for each group.

The low-redshift end of each massive group ($\sigma > 250 \text{ km s}^{-1}$) shows a steep slope. These stacks have much larger age uncertainties than those at higher redshift, so the cut $\text{S}/\text{N}(t) > 5.0$, applied uniformly to all the stacks, removes only them. If data with large uncertainties were randomly distributed in redshift, their effect on the cosmographic fit would be minor, but their concentration at one end of the redshift distribution biases the cosmographic parameters, in particular $H_{z_0}$ (if measured locally) and $q_{z_0}$.

\begin{figure}[H]
	\includegraphics[width=\columnwidth]{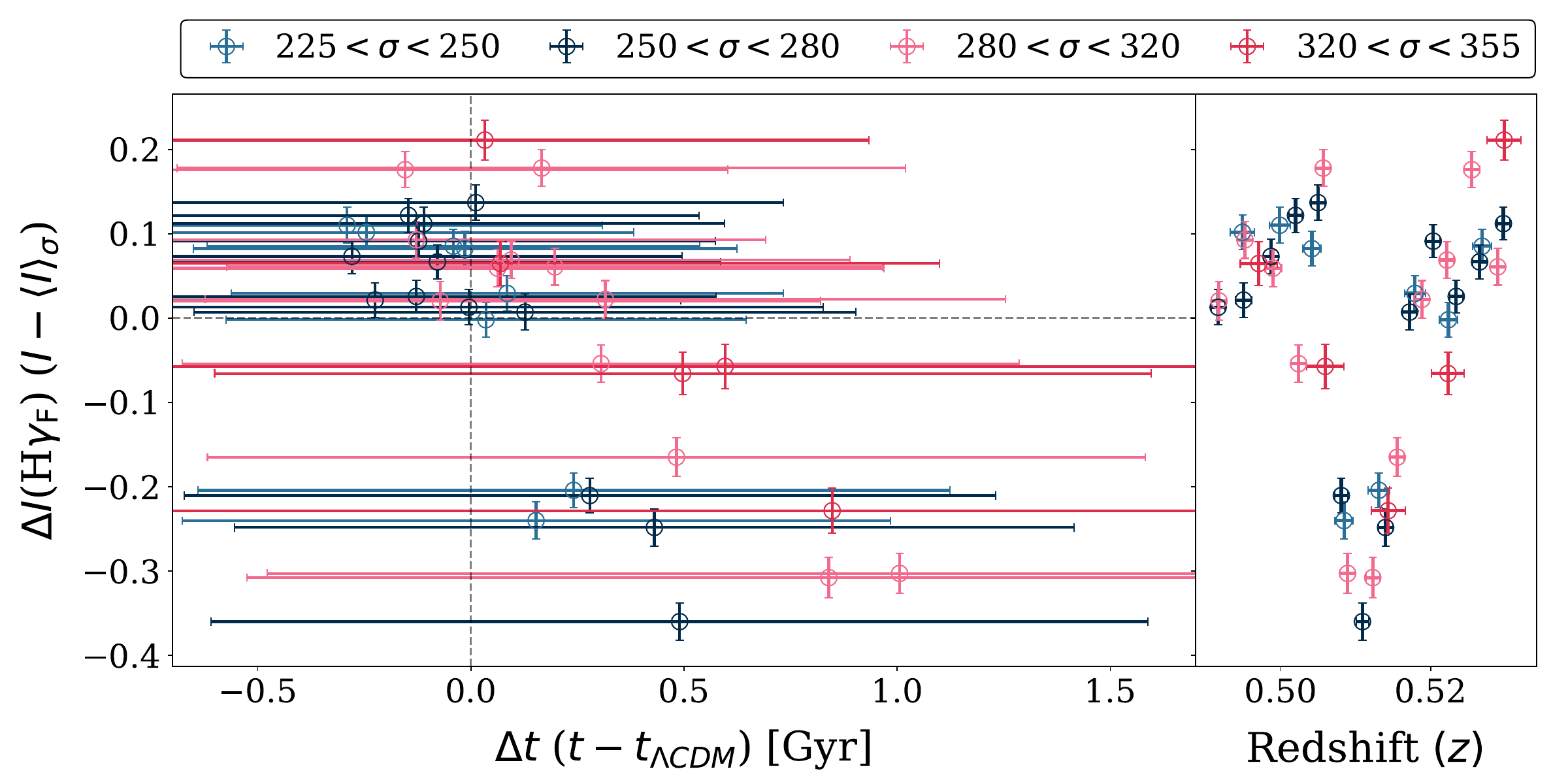}
    \caption{Relation between the $t-z$ oscillation at $z\sim 0.5$ and the H$\gamma_\text{F}$ index. Left: change in the index strength with respect to its average value in the window against the deviation of the estimated age from the $\Lambda$CDM prediction, for each velocity dispersion group. Right: the same change in the index strength against redshift.}
    \label{fig:HgF}
\end{figure}

\begin{figure}
    \centering
    \includegraphics[width=\linewidth]{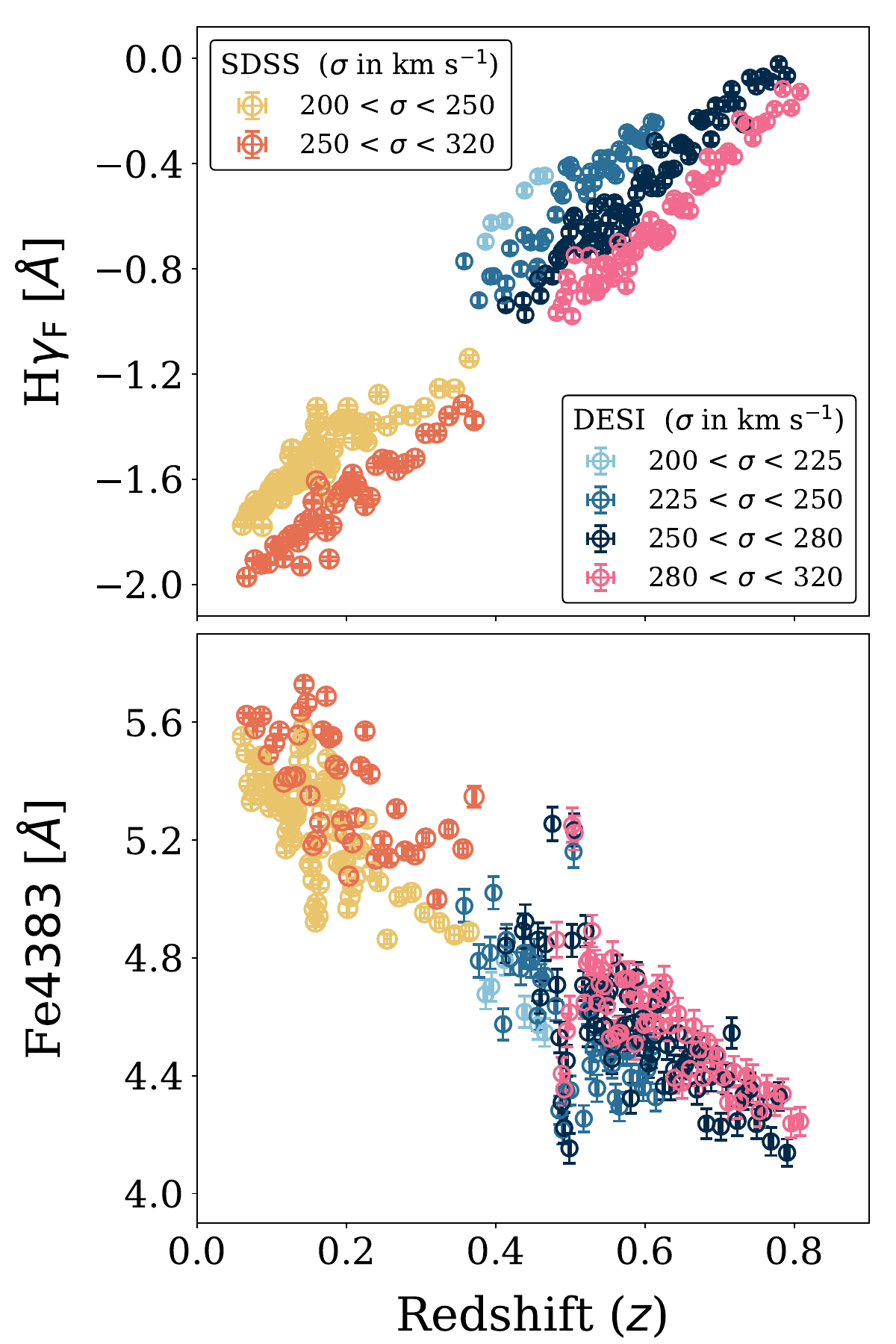}
    \caption{Redshift evolution of the H$\gamma_\text{F}$ (top) and Fe$4383$ (bottom) indices measured on the SDSS stacks of \protect\citetalias{alvarez25} (warm colours, $z\lesssim0.37$) and on the DESI stacks of this work (blue and pink tones, $z\gtrsim0.36$), corrected to zero velocity dispersion and MILES resolution, for the velocity dispersion range common to both works ($200<\sigma\ [\text{km s}^{-1}]<320$). The groups of \protect\citetalias{alvarez25} are twice as wide as ours, so each SDSS group should be compared with the two corresponding DESI groups. DESI stacks excluded from the baseline by the quality flags (Sec.~\ref{sec:QFs}) are also shown, so that the oscillation at $z\approx0.5$ is visible.}
    \label{fig:indicesSDSSDESI}
\end{figure}

An unexpected side effect of the quality flags is the complete removal of the stacks in the short redshift window $0.49\lesssim z\lesssim 0.53$. The behaviour there is similar to that presented at the end of the results of \citetalias{alvarez25} and in appendix B of \cite{moresco16a}. As noted in \citetalias{alvarez25}, the oscillation in the $t-z$ relations is tightly related to the behaviour of a few indices. The index that revealed this oscillating pattern in the SDSS dataset, H$\gamma_\text{F}$, shows the same pattern here (Fig. \ref{fig:HgF}). We observed similar behaviours for other indices, such as H$\delta_\text{F}$ and, most prominently, Fe$4383$ (Fig.~\ref{fig:indicesSDSSDESI}).

The oscillation is accounted for in the systematic uncertainty through the cases in which the quality flags are removed. In addition, we performed a test fit including all the stacks at $z>0.49$, regardless of whether they passed the cuts: the cosmographic fit was essentially insensitive to this local oscillation, and the posteriors remained virtually unchanged with respect to the baseline.

Finally, we checked the consistency of our index measurements with those of \citetalias{alvarez25}. Figure~\ref{fig:indicesSDSSDESI} shows the redshift evolution of H$\gamma_\text{F}$ and Fe$4383$ measured on the SDSS stacks of \citetalias{alvarez25} and on our DESI stacks. Note that the colours correspond to those used in the $t-z$ figures in \citetalias{alvarez25} (figure 6) and this work (Fig. \ref{fig:tzmain}). We chose H$\gamma_\text{F}$ and Fe$4383$ because, as shown in section 5.2 of \citetalias{alvarez25}, their variations correlate most strongly with changes in the estimated ages. The two surveys sample almost contiguous redshift ranges, which join at $z\approx0.36-0.37$. Taking into account that each SDSS group combines two of our DESI groups, the indices show good continuity across the two datasets: the ordering of the groups with velocity dispersion is preserved (weaker H$\gamma_\text{F}$ and stronger Fe$4383$ for higher $\sigma$), and the redshift gradients are similar on both sides. Fe$4383$ connects smoothly, whereas for H$\gamma_\text{F}$ the $200<\sigma\ [\text{km s}^{-1}]<250$ SDSS group lies slightly below what would be needed to match the DESI stacks perfectly; we recall that all these indices are corrected for velocity dispersion and taken to the MILES resolution. The figure also shows that the oscillation at $z\approx0.5$ is already present in the observed indices. It is particularly evident in Fe$4383$, which departs from its mean trend by up to $\sim0.4-0.5$ \AA\ (comparable to the total variation of the index over $0.4<z<0.8$), whereas in H$\gamma_\text{F}$ the deviations are much smaller relative to its redshift trend, but still perceptible (see also Fig. \ref{fig:HgF}).

\section{Age scaling relation at $z \approx 0.57$ \label{sec:app_scalrel}}

Age scaling relations (also known as stellar archaeology) are usually analysed separately, using galaxy data from the local Universe. Although we do not intend to measure these relations in this work, the pivotal-redshift ages can be used to obtain an indirect measurement of the age scaling relation, which traces the progression in the formation of the stellar populations of ETGs. Indeed, the parameters of the joint cosmographic fit of Sec. \ref{sec:jointcos} include the ages of the velocity dispersion groups at the pivotal redshift. A simple inspection of these ages supports the downsizing scenario for the formation of ETGs, and our data provide yet another independent measurement of how galaxies were assembled, as seen by an observer at $z \approx 0.57$.

\begin{figure}
	\includegraphics[width=\columnwidth]{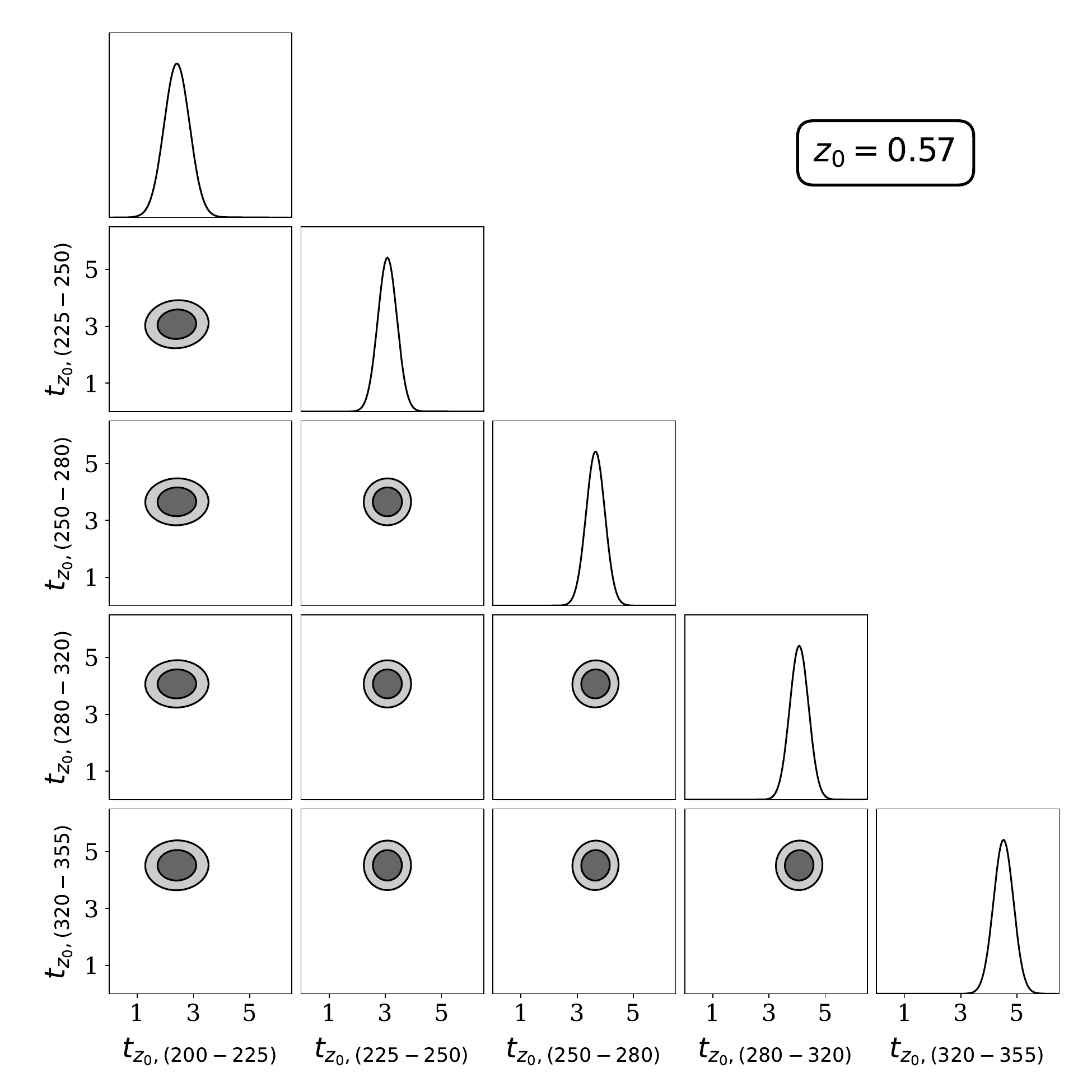}
    \caption{Posterior probability distribution for the pivotal redshift ages derived from the full joint fit of all velocity dispersion groups shown in Fig. \ref{fig:tzmain}.}
    \label{fig:posteriorsjointage}
\end{figure}

Figure \ref{fig:posteriorsjointage} shows the rest of the posterior distribution of the parameters of the cosmographic fit (eq. \ref{eq:tvdez}), whose part corresponding to the kinematic parameters, $\left\{H_{z_0}, q_{z_0}, j_{z_0}\right\}$, is shown in Fig. \ref{fig:posteriorsjoint}. We do not show the corner plots that cross $\left\{H_{z_0}, q_{z_0}, j_{z_0}\right\}$ with the $\{t_{z_0; v}\}$, since they show no correlation, which reinforces the idea that the scaling relations and the cosmological expansion are separable. The full MCMC chain of the cosmographic fit, including all its parameters, is publicly available (see Data availability).

\begin{figure}[H]
	\centering\includegraphics[width=\columnwidth]{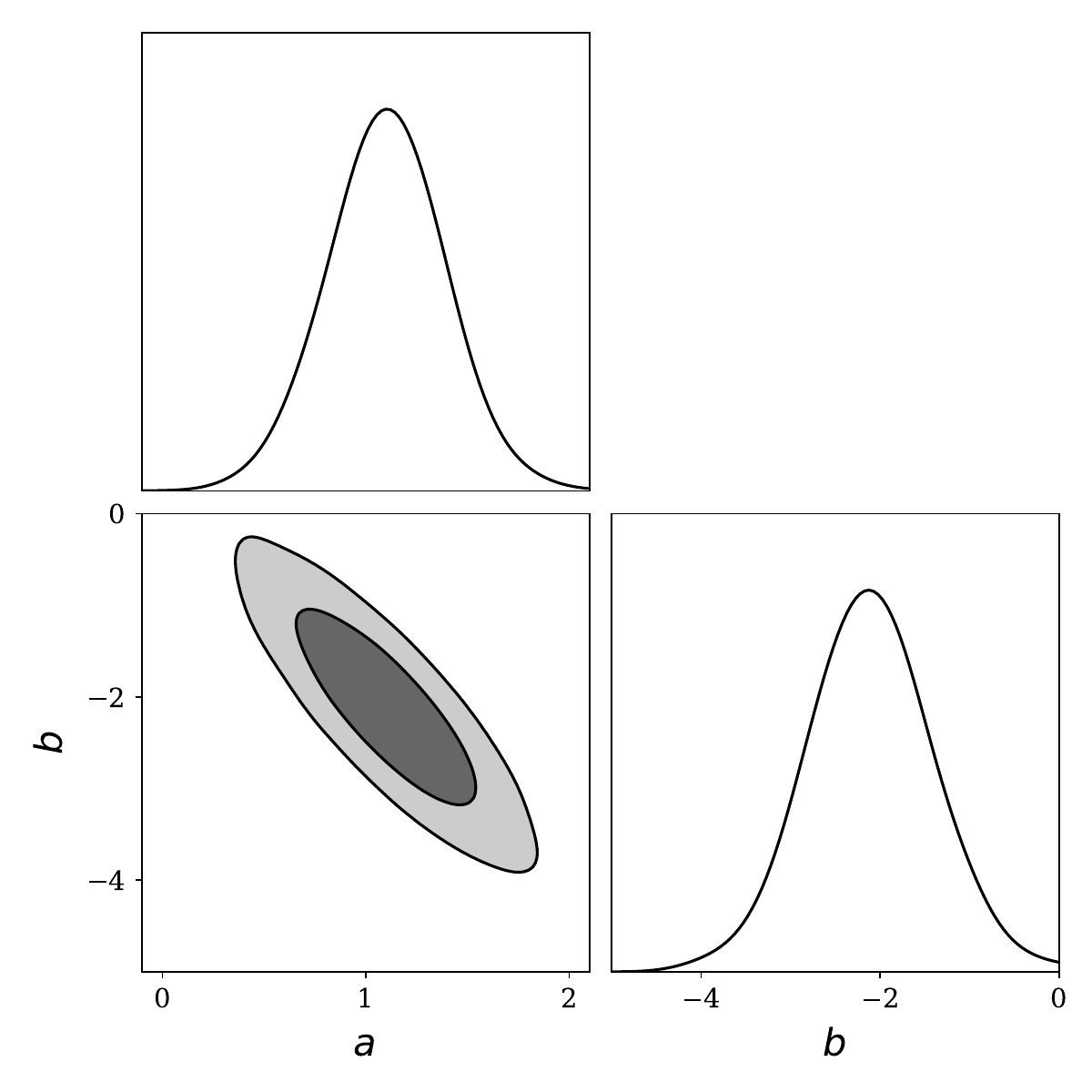}
    \caption{Posterior probability distribution for the age scaling relation parameters $a$ and $b$.}
    \label{fig:posteriorsab}
\end{figure}

We estimate the parameters of the $t-\sigma$ relation, assumed to be linear in the logarithm of both quantities, from the pivotal-redshift ages and their statistical uncertainties, by maximising the log-likelihood
\begin{equation}
    \log \mathcal{L} = -\sum_{v} \frac{1}{2}\left(\frac{t_{z_0; v} - \hat{t}_{z_0}(\sigma_{v}; a, b)}{\delta t_{z_0; v}}\right)^2,
    \label{eq:loglike_tscalrel}
\end{equation}
where $t_{z_0; v}$ and $\delta t_{z_0; v}$ are the pivotal-redshift ages and their uncertainties, and the model $\hat{t}(\sigma_v)$ takes the form
\begin{equation}
    t(\sigma) = 10^{a\cdot \log(\sigma) + b},
    \label{eq:funfit_tscalrel}
\end{equation}
which is customary in the literature \citep{clemens06, clemens09, thomas10}. We find well-closed posteriors for the parameters $\{a, b\}$ with uniform priors $a\in[0, 2]$ and $b\in[-5, 0]$, as shown in Fig. \ref{fig:posteriorsab}.

If we only take into account the statistical uncertainties, we can read $\log t =-2.1^{+0.7}_{-0.6} + (1.1^{+0.3}_{-0.3}) \log \sigma$ as the age scaling relation from the marginal posterior distribution of $a$ and $b$. The slope agrees with previous measurements \citep{thomas10, johansson12, alvarez25}, which is the most relevant result from this appendix. Indeed, both the tightness of the contours and the good match of the progression of ages with formed stellar mass, given by the slope, $a$, support the downsizing scenario at intermediate redshifts. This is further evidence that the underlying physics of galaxy assembly, following the downsizing scenario, remains consistent across different cosmic epochs. On the other hand, we see a slight shift toward a smaller intercept, $b$, compared to previous studies. In the context of eq. (\ref{eq:funfit_tscalrel}), this represents a global factor in the $t-\sigma$ relation, which physically means a smaller separation between the formation times of different mass groups. In other words, the smaller value found here for $b$ suggests that, given our estimated average ages for the different velocity dispersion groups at the pivotal redshift ($z_0\approx0.57$) in the joint cosmographic case (Sec. \ref{sec:jointcos}), ETGs appear more compressed in age range than what might be extrapolated from local universe scaling relations.

Regarding the statistical uncertainties, our credible intervals are significantly larger than those of previous works, as a direct consequence of extracting these values indirectly from the cosmographic fit. This reflects the trade-off between measuring universal expansion and individual galaxy evolution.

\end{appendix}

\end{document}